\documentclass[fleqn,usenatbib]{mnras}

\usepackage[T1]{fontenc}

\DeclareRobustCommand{\VAN}[3]{#2}
\let\VANthebibliography\thebibliography
\def\thebibliography{\DeclareRobustCommand{\VAN}[3]{##3}\VANthebibliography}

\usepackage{graphicx}	% Including figure files
\usepackage{amsmath}	% Advanced maths commands
\usepackage{amssymb}	% Extra maths symbols

\usepackage{tabularx}

\newcommand{\bs}{\boldsymbol}
\newcommand{\percent}{\,\mathrm{per\,cent}}
\newcommand{\feh}{[\mathrm{Fe}/\mathrm{H}]}

\usepackage[table]{xcolor}
\definecolor{lightgray}{gray}{0.9}
\definecolor{darkred}{rgb}{0.76, 0.23, 0.13}

\usepackage{newtxtext,newtxmath}
\title[Mira variable period--luminosity relations]{The period--luminosity relation for Mira variables in the Milky Way using Gaia DR3: a further distance anchor for $H_0$}

\author[J.~L. Sanders]{Jason L. Sanders\thanks{jason.sanders@ucl.ac.uk}
\\
Department of Physics and Astronomy, University College London, London WC1E 6BT, UK\\
}

\date{Accepted XXX. Received YYY; in original form ZZZ}
\pubyear{2021}

\begin{document}
\label{firstpage}
\pagerange{\pageref{firstpage}--\pageref{lastpage}}
\maketitle

% Abstract of the paper
\begin{abstract}
Gaia DR3 parallaxes are used to calibrate preliminary period--luminosity relations of O-rich Mira variables in the 2MASS $J$, $H$ and $K_s$ bands using a probabilistic model accounting for variations in the parallax zeropoint and underestimation of the parallax uncertainties.
The derived relations are compared to those measured for the Large and Small Magellanic Clouds, the Sagittarius dwarf spheroidal galaxy, globular cluster members and the subset of Milky Way Mira variables with VLBI parallaxes. 
The Milky Way linear $JHK_s$ relations are slightly steeper and thus
fainter at short period than the corresponding LMC relations suggesting population effects in the near-infrared are perhaps larger than previous observational works have claimed.
Models of the Gaia astrometry for the Mira variables suggest that, despite the intrinsic photocentre wobble and use of mean photometry in the astrometric solution of the current data reduction, the recovered parallaxes should be on average unbiased but with underestimated uncertainties for the nearest stars.
The recommended Gaia EDR3 parallax zeropoint corrections evaluated at $\nu_\mathrm{eff}=1.25\,\mu\mathrm{m}^{-1}$ require minimal ($\lesssim5\,\mu\mathrm{as}$) corrections for redder five-parameter sources, but over-correct the parallaxes for redder six-parameter sources, and the parallax uncertainties are underestimated, at most by a factor $\sim1.6$ at $G\approx12.5\,\mathrm{mag}$.
The derived period--luminosity relations are used as anchors for the Mira variables in the Type Ia host galaxy NGC 1559
to find $H_0=(73.7\pm4.4)\,\mathrm{km\,s}^{-1}\mathrm{Mpc}^{-1}$. 
\end{abstract}


%%%%%%%%%%%%%%%%%%%%%%%%%%%%%%%%%%%%%%%%%%%%%%%%%%

%%%%%%%%%%%%%%%%% BODY OF PAPER %%%%%%%%%%%%%%%%%%
\section{Introduction}
Mira variables are thermally pulsating asymptotic giant branch (AGB) stars with characteristic periods of between $100$ and $1000$ days, and high amplitudes \citep[$\gtrsim2.5$ in $V$ and between $\sim0.3$ and $\sim1$ in $K_s$,][]{Matsunaga2009,Catelan2015}. Primarily from their study in the Large Magellanic Cloud  \citep[LMC,][]{GlassLloydEvans1981,Wood1999,Soszynski2009}, they are known to follow period--luminosity relations
\citep[with a typical scatter of $\sim0.2\,\mathrm{mag}$ from single-epoch $K_s$ data and $\sim0.1\,\mathrm{mag}$ for mean $K_s$ measurements,][]{Yuan2018}.
As AGB stars, Mira variables have chemistry dominated by either carbon-rich or oxygen-rich species as determined by the strength of dredge-up episodes, largely a reflection of their initial mass and composition \citep{Hofner2018}. Both C-rich and O-rich Mira variables satisfy period--luminosity relations \citep[e.g. the recent calibrations from][]{Iwanek2021} although the O-rich relations are typically tighter than the C-rich relations in the near-infrared due to the presence of significant circumstellar dust in the C-rich Mira variables \citep{Ita2011}. This makes O-rich Mira variables powerful distance tracers for both Galactic and cosmological studies.

The need for reliable well-calibrated distance indicators has received significant recent interest in light of the `Hubble tension'. The current expansion rate of the Universe, the Hubble constant $H_0$, can be measured using Type Ia supernovae in nearby galaxies or alternatively extrapolated from the early Universe using the best-fitting $\Lambda$CDM model of the cosmic microwave background radiation \citep{Planck}. An absolute calibration, or anchor, of the Hubble diagram is required to utilise the Type Ia supernovae, and traditionally the most precise and well-studied calibrators have been the classical Cepheids \citep{Freedman2001,Riess2011,Riess2021,Riess2022}. The problem then becomes anchoring the Cepheid scale, which can be done with local Cepheids using Gaia parallax measurements \citep{GaiaEDR3} of individual Cepheids \citep{Riess2021} or those of their host cluster \citep{Riess2022_CLUSTER}, eclipsing binaries in the Magellanic Clouds \citep{Pietrzynski2019,Graczyk2020} or the water maser in NGC 4258 \citep{Reid4258}. The latest estimates of the Hubble constant from \cite{Riess2022,Riess2022_CLUSTER} using classical Cepheids with a combination of all three anchors are in tension at the $\sim 5\sigma$ level with the early Universe extrapolation from \cite{Planck} possibly pointing towards new physics beyond the standard cosmological model \citep{H0review}. However, the discrepancy could also arise from systematics in the use of Cepheids \citep[e.g.][]{Efstathiou}. There have been many proposed and applied alternatives to classical Cepheids such as the tip of the giant branch \citep[e.g.][]{Freedman2021} which produces a more intermediate result between that of \cite{Planck} and \cite{Riess2021}, the J-AGB method \citep{JAGB}, gravitational lensing \citep{Wong2020} and masers \citep{Pesce2020}. Mira variables offer another interesting alternative to the usual classical Cepheid variables as (i) they are less biased to young populations so are present in a broad range of galaxies, in particular the full range of Type Ia supernovae hosting galaxies, (ii) as intermediate age tracers they are likely not in crowded or dust-obscured regions of their host galaxies so the photometric systematics are weaker, and (iii) they can be brighter than Cepheid variables in the infrared so can be utilised in more distant galaxies, especially in the era of the James Webb Space Telescope. Recently \cite{Huang2020} have used a sample of Mira variables in NGC 1559 anchored to Mira variables in the LMC and/or NGC 4258 to estimate the distance to SN 2005df and measure $H_0=(73.3 \pm4.0)\,\mathrm{km\,s}^{-1}\mathrm{kpc}^{-1}$ in good agreement with other local measurements \citep[as well as the early Universe extrapolated value from][at the $\sim1.5\sigma$ level]{Planck}.

Mira variables have also found significant use as a tracer of Galactic and Local Group structure. Thanks to their brightness in the infrared and their representation across a range of intermediate age populations, they are useful probes of structure across the Galactic disc \citep{FeastWhitelock2000,Grady2019,Grady2020}, the Galactic bulge \citep{Catchpole2016}, the heavily-extincted nuclear stellar region \citep{Glass2001,Matsunaga2009,Sanders2022} and the Magellanic Clouds \citep[e.g.][]{Deason2017}. Furthermore, their periods are linked to their age (and possibly metallicity), as confirmed empirically by variations of velocity dispersion with period \citep{FeastWhitelock2000} and demonstrated theoretically in non-linear pulsation calculations \citep{Trabucchi2022}. Recently, \cite{Grady2020} have used the empirical period--age relation for O-rich Mira variables to map the age structure of the Milky Way's bar-bulge and disc.

Typically, the period--luminosity relation of Mira variables has been calibrated using Mira variables in the LMC \citep[][]{GlassLloydEvans1981,Feast1989,Ita2004,Groenewegen2004,Fraser2008,Riebel2010,Ita2011,Yuan2017,Yuan2018,Bhardwaj2019,Iwanek2021b}. However, population effects (e.g. metallicity and age variations) can alter the period--luminosity relation \citep[e.g.][]{Qin2018}. For both extragalactic and Galactic studies, a calibration based on the perhaps more representative Milky Way Mira variables could be preferable. \cite{Whitelock2008} used a sample of $184$ O-rich Mira variables observed by the Hipparcos satellite in combination with Mira variables in globular clusters and those observed with VLBI to derive a near-infrared period--luminosity relation of $M_K=(-7.25\pm0.07)+3.50(\log_{10}P-2.38)$, within $\sim0.02$ of their derived LMC relation \citep[correcting for the updated LMC distance modulus from][]{Pietrzynski2019}. This already suggests the population effects on the ($K_s$-band) period--luminosity relation are small.

The arrival of data from the Gaia satellite \citep{Gaia1,Gaia2,GaiaEDR3} has opened up the possibility of an updated fully geometric calibration of the Mira period--luminosity relation, particularly as Gaia's multi-epoch observations have enabled all-sky catalogues of Mira variables to be extracted from the data \citep{Mowlavi2018,Lebzelter2022}. However, significant care must be taken when using astrometric data. Large parallax uncertainties can introduce a Lutz-Kelker bias when converting parallax measurements to distances which must be avoided with more careful probabilistic inversions \citep[e.g.][]{Luri2018, BailerJones2018}. Furthermore, systematic variations in the Gaia parallax zeropoint are present at the $\sim10\,\mu\mathrm{as}$ level and vary with magnitude, colour, on-sky location and other more subtle variables \citep{EDR3_ZPT}, and the formal parallax uncertainties from Gaia EDR3 are believed to be underestimated particularly at the bright end by a few $10$s of percent \citep{ElBadryRix2021,MaizApellaniz2021}. However, the existence of period--luminosity relations for certain stellar types opens up the possibility of measuring these systematic effects \citep[e.g.][]{Ren2021} and indeed a fully probabilistic model can simultaneously calibrate the properties of standard candles \emph{and} measure systematic issues with the data \citep[e.g.][]{Sesar2017,ChanBovy2020}.

In this paper, new period--luminosity relations for O-rich Mira variables in the Milky Way are provided using data from Gaia Data Release 3. The relations are derived using a probabilistic model incorporating distance priors and a model for Gaia parallax systematics. The new relations are then used to estimate $H_0$ using Mira variables in the Type Ia supernova host galaxy, NGC 1559. Section~\ref{section::data} describes the dataset employed in this work to measure the period--luminosity relations before the methodology is described in Section~\ref{section::model}. The new O-rich Mira variable period--luminosity calibrations are presented and discussed in Section~\ref{section::results} before they are utilised for the estimation of the Hubble constant in Section~\ref{section::h0}, taking into account the non-negligible C-rich contamination. The conclusions are presented in Section~\ref{section::conclusions}. In three appendices, the approximate completeness of the Gaia DR3 Mira variable catalogue is discussed (Appendix~\ref{appendix::completeness}), the expected Gaia performance for pulsating AGB stars is presented (Appendix~\ref{appendix::gaia_scanning}) and the period--luminosity relations for the LMC, SMC and the Sgr dwarf spheroidal galaxy are estimated (Appendix~\ref{appendix::lmc}).

\section{O-rich Mira variables in Gaia DR3}\label{section::data}
\begin{figure*}
    \centering
    \includegraphics[width=\textwidth]{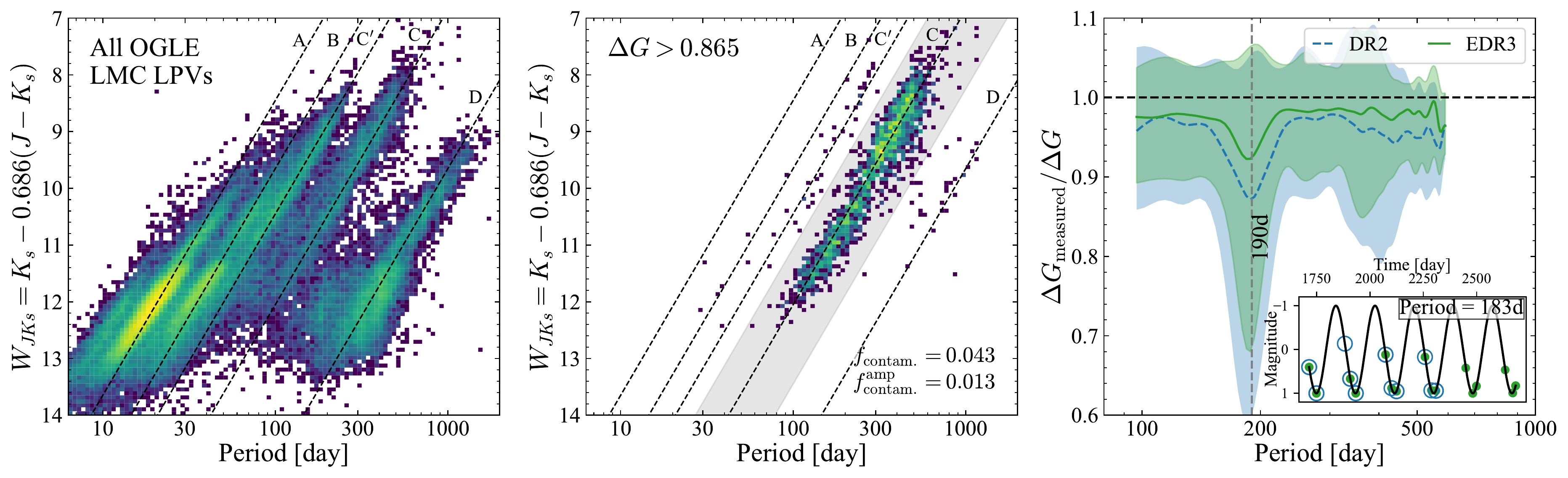}
    \caption{
    Period--magnitude diagram for the OGLE long-period variable sample in the LMC \protect\citep{Soszynski2009}. The left panel shows the logarithmically coloured density of the full sample and the central panel highlights those stars with $\Delta G>\Delta G_\mathrm{thresh} = 0.865\,\mathrm{mag}$. The different sequences from \protect\cite{Wood1999}, \protect\cite{Wood2000} and \protect\cite{Ita2004} are marked as dashed lines and labelled (note `C' sequence should not be confused with C-rich). In the central panel, a fraction $f_\mathrm{contam.}=0.043$ of the selected sources fall off the C sequence (defined by the grey shaded area). Restricting further to those with $\Delta G_\mathrm{Fourier}>0.865$ produces a contamination fraction of $f^\mathrm{amp}_\mathrm{contam.}=0.013$. The right panel shows the (median and $\pm1\sigma$) ratio of measured to true $\Delta G$ for a set of simulated sinusoidal light curves with periods assigned from the dataset used in this paper and randomly sampled phases sampled using the EDR3 photometric scanning law. Results for both the DR2 and DR3 sampling period are shown. $\Delta G$ is predominantly biased low, particularly at the aliasing period of $190\,\mathrm{day}$. This is illustrated in the inset for the simulated sampled light curve of a $183\,\mathrm{day}$ source (with the solid dots the DR3 measurements and circles the DR2 measurements).
    }
    \label{fig:delta_G_selection}
\end{figure*}

\begin{figure}
    \centering
    \includegraphics[width=\columnwidth]{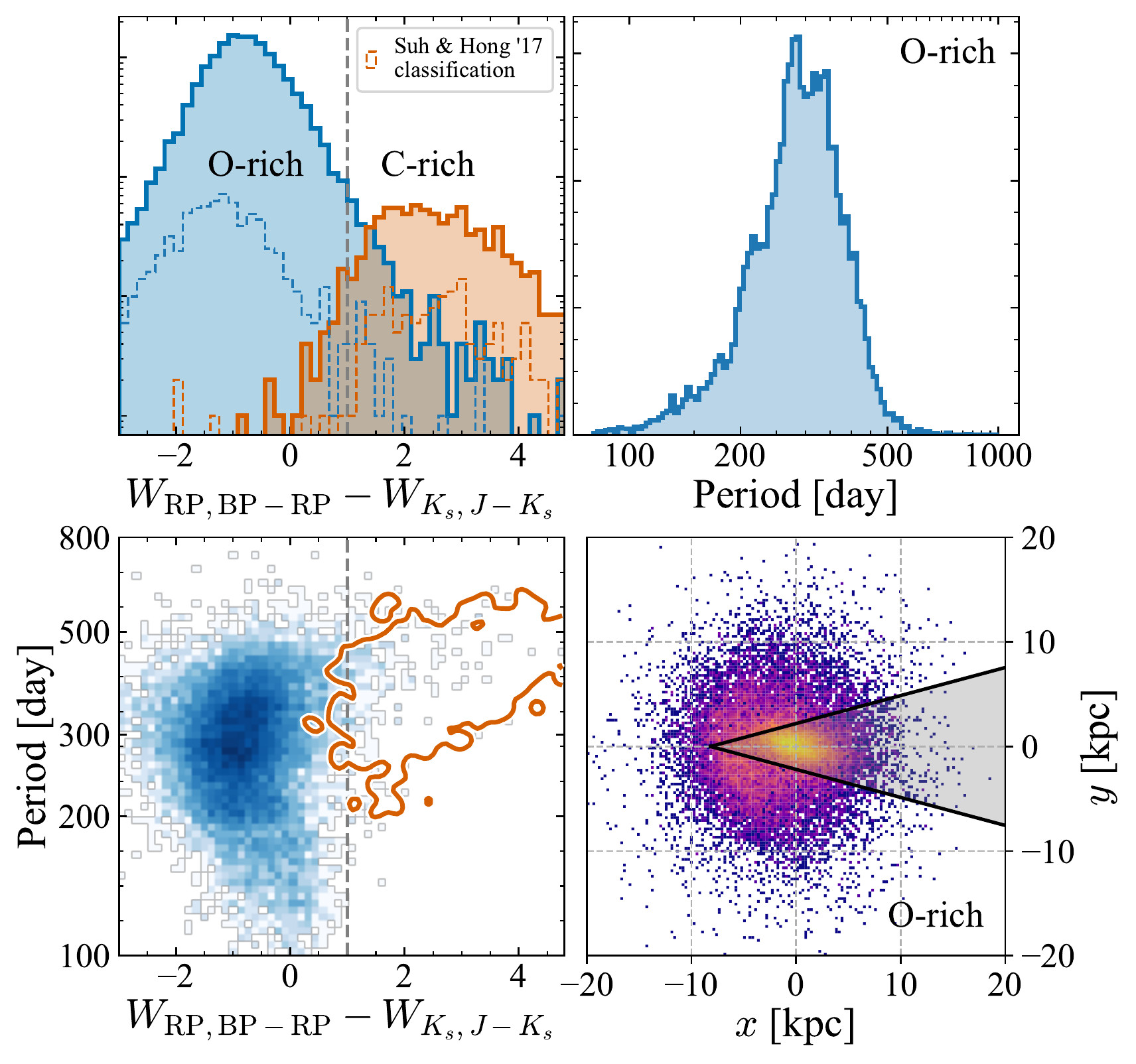}
    \caption{Properties of the O-rich Mira sample: the top left panel shows the distribution of the Wesenheit index difference from \protect\cite{Lebzelter2018} for all stars classified here as Mira variables split by O-rich or C-rich classification (the thin dotted histograms show the distribution of those sources cross-matched to the \protect\cite{SuhHong2017} catalogue using their classifications). The lower left panel shows the distribution of the Wesenheit index difference vs. period with the blue shading showing the logarithmic density of the O-rich Mira variables and the orange contour containing $90\percent$ of C-rich Mira variables. The top left panel shows the period distribution of the O-rich Mira variables (the small depletion around $\sim190$ days is due to Gaia's scanning strategy). The lower right panel shows the top-down Galactocentric view (from the North Galactic Pole) of the O-rich Mira variables using the LMC period--luminosity relation. In the analysis, stars in the midplane ($|b|<3\,\mathrm{deg}$) and those in the bulge region (shown in projection by the grey wedge) are also removed.}
    \label{fig:sample_properties}
\end{figure}
The primary data source is the long period variable (LPV) candidate catalogue \citep{Lebzelter2022} from Gaia DR3 \citep{Gaia1,Gaia3}. Gaia DR3 includes $34$ months of data with a mean number of observations per source of $43$. The Gaia variability processing consists of two stages: an initial classification of all likely variable sources \citep{Holl2018,Rimoldini2019,Rimoldini2022} and then a series of specific object studies (SOS) that further process each variability class. The initial classification was performed on all sources with at least $5$ Gaia field-of-view transits in their processed and cleaned photometric time series, and that were classified as likely variable when comparing to the variability of literature variable objects and the $75\%$ least variable Gaia sources at each magnitude. Classification into separate variability classes was then performed using features including time-series summary statistics, Lomb-Scargle periods, colours and parallax, and a training set composed of literature classifications. Gaia DR3 published all stars classified as LPV with $G$ $5$th-$95$th percentile greater than $0.1\,\mathrm{mag}$, $G_\mathrm{BP}-G_\mathrm{RP}>0.5$, at least $10$ \texttt{visibility\_periods\_used}, a reported renormalized unit weight error (RUWE), more than $9$ $G$ observations and a ratio of number of $G_\mathrm{RP}$ measurements in the cleaned time series to number of $G$ measurements $>0.5$. Of these a stricter subset with more than $12$ $G$ observations and number of $G_\mathrm{RP}$ measurements to number of $G$ measurements ratio of $>0.8$ were considered in the SOS (along with $522$ sources that satisfy all SOS LPV requirements but were mostly classified as symbiotic stars).
Periods were found using a generalised Lomb-Scargle method and were published if the period was $>35\,\mathrm{day}$ and shorter than the time series duration, the $G$ band signal-to-noise was greater than $15$ and no strong correlation was detected between the photometric time-series and the image parameter determination time series. This resulted in $392\,240$ LPV candidates with published periods from $2\,326\,297$ sources in Gaia DR3 classified as LPV. The completeness of the full LPV candidates catalogue and the subset with published periods is briefly assessed in Appendix~\ref{appendix::completeness}. In conclusion, the completeness of the Milky Way sample with periods is $\gtrsim90\percent$ for $|b|>3\,\mathrm{deg}$ and $\Delta G>\Delta G_\mathrm{thresh}$ with respect to the full Gaia DR3 source catalogue.

Due to Gaia's scanning strategy, periods around $\sim190$ days and below $120$ days are susceptible to aliasing.
Cross-matching those LPVs later defined as Mira variables with the AAVSO International Variable Star Index \citep[VSX,][downloaded 30th April 2022]{Watson2006}, ASAS-SN \citep{Jayasinghe2018,Jayasinghe2019b,Jayasinghe2019a}, and the OGLE LPV sample \citep{Soszynski2009,Iwanek2022}, the fraction of likely aliases (periods disagreeing by more than $25\percent$ -- the approximate width of the one-to-one relations upon cross-matching) is $3.4\percent$, $3.8\percent$ and $0.8\percent$ respectively, indicating aliasing is a minor issue and largely the Gaia periods are accurate \citep{Lebzelter2022}.
\cite{Mowlavi2018} report that the Gaia DR2 LPV catalogue is contaminated at the few percent level by young stellar objects (YSO) and the same is expected for Gaia DR3. With some parallax information, these can be identified as intrinsically fainter than the Mira variables. A conservative cut is employed by removing a handful of sources with $G-5\log_{10}(100\,\mathrm{mas}/(\varpi-3\sigma_\varpi))>1.75(G_\mathrm{BP}-G_\mathrm{RP})-3$. After this cut, the vast majority of the sample cross-matched to VSX and ASAS-SN are classified by these collections as LPVs (Mira variables, semi-regular variables or otherwise) with the largest contaminant being YSOs but only at the $\lesssim0.2\percent$ level.

The Gaia DR3 catalogue of candidate LPVs is complemented with variables from VSX \citep[][downloaded 30th April 2022]{Watson2006}. VSX is a compilation of variables initially built from the General Catalogue of Variable Stars \citep{GCVS}. All sources labelled as type `M' (Mira), `M:' (ambiguous Mira), `SR' (semi-regular), `SRA' (semi-regular variables similar to Mira but with small amplitude) and `LPV' (long-period variable) are selected and cross-matched to Gaia DR3, removing variables already in the Gaia DR3 LPV catalogue. This adds $2587$ stars to our Mira variable sample and $632$ to our more restricted sample used for fitting defined later.

Only LPVs with 2MASS photometry are used (cross-matched within $1\,\mathrm{arcsec}$ using proper motions to account for the epoch difference).
2MASS observations are single-epoch so will produce additional scatter about any fitted period--luminosity relation. However, the relations should be unbiased representations of the arithmetic mean magnitude period--luminosity relations (as required in the later $H_0$ analysis). Mean $J$, $H$ and $K_s$ magnitudes could be estimated using the Gaia light curves. However, the epoch difference ($\sim17$ year) is large enough that the typical Gaia frequency uncertainties ($\Delta\nu\approx0.05\,\mathrm{year}^{-1}$, $\Delta\nu/\nu\approx4\percent$) produce $\Delta\phi/\phi \approx \Delta\nu\times(17\,\mathrm{year})\approx85\percent$ uncertainties in the phase at the 2MASS epoch. This simple consideration does not account for uncertainties in the light curve fits at fixed period, the uncertainty in the amplitude ratios between the $JHK_s$ and $G$ bands, or any stochastic cycle-to-cycle variation that can be observed in Mira variables \citep{Ou2022,Iwanek2022}. Therefore, it appears with the current data any attempt to find the mean magnitudes from the single-epoch data will only add noise. For this reason, only the single-epoch measurements are used here.

\subsection{Selecting Mira variables}
To isolate a sample of Mira variables from the combined Gaia DR3 and VSX LPV candidates catalogue, two amplitude measures are combined: $\Delta G_\mathrm{Fourier}$, the amplitude derived from a Fourier fit provided in the Gaia DR3 LPV candidate catalogue (the \texttt{amplitude} column gives the $G$-band semi-amplitude from the Fourier fit i.e. half the required value) and $\Delta G$, the $G$-band amplitude measure computed from the reported Gaia uncertainties \citep{Belokurov2016}. This latter quantity is defined as
\begin{equation}
\Delta G = \frac{5\sqrt{2}}{\ln 10}\frac{\sqrt{\texttt{phot\_g\_n\_obs}}}{\texttt{phot\_g\_mean\_flux\_over\_error}}.
\label{eqn::delta_G}
\end{equation}
For light curves that are near sinusoidal and sampled fairly over period, this measure will be equal to the Fourier amplitude. Although the two measures correlate strongly with each other \citep{Sanders2023}, both quantities are used for defining Mira variables as they behave differently for poorly sampled lightcurves. If the lightcurve is undersampled, $\Delta G_\mathrm{Fourier}$ overestimates the amplitude as only a limited range of phases are used in the fit. However, as $\Delta G$ is a measure of the data scatter, it will be underestimated in this regime. In the right panel of Fig.~\ref{fig:delta_G_selection} the ratio of the measured to true $G$-band amplitudes is shown for a set of simulated sinusoidal light curves with periods assigned from the dataset used in this work and randomly drawn phases. The light curves are sampled according to the DR2 and DR3 photometric scanning laws \citep[using the Gaia DR2 scanning law from \citealt{Boubert2021} and the DR3 nominal scanning law both as part of the \texttt{scanninglaw} package,][with the data-taking gaps from \citealt{Riello2021}]{Green2018,Boubert2020,Everall2021}. Although $\Delta G$ can be underestimated, particularly around the troublesome $190$ day period or for stars that were part of the ecliptic pole scanning law which had approximately two thirds of their observations taken within a month, it is rarely significantly overestimated so when selecting using $\Delta G$ very low contamination from lower amplitude non-Mira variables is expected. Furthermore, when combined with $\Delta G_\mathrm{Fourier}$ it is quite certain that the LPVs are high amplitude.

Following \cite{Grady2019}, the cut $\Delta G > \Delta G_\mathrm{thresh}$ and $\Delta G_\mathrm{Fourier} > \Delta G_\mathrm{thresh}$ where $ \Delta G_\mathrm{thresh}=(5\sqrt{2}/\ln10)10^{-0.55}\approx0.865\,\mathrm{mag}$ is employed to isolate Mira variable stars. For the small set of stars from VSX without counterparts in the Gaia DR3 LPV catalogue, $\Delta G_\mathrm{Fourier}$ is not measured so only the $\Delta G > \Delta G_\mathrm{thresh}$ cut is employed.
In Fig.~\ref{fig:delta_G_selection}, the sample of OGLE LMC LPV stars from \cite{Soszynski2009} is shown along with those that satisfy $\Delta G>\Delta G_\mathrm{thresh}$. These selected stars predominantly lie along the `C' sequence associated with fundamental mode pulsation \citep{Wood1999,Wood2000,Ita2004} with only a fraction $f_\mathrm{contam.}=0.04$ consistent with membership of a different sequence. $21\percent$ of the $\Delta G$--selected OGLE LPVs are classified as semi-regular variables by \cite{Soszynski2013} on the basis of their $I$ amplitudes but as acknowledged by these authors and \cite{Trabucchi2021b} the traditional definitions of Mira variables are possibly not appropriate as lower amplitude variables or irregular Mira variables follow the same period--luminosity relation (as evident from Fig.~\ref{fig:delta_G_selection}) and are probably governed by the same physics. If the set of OGLE LPVs with Gaia DR3 Fourier amplitudes is considered, cutting on both $\Delta G>G_\mathrm{thresh}$ and $\Delta G_\mathrm{Fourier}>G_\mathrm{thresh}$ reduces the contamination from non-C-sequence stars to $f_\mathrm{contam.}=0.01$. 

\subsection{Separation of O-rich and C-rich Mira variables}
LPVs exhibit oxygen-rich and carbon-rich chemistry depending on the initial mass and metallicity of the star \citep{Hofner2018}. Of these two populations, the O-rich subset are more useful as they follow a tighter period--luminosity relation \citep{Ita2011}. Although significant within the LMC, C-rich Mira variables are rarer within the Galactic disc \citep{Blanco1984} and tend to be confined to the outer disc. C-rich Mira variables are typically redder and dustier than their O-rich counterparts. \cite{Lebzelter2018} showed that O-rich and C-rich Mira variables within the LMC can be separated in the plane of $W_\mathrm{RP,BP-RP}-W_{Ks,J-Ks}$ vs. $K_s$. Here the two Wesenheit indices are $W_\mathrm{RP,BP-RP}=G_\mathrm{RP}-1.3(G_\mathrm{BP}-G_\mathrm{RP})$ and $W_{Ks,J-Ks}=K_s-0.686(J-K_s)$. The boundary employed by \cite{Lebzelter2018} is slightly curved in `colour'-magnitude space but the curvature is weak and a pure $W_\mathrm{RP,BP-RP}-W_{Ks,J-Ks}$ cut performs similarly.

\cite{Lebzelter2022} have discussed how O-rich and C-rich LPVs can be distinguished using the Gaia DR3 BP/RP spectra due to the distinct separation of a set of bandheads arising from TiO for O-rich stars and CN for C-rich stars. As acknowledged by \cite{Lebzelter2022}, the bandhead separation diagnostic performs poorly for very red sources leading to the misclassification of many O-rich sources as C-rich. \cite{Sanders2023} utilised an unsupervised classification approach using the BP/RP spectra that uses the UMAP \citep[Uniform Manifold Approximation and Projection,][]{McInnes2018} algorithm on the normalized coefficients. This approach performs better than the published Gaia DR3 classifications for highly-extincted stars. For those stars without BP/RP spectra, \cite{Sanders2023} used a supervised classification algorithm \citep[XGBoost]{xgboost} trained on Gaia and 2MASS photometric data, periods and amplitudes for the stars with unsupervised classifications. This produces a $95\percent$ purity C-rich sample and $99.5\percent$ purity O-rich sample (due to the dominance of O-rich sources in the sample). Here the BP/RP unsupervised classifications are used when available, falling back to the supervised photometric classifications when no BP/RP spectra is provided in Gaia DR3.
Fig.~\ref{fig:sample_properties} shows the distribution of the Mira variable sample in the Wesenheit `colour' vs. period where the separation of the O-rich and C-rich populations is clear. A simple cut of $W_\mathrm{RP,BP-RP}-W_{Ks,J-Ks}<1$ would remove most C-rich sources but would also remove some longer period O-rich sources. The stars in the final sample that are also in the catalogue of \cite{SuhHong2017} are shown by the dotted histogram separated using these authors' classification. The classifications are a combination of low-resolution spectroscopic, maser and photometric classifications. Using our classifications to isolate O-rich stars results in only $5$ of the $783$ matches ($0.6\percent$) with the \cite{SuhHong2017} catalogue being classified by them as C-rich (using the updated IRAS PSC catalogue of \cite{Suh2021} results in $12$ of $867$ matches classified as C-rich, $1.4\percent$).

\subsection{Summary of selections}
In summary, the Gaia DR3 LPV candidates with reported periods have been combined with additional LPVs from VSX. Mira variables have been isolated by cutting on the $G$-band Fourier amplitude, $\Delta G_\mathrm{Fourier}$, and $G$-band scatter, $\Delta G$, and potential YSO contaminants have been removed with a parallax cut. O-rich and C-rich separation has been performed using the BP/RP spectra where available, and otherwise using broadband Gaia and 2MASS photometry combined with periods and amplitudes. Considering the issues of period aliasing, YSO contamination, non-Mira LPV contamination and C-rich contamination altogether, it seems the cuts defined here produce a O-rich Mira variable catalogue with a reliability upwards of $95\percent$.

For fitting the period--luminosity relations, only stars with $G<17$, $G_\mathrm{BP}-G_\mathrm{RP}>1.9$, distances $<25\,\mathrm{kpc}$ (as estimated a priori using the LMC period--luminosity relations in Appendix~\ref{appendix::lmc}), periods more than $100$ days and less than $1000$ days, period uncertainties $<50\percent$ (the median period uncertainty is $5\,\percent$ and the 95th percentile is $11\percent$) and Gaia EDR3 RUWE $<1.4$ (see next section) are retained. Furthermore, stars in the bulge region ($|\ell|<15,|b|<10$), those in the Galactic mid-plane ($|b|<3\,\mathrm{deg}$), those within $15$ deg of the LMC, those within $10$ deg of the SMC and those at distances greater than $18\,\mathrm{kpc}$ within $15$ deg of the Sgr dSph are removed. These on-sky selections are visualized in Fig.~\ref{fig:completeness}. With this set of cuts, there remain
$15159$ O-rich and $875$ C-rich Mira variables (from an initial catalogue of $86477$ stars with the $|b|$ and bulge cuts most severely reducing the sample). The lower right panel of Fig.~\ref{fig:sample_properties} shows the view of the sample from the Galactic North Pole using the LMC period--luminosity relation.

\section{Astrometric data quality}\label{section:photocentre}
\subsection{Initial considerations}
\begin{figure*}
\centering
\includegraphics[width=\textwidth]{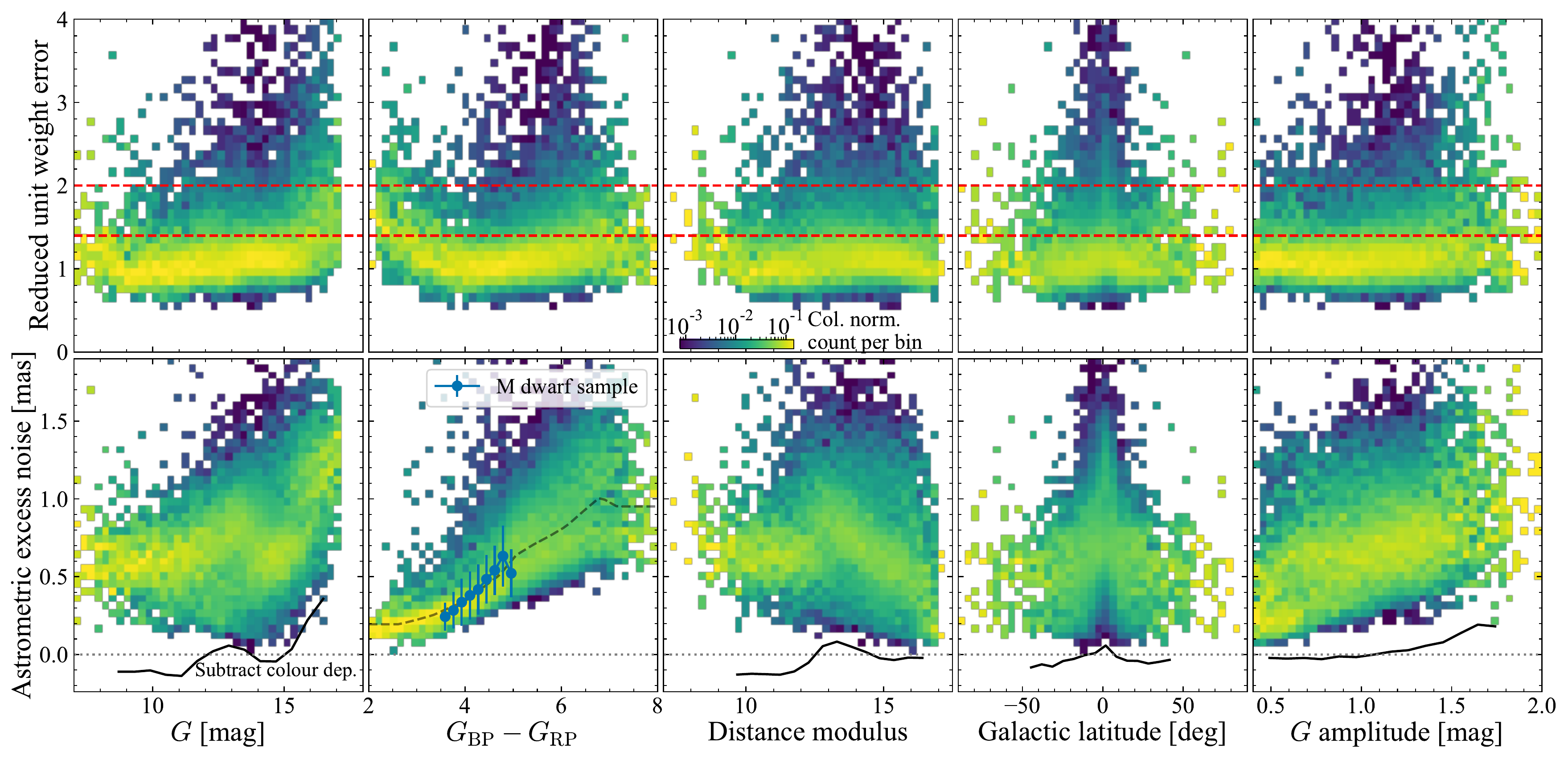}
\caption{Column-normalized distributions of the reduced unit weight error (RUWE, top) and the astrometric excess noise (bottom) against various quantities for our Milky Way Mira variable sample. The two horizontal lines in the top panels show the RUWE cuts ($1.4$ and $2$) employed in this work. The blue points in the second lower panel shows the binned distribution (median $\pm1\sigma$) of an M dwarf sample defined in the text. The dashed line in the second lower panel is the median trend subtracted in each of the other lower panels to produce the black line.}
\label{fig:ruwe_aen}
\end{figure*}
To confidently use the Gaia EDR3 astrometric data for period--luminosity calibration, their quality must be assessed (note Gaia DR3 did not update the astrometry so EDR3 and DR3 astrometry refer to the same thing). This is a particular concern for Mira variables as they are some of the reddest sources observed by Gaia. Additionally, their variability (in both colour and magnitude) makes the astrometry challenging, and as discussed in \cite{Mowlavi2018} in the current Gaia data releases epoch photometry is not utilised in the astrometric solution \citep{Lindegren2021} which could lead to errors for variable sources \citep[][see Appendix~\ref{appendix::gaia_scanning}]{Pourbaix2003} . There are a number of recommended quality cuts for handling Gaia data \citep{Fabricius2021} but the only quality criterion used here is the renormalized unit-weight error (RUWE) from Gaia EDR3 by ensuring all stars have RUWE $<1.4$ (a test with $<2$ is also run). Although nearly all of the sample has significant ($>3$) astrometric excess noise, \cite{Lindegren2021} caution against using astrometric excess noise for very red sources ($G_\mathrm{BP}-G_\mathrm{RP}>3$) as it likely reflects shortcomings of the instrument and attitude modelling. Also, a large fraction of the sample have \texttt{ipd\_gof\_harmonic\_amplitude} $>0.2$ ($62\percent$) and \texttt{ipd\_frac\_multi\_peak} $>2$ ($33\percent$) which is indicative of poor LSF/PSF fits due to possible binarity \citep{Lindegren2021}. However, the LSF/PSF calibrations \citep{Rowell2021} have only been performed down to $\nu_\mathrm{eff}=1.24\,\mu\mathrm{m}^{-1}$ so it is anticipated that redder sources will not have well fitting LSF/PSFs. Furthermore, for six-parameter solutions a default LSF/PSF at $\nu_\mathrm{eff}=1.43\,\mu\mathrm{m}^{-1}$ is utilised which perhaps makes the IPD statistics unreliable for the significantly redder sources. Finally, these sources typically fall outside the advised adjusted BP/RP excess factor range as a function of magnitude but this is probably due to their variability \citep[as already highlighted by figure 21 of][]{Riello2021}.

In Fig.~\ref{fig:ruwe_aen} the column-normalized distributions of RUWE and astrometric excess noise are shown against various other quantities for the Mira variable sample. The RUWE distributions are largely flat with all plotted quantities except for an enhancement in the Galactic midplane and a small uptick at bluer $(G_\mathrm{BP}-G_\mathrm{RP})$. There is some slight evidence of an increase in RUWE for nearby, brighter sources. The astrometric excess noise shows strong trends, particularly with colour. However, the astrometric excess noise vs. $(G_\mathrm{BP}-G_\mathrm{RP})$ is shown for a sample of M dwarf stars from Gaia DR3 with $(G_\mathrm{BP}-G_\mathrm{RP})>3.5$, RUWE $<1.4$ and $\varpi>10\,\mathrm{mas}$. This traces the trend in the Mira variables in the overlapping region. It therefore appears the astrometric excess noise trend arises from poor characterisation of the instrument performance, rather than anything intrinsic. In other panels the trends can be related to the fundamental trend in colour i.e. redder stars are fainter, typically higher amplitude and found more often in the midplane. This is corroborated by the black lines which depict the median trends after subtracting the median colour dependence (shown as a black dashed line in the second lower panel). 

The use of mean photometry in the astrometric solution leads to two effects: (i) the centroids have a residual uncorrected offset due to the colour variation of the sources and (ii) an average astrometric error instead of the epoch astrometric errors is used. Using the pseudocolour uncertainties for the six-parameter solutions, the typical centroid shift with effective wavenumber is estimated as $\sim 2\,\mathrm{mas}\,\mu\mathrm{m}$ \citep[see also][]{deBruijne,Lindegren2021} which using the typical amplitudes and colours of the Mira variable sample is $\sim6\percent$ of the reported uncertainties in the median. The variation of the epoch uncertainties due to the typical $G$ and $(G_\mathrm{BP}-G_\mathrm{RP})$ amplitudes of the sample is $\sim20\percent$. The full analysis presented in Appendix~\ref{appendix::gaia_scanning} demonstrates that in combination these effects lead to a modest underestimate of the astrometric uncertainties of at most $10\percent$ with the largest underestimates arising from the highest amplitude stars.

\subsection{Intrinsic photocentre wobble}
A further concern is that AGB stars have large radii, $\sim1\,\mathrm{AU}$, and complex surface dynamics and, as highlighted recently by \cite{Chiavassa2018} \citep[see also][]{vanBelle2002}, the motion of the atmosphere can lead to shifts of the photocentre typically of order $5-10\percent$ of the radius. Appendix~\ref{appendix::gaia_scanning} investigates this issue in considerable detail and here only simple arguments as to its impact on the Gaia EDR3 astrometry are presented.
The previous comparison with the M-dwarf sample suggests the quality of the astrometry for the Mira variables arises from Gaia's performance rather than any intrinsic noise, but this is validated further here.

\cite{Chiavassa2011} presented a simulation of the red supergiant Betelgeuse finding a $G$-band photocentre wobble of $0.065\,\mathrm{AU}$, about $2\percent$ of its radius, whilst \cite{Chiavassa2018} presented $8$ simulations of AGB stars with typical $G$-band photocentre wobbles of $5-10\percent$ with longer period stars (or more precisely longer pressure scaleheight) having a larger wobble. In the optical, the photocentre wobble is composed of long variations on the order of years due to large convective cells covering of order $\sim1/3$ the stellar radius (more evident in infrared observations) with shorter variations on the order of months due to smaller convective cells in the upper atmospheres of size $10\percent$ the stellar radius. For nearby AGB stars, this photocentre wobble can be a significant observable effect presenting a fundamental error floor for the astrometry.
However, due to the stochasticity of the AGB photocentre wobble and the lack of preferred direction relative to the parallax ellipse and proper motion vector, it is expected that over long enough timespans (or averaged over many stars) the wobble should manifest as an additional random uncertainty and the astrometric parameters will be unbiased but possibly with poorly estimated uncertainties \citep{Chiavassa2011}.

\begin{figure}
    \centering
    \includegraphics[width=\columnwidth]{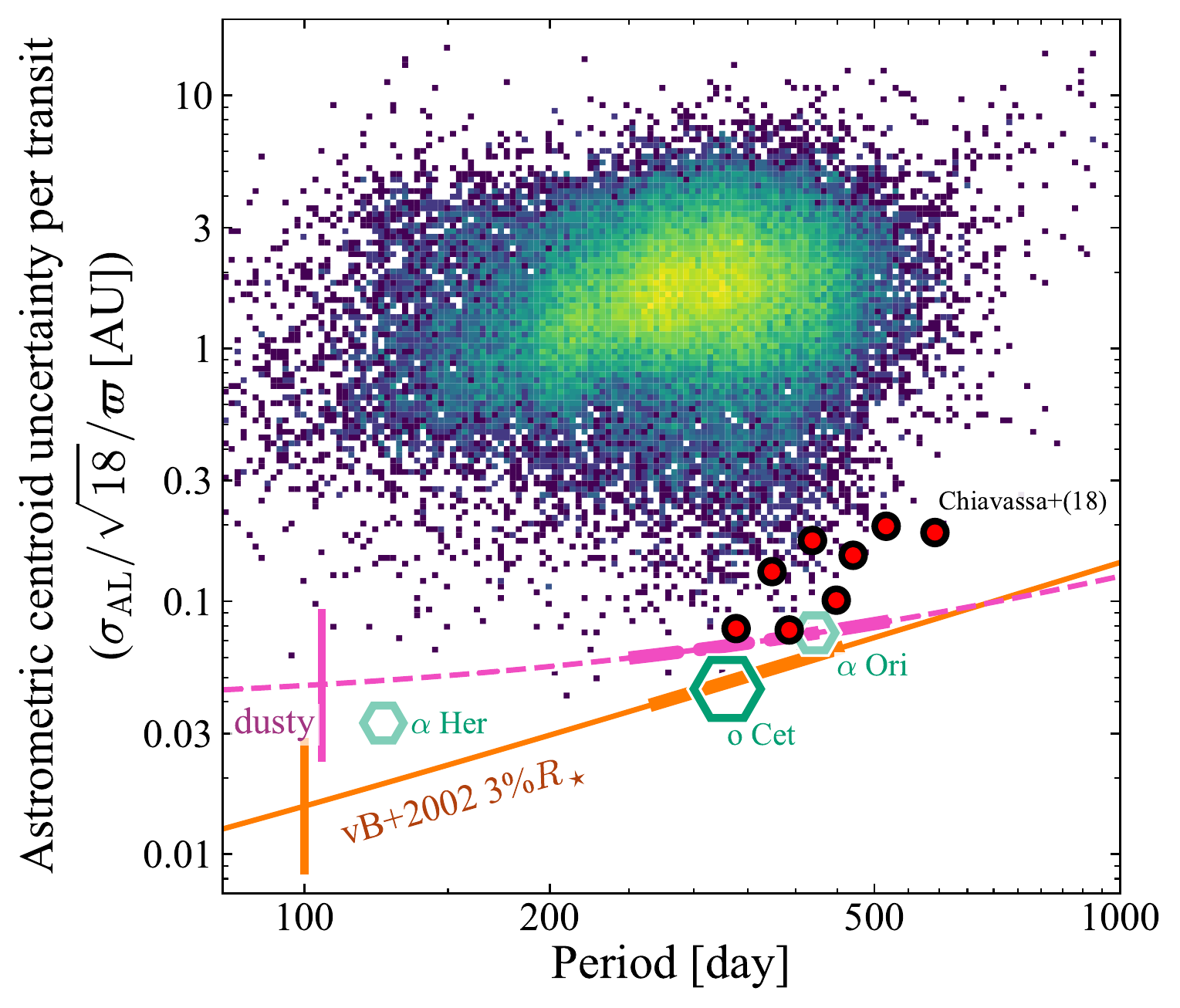}
    \caption{
    Approximate single-epoch astrometric uncertainty from Gaia for the O-rich Mira variable sample (in AU). The along-scan astrometric uncertainty, $\sigma_\mathrm{AL}$, is approximated from the Gaia parallax uncertainty, $\sigma_\varpi$, as $0.53\sqrt{N}\sigma_\varpi$. There are $18$ observations per transit ($9$ CCD observations in each field-of-view) such that, ignoring systematics, the single-epoch astrometric uncertainty is $\sigma_\mathrm{AL}/\sqrt{18}$. This is transformed into AU using the parallax computed from the LMC period--luminosity relation.
    The expected photocentre wobble for `normal' and dusty Mira variables is shown by the orange solid and pink dashed lines assuming the period--radius relations from \protect\citet[][vB+2002]{vanBelle2002} and a $3\percent$ radial variation (the errorbars give the amplitude of the uncertainty in the relations and the thicker parts of the lines are the regions over which vB+2002 had data). The AGB models from \protect\cite{Chiavassa2018} are shown as red points. The measured photocentre wobble for o Cet (Mira, large hexagon) and $\alpha$ Her and $\alpha$ Ori (small faint hexagons) are shown (note the latter two stars are red supergiants and their periods have been used for convenience to place them in the plot).
    Essentially all of the sample lies above the models suggesting the astrometric uncertainties are not dominated by photocentre wobble and the parallaxes are reliable.}
    \label{fig:centroid_wobble}
\end{figure}

Photocentre wobble is only detectable when it is similar to or greater than the Gaia single-epoch astrometric uncertainty. \cite{Lindegren2021} provides the median along-scan astrometric uncertainty in Gaia EDR3, $\sigma_\mathrm{AL}$, as a function of $G$ but with no information on the uncertainty as a function of colour. \cite{BelokurovBinary} have demonstrated that $\sigma_\mathrm{AL}$ is approximately related to the reported parallax uncertainty, $\sigma_\varpi$, as $\sigma_\mathrm{AL}\approx0.53\sqrt{N}\sigma_\varpi$ where $N$ is the number of observations (\texttt{astrometric\_n\_good\_obs\_al}) allowing for the estimation of $\sigma_\mathrm{AL}$ for a range of different magnitudes, colours, on-sky positions etc. Gaia typically makes $18$ astrometric observations in a short timespan ($9$ CCDs for each of the two fields of view) such that in the absence of systematic uncertainties, the single-epoch along-scan astrometric uncertainty is $\sim\sigma_\mathrm{AL}/\sqrt{18}$. It is this uncertainty that must be compared with the expected AGB photocentric wobble. Using the parallax to transform this astrometric uncertainty into a physical scale gives $\sim\sqrt{N/350}(\sigma_\varpi/0.1\mathrm{mas})(\mathrm{mas}/\varpi)(0.24\mathrm{AU})$ where typical values for $N$ and $\sigma_\varpi$ for the sample are used. For a star at $1\,\mathrm{kpc}$, the single-epoch astrometric uncertainty is larger than the expected $5-10\percent$ of the radius wobble (assuming the radius is $1\,\mathrm{AU}$). In Fig.~\ref{fig:centroid_wobble} the single-epoch astrometric uncertainty, $(1/\sqrt{18})0.53\sqrt{N}\sigma_\varpi/\varpi$ in AU, is displayed for the full O-rich Mira variable sample with RUWE$<1.4$ and $G<17$ using $\varpi$ estimated from the LMC period--luminosity relation (Appendix~\ref{appendix::lmc}). This can be compared to the AGB models from \cite{Chiavassa2018}, the measured photocentre wobble from o Cet (Mira) and the two supergiants, $\alpha$ Her and $\alpha$ Ori, \citep{Chiavassa2011} and a simple model of the photocentre wobble using the period-radius models from \cite{vanBelle2002} and a $3\percent$ radius wobble that fits the Mira observation well.

We see that because the bulk of the sample has $\varpi\ll1\,\mathrm{mas}$, the physical scale Gaia is capable of probing for the sample is significantly greater than $0.1\,\mathrm{AU}$ and the uncertainty budget is dominated by Gaia's limitations. If the astrometric excess noise is instead used as the measure of along-scan astrometric uncertainty a similar result is found. This gives confidence that for the majority of the considered sample the astrometry should be free from any effects arising from intrinsic photocentre wobble and that Gaia is capable of providing precision measurements for this type of star. However, this may be more of a concern with future data releases with improved astrometric uncertainties for red stars. For example, \cite{Chiavassa2011} estimated that the photocentre wobble should be a measurable effect from the Gaia uncertainties for stars within $4.4\,\mathrm{kpc}$ assuming the predicted wobble from models of Betelgeuse. Their analysis assumed final Gaia parallax uncertainties of $7.8\,\mu\mathrm{as}$ whilst the typical uncertainty for the present sample is an order of magnitude larger around $0.1\,\mathrm{mas}$. However, it should be stressed that improved measurements over longer baselines will likely not produce on average biased astrometric results, but more affect the reported uncertainties.

One caveat here is that the AGB model expectation might be very wrong and $\sigma_\mathrm{AL}$ in fact does reflect the photocentre wobble rather than any limitation on Gaia's performance. This is unlikely considering for the bulk of stars the single-epoch astrometric uncertainty is of order the radius of the star and also that the measured photocentre wobble of Mira suggests if anything the AGB models of \cite{Chiavassa2018} produce too large a photocentre wobble. Furthermore, a comparison with M dwarf stars at similar colours and magnitudes shows similar $\sigma_\mathrm{AL}$ and astrometric excess noise to the sample used here (see Fig.~\ref{fig:ruwe_aen} and Fig.~\ref{fig:astrometric_colour}). No photocentre wobble is expected for these sources suggesting in the majority of cases $\sigma_\mathrm{AL}$ is governed by Gaia's limitations.

A much fuller analysis of the expected Gaia performance for AGB stars is presented in Appendix~\ref{appendix::gaia_scanning} and reaches the same conclusions as the simpler considerations presented here. The analysis demonstrates that on average the astrometric parameters for the sample of stars used in this work are unbiased but the uncertainties are underestimated for $G\lesssim11$ and $\varpi>0.5\,\mathrm{mas}$ (a small fraction of the total sample).

\section{Period--luminosity relation for O-rich Mira variables}\label{section::model}

A probabilistic model is introduced to measure the period--luminosity relation for the sample of O-rich Mira variable stars from Gaia DR3. This allows the inclusion of uncertainties in the data and a prior when transforming from the uncertain parallax measurements to absolute magnitudes \citep{BailerJones2018,Luri2018}. Furthermore, the fact the Mira variables appear to follow a period--luminosity relation can be used to simultaneously calibrate this relation and measure the parallax zeropoint and parallax uncertainties of the sample \citep[e.g.][]{Sesar2017,ChanBovy2020}. As highlighted previously, variations in the accuracy of the astrometry with both colour and magnitude are anticipated. This is particularly important for the Mira variables as they are some of the reddest sources observed by Gaia and many fall outside the effective wavenumber range covered by previously-published zeropoint corrections \citep{EDR3_ZPT}. 

\begin{figure}
    \centering
    \includegraphics[width=\columnwidth]{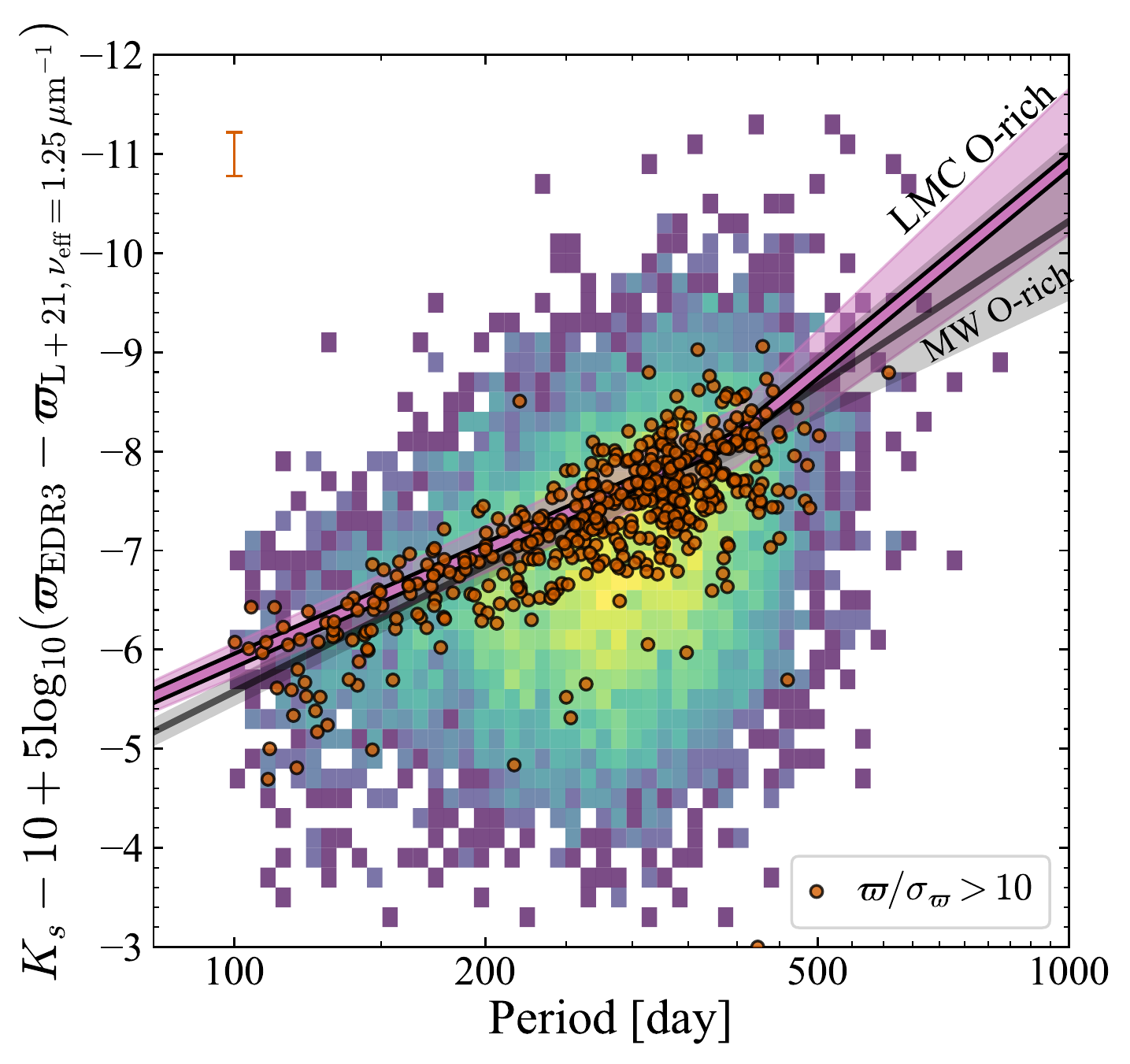}
    \caption{Period--luminosity distribution for the Milky Way O-rich Mira sample. The background shows a log-scaled histogram of the subsample with parallax signal-to-noise greater than $1$ (the large scatter is due primarily to the parallax uncertainties). Absolute magnitudes have been computed using the Gaia EDR3 parallaxes corrected by the \protect\cite{EDR3_ZPT} zeropoint corrections evaluated at $\nu_\mathrm{eff}=1.25\,\mu\mathrm{m}^{-1}$ as described in Section~\ref{sec::zpt} and using the extinction corrections described in Section~\ref{section::ext_corr}. The orange points are the subset with parallax errors smaller than $10\percent$ (note this selection has the effect of biasing the measurements towards higher parallaxes and absolute magnitudes). The errorbar shows the minimum formal uncertainty from the parallax measurements alone. The pink line shows the LMC O-rich Mira relation as derived in Appendix~\ref{appendix::lmc} along with its scatter (a combination of that due to single epoch observations and any intrinsic scatter due to population variations). The black line shows similar for the MW relation derived in this work.}
    \label{fig:plr_example}
\end{figure}

As a preliminary illustration of the sample and an indication of the ability to measure the period--luminosity relation accurately, the $K_s$ absolute magnitude computed from the Gaia EDR3 parallax vs. period is shown in Fig.~\ref{fig:plr_example}. The $K_s$ magnitudes have been corrected for extinction as described later in Section~\ref{section::ext_corr} and the Gaia EDR3 parallaxes have been zeropoint corrected using the \cite{EDR3_ZPT} corrections evaluated at $\nu_\mathrm{eff}=1.25\,\mu\mathrm{m}^{-1}$ as described later in Section~\ref{sec::zpt}. For comparison, the period--luminosity relation for the LMC as derived in Appendix~\ref{appendix::lmc} is shown. The subset of stars with parallax uncertainties better than $10\percent$ align nicely with the LMC relation, possibly falling slightly under in the mean, although this effect is partly due to the selection on parallax errors biasing the measurements towards higher parallaxes and hence higher absolute magnitudes. The fuller sample shows a significant scatter about the expected period--luminosity relation due to the parallax uncertainties.
In the following sections, the model for the data is introduced, before the handling of the parallax zeropoint modelling is described in more detail.

\subsection{Probabilistic model}
The joint single-star likelihood of the Gaia EDR3 parallax $\varpi$ and a magnitude $m$ given the $G$ magnitude, effective wavenumber $\nu_\mathrm{eff}$ (pseudo-colour for six-parameter astrometric solutions), period $P$ and on-sky location $(\ell,b)$ (and corresponding uncertainties) is expressed as
\begin{equation}
\begin{split}
    p(\varpi,m|G,\nu_\mathrm{eff},P,\ell,b) = \int\mathrm{d}s\,p(\varpi|&s,G,\nu_\mathrm{eff},\ell,b)\\\times &p(m|s,P)p(s|\ell,b),
\end{split}
\label{eqn::zeropoint_model}
\end{equation}
where
\begin{equation}
    \begin{split}
        p(\varpi|s,G,c,\ell,b) &= \mathcal{N}(\varpi|1/s+\varpi_0
        ,
        f^2_\varpi
        \sigma_\varpi^2+\sigma^2_{\varpi,0}),\\
        p(m|s,P) &=\sum_{j=1}^{j=2}\vartheta_j\mathcal{N}(m|m_\mathrm{abs}(P)+\mu,\sigma_{m,j}^2(P)),
        \\
        p(s|\ell,b) &= \frac{s^2}{2L^3(\ell,b)} \mathrm{e}^{-s/L(\ell,b)}.
    \end{split}
    \label{eqn::model_spec}
\end{equation}
$\mathcal{N}(x|\mu,\sigma^2)$ is a normal distribution with mean $\mu$ and standard deviation $\sigma$.
Here $s$ is the true distance (with corresponding distance modulus $\mu$), $\varpi_0(G,\nu_\mathrm{eff},\ell,b)$ is a colour-,  magnitude- and spatially-dependent parallax zeropoint offset (described in a later subsection), $\sigma_\varpi$ are the reported parallax uncertainties with $f_\varpi(G,\nu_\mathrm{eff})$ a colour- and magnitude-dependent scaling (again specified later) and $\sigma_{\varpi,0}$ an additional systematic error floor. A two-component Gaussian mixture model is employed for the magnitude distribution about the predicted magnitude $m_\mathrm{abs}(P)+\mu$ where $m_\mathrm{abs}(P)$ is the period--luminosity relation. This mixture model accounts for possible outliers using a mixing simplex $\vartheta_j$ ($\vartheta_1+\vartheta_2=1$). In Section~\ref{section::data} and Fig.~\ref{fig:sample_properties} the contamination was estimated to be at the few per cent level. $p(s|\ell,b)$ is the distance prior. The various modelling choices are discussed in the following subsections.

\subsubsection{Period--magnitude relation}
The adopted period-magnitude relation $m_\mathrm{abs}(P)$ for magnitude $m$ is
given by
\begin{equation}
m_\mathrm{abs}(P) = a_m +
\begin{cases}
b_m(\log_{10}P-2.3),&\mathrm{if} \log_{10}P\leq2.6,\\
0.3b_m+c_m(\log_{10}P-2.6),&\mathrm{otherwise}.
\end{cases}
\label{eqn::period_luminosity_split}
\end{equation}
Period--luminosity relations for O-rich Mira variables have been computed using those stars in the LMC \citep[e.g.][]{Ita2011,Yuan2017,Yuan2018}. Typically a linear relation is appropriate for $P<400\,\mathrm{days}$ beyond which a break occurs and the period--luminosity relation is steeper \citep{Ita2011,Bhardwaj2019}. This is often attributed to additional luminosity arising from the onset of hot-bottom burning for stars with $P>400\,\mathrm{days}$ \citep{Whitelock2003}. Following \cite{Ita2011}, a break is placed at $\log_{10}P=2.6$, which is validated by fits to the LMC (see Appendix~\ref{appendix::lmc}) although \cite{Bhardwaj2019} advocate for a slightly lower break at $300$ days. Often the entire period range is modelled with a quadratic relation \citep{Yuan2018}. Quadratic relations are weakly disfavoured over broken linear relations for the LMC data (see Appendix~\ref{appendix::lmc}) and also have the tendency to bias the relation for short periods when attempting to fit the curvature at long periods. Furthermore, the broken linear relation are made continuous \citep[c.f.][]{Ita2011} as this form is perhaps more physically motivated and reduces the number of parameters by one.

In Appendix~\ref{appendix::lmc} period--luminosity relations for O-rich and C-rich Mira variables in the LMC are provided using the form of the period-magnitude relation in equation~\eqref{eqn::period_luminosity_split}. The resulting $(b_m,c_m)$ posterior distributions are used as priors for the Milky Way O-rich sample.

\subsubsection{Period--amplitude relation}
The scatter about the period-magnitude relation for each component is given by
\begin{equation}
    \sigma_{m,j}^2(P)=\sigma^2_\mu(P)+\mathrm{Var}(m_\mathrm{abs,LMC}(P))+\sigma_{m,\mathrm{obs}}^2+\sigma^2_{\mu,0,j}.
    \label{eqn::scatter}
\end{equation}
The scatter consists of four terms: the first term $\sigma^2_\mu(P)$ gives the intrinsic scatter about the period--amplitude relation. The bulk of the spread arises from using single-epoch observations. Longer period variables have larger amplitudes so a model of the form
\begin{equation}
\sigma_\mu(P) = \sigma_{2.3} +
\begin{cases}
m_{\sigma-}(\log_{10}P-2.3),&\mathrm{if} \log_{10}P\leq2.6,\\
0.3m_{\sigma-}+m_{\sigma+}(\log_{10}P-2.6),&\mathrm{otherwise},
\end{cases}
\label{eqn::scatter_period_luminosity_split}
\end{equation}
is employed.
The choice here mirrors the period--magnitude relation of equation~\eqref{eqn::period_luminosity_split} as the break in the period--luminosity relation potentially due to hot-bottom burning is accompanied by a break in the period--amplitude relation \citep[e.g.][]{Matsunaga2009}. Even if multi-epoch data from which accurate mean magnitudes could be estimated were available, some intrinsic scatter might be expected due to other hidden dependencies \citep[e.g. age and metallicity,][]{Qin2018} so $\sigma_\mu(P)$ is considered as a quadrature sum of the single-epoch scatter and intrinsic scatter. Again the posterior distributions for fits to the (single-epoch) LMC data (Appendix~\ref{appendix::lmc}) are used as priors for $F\equiv(\sigma_{2.3},m_{\sigma-},m_{\sigma+})$.

The second term in the scatter is $\mathrm{Var}(m_\mathrm{abs,LMC}(P))$ which gives the variance arising from the uncertainty in the period. For simplicity, the additional spread in the magnitude $\mathrm{Var}(m_\mathrm{abs,LMC}(P))$ is then computed using the fitted period-magnitude relations for the LMC (see Appendix~\ref{appendix::lmc}). For large period uncertainties, the prior understanding of the width of the period distribution is also important. The Gaussian in $\log_{10}P$ with mean $q$ and width $\sigma_q$ fitted to the LMC data in Appendix~\ref{appendix::lmc} is used. The uncertainty in the period is then computed by combining with the prior distribution. The third term in equation~\eqref{eqn::scatter}, $\sigma_{m,\mathrm{obs}}^2$, is the variance arising from the photometric uncertainties, uncertainties in the extinction and uncertainties in the extinction coefficients. The final term $\sigma^2_{\mu,0,j}$ is an additional residual only employed for the outlier component such that $\sigma^2_{\mu,0,1}=0$.

\subsection{Extinction corrections}\label{section::ext_corr}
The magnitudes $m$ must be corrected for the effects of extinction. When available, the \citet[][Bayestar 2019]{Green2019} extinction estimates and their uncertainties are evaluated at the distance of each Mira variable using the LMC Wesenheit period--luminosity relations of Appendix~\ref{appendix::lmc}. The reported extinctions are assumed to be exactly equal to $E(B-V)$ on the \cite{SFD} scale (validated as $E(B-V)=1.02E(g_\mathrm{PS}-r_\mathrm{PS})$ from \citealt{WangChen2019}) so must be adjusted to account for the $14\,\percent$ reduction reported by \cite{SchlaflyFinkbeiner2011} and to convert to `true' $E(B-V)$. The extinction estimates are flagged as possibly unreliable if stars are beyond the faintest main-sequence star in Pan-STARRS at a given on-sky location. \cite{Green2019} also use giant stars in their extinction estimates such that the extinctions beyond the faintest main-sequence star can be constrained. However, the giant models are less certain than the main sequence models. To account for this, the extinction uncertainties are arbitrarily inflated by a factor two for the estimates flagged as unreliable. Outside the Pan-STARRS footprint, the extinction map from \cite{SFD} is used accounting for the recalibration from \cite{SchlaflyFinkbeiner2011} and a $16\percent$ uncertainty is employed.

For computing the extinction in a general band, the extinction coefficients from \cite{WangChen2019} are used and their provided uncertainties in the extinction coefficients are propagated. An alternative to explicitly correcting for extinction is to use the Wesenheit magnitudes given by $m= W_{x,y-x}\equiv x-e(y-x)$ where the extinction coefficient $e\equiv A(x)/E(y-x)$ from \cite{WangChen2019} \citep[or from][as a model variant]{Yuan2013} and $x$ and $y$ are the observed extincted magnitudes. Variations in $e$ are somewhat degenerate with changes to the period--luminosity relation so it is preferable to keep $e$ fixed although it is allowed to vary in one model variant.

\subsection{Distance prior}
$p(s|\ell,b)$ in equation~\eqref{eqn::model_spec} is the prior on distance. \cite{Luri2018} emphasised the importance of using an appropriate prior when working with parallax data. Whilst it is tempting to simultaneously constrain a Galactic density model prior alongside calibrating the parallax data and period--luminosity relation, this is non-trivial as the sample is subject to complex selection effects (see Appendix~\ref{appendix::completeness}). For example, the effects of the Gaia scanning law are visible on small scales. More severe, however, is the incompleteness in the plane due to extinction. \cite{AstraatmadjaBJ2016b} explored using a fixed Galactic prior for finding distances from parallaxes with and without photometric information, but find that the simple exponentially decreasing space density prior $p(\bs{x})\propto\exp(-s/L)$ produces similar (but in the case of the Galactic centre regions significantly less biased) distance estimates and is significantly simpler to work with. \cite{BailerJones2018} used the exponentially decreasing space density prior to estimate distances from the full Gaia DR2 dataset, adopting a scalelength $L(\ell,b)$ that varies with on-sky position. The adopted functional dependence is determined in on-sky bins from a Gaia mock catalogue and fitted with a spherical harmonic series. \cite{BailerJones2021} updated this procedure for Gaia EDR3 by introducing an additional parameter into the prior ($p(\bs{x})\propto s^{\beta-2}\exp(-(s/L)^\alpha)$ with $\alpha$, $\beta$ and $L$ all functions of on-sky position. The simpler single-parameter exponentially decreasing prior is chosen adopting a spherical harmonic series in $\ln L$ given by
\begin{equation}
    \ln L/L_0 =
    \sum_{n=1}^{n_\mathrm{max}}\sum_{m=0}^{n}\Big[
    s_{nm}P_n^m(\sin b)\sin m\ell
    +
    c_{nm}P_n^m(\sin b)\cos m\ell
    \Big].
\end{equation}
Here $P_n^m(x)$ are associated Legendre polynomials and $s_{n0}=0$. A QR re-parametrization for this series is used which significantly improves sampling\footnote{Stan Development Team. 2018. Stan Modeling Language Users Guide and Reference Manual, Version 2.18.0. \href{http://mc-stan.org}{http://mc-stan.org}.}. The $P_n^m(\sin b)\sin m\ell$ and $P_n^m(\sin b)\cos m\ell$ terms are combined into a single matrix $\mathbfss{M}$ of dimensions $(N_\mathrm{data},N_\mathrm{series})$ where $N_\mathrm{series}=n_\mathrm{max}(n_\mathrm{max}+2)$, and the coefficients $s_{nm}$ and $c_{nm}$ into a vector $\bs{S}$ of length $N_\mathrm{series}$. $\mathbfss{M}=\mathbfss{QR}$ is decomposed into the thin QR decomposition and then samples are taken in the transformed vector $\bs{\tilde S}=\mathbfss{R}\bs{S}$. A shrinkage prior is placed on $\bs{\tilde S}\sim\mathcal{N}(0,\tau)$ where $\tau$ follows a unit half-Cauchy prior. The prior scalelength for datum $i$ is $\ln L/L_0=(\mathbfss{Q}\bs{\tilde{S}})_i$. $n_\mathrm{max}$ is set to $10$. The bar--bulge region is not used in the modelling to avoid biases introduced by an inappropriate prior for this region.

\subsection{Parallax zeropoint model}\label{sec::zpt}

As reported initially by \cite{Lindegren2018} for Gaia DR2 and by \cite{EDR3_ZPT} and \cite{Fabricius2021} for Gaia EDR3, the reported Gaia parallaxes and proper motions have zeropoint offsets and typically underestimated uncertainties due to limitations in the instrument and attitude modelling. \cite{EDR3_ZPT} reported an approximation for the zeropoint offset of the Gaia EDR3 parallaxes using samples of quasars, binaries and stars in the LMC. The sources with five- and six-parameter astrometric solutions were treated separately. The zeropoint correction was approximated as a function of $G$ magnitude, ecliptic latitude and colour (using $\nu_\mathrm{eff}$ for the five-parameter solutions and the pseudo-colour for the six-parameter solutions). The implementation is available at \url{https://gitlab.com/icc-ub/public/gaiadr3_zeropoint}. Several works \citep{Riess2021,Zinn2021,Huang2021} have validated the \cite{EDR3_ZPT} corrections, typically with some adjustment needed for bright stars ($G\lesssim11$). \cite{Groenewegen2021} presented an independent analysis of the Gaia EDR3 parallax zeropoint using a sample of quasars and wide binaries. This analysis differed from that presented by \cite{EDR3_ZPT} by not separating five- and six-parameter solutions, and using on-sky bins rather than polynomials to capture the spatial dependence of the zeropoint. \cite{MaizApellaniz2021} carried out a similar investigation of the Gaia EDR3 zeropoint to \cite{EDR3_ZPT} using a sample of open clusters, globular clusters and Magellanic Cloud data finding agreement with \cite{EDR3_ZPT} for faint objects ($G>13$) but some discrepancy for the brighter objects.

In summary, these previous analyses have shown that the Gaia EDR3 parallax zeropoint, $\varpi_0$, is observed to vary at the $\sim30\,\mu\mathrm{as}$ level as a function of colour, magnitude, on-sky position and the type of astrometric solution \citep{EDR3_ZPT}. Ideally, all possible variations would be included in the modelling here and the parallax zeropoint behaviour simultaneously constrained. However, initial tests demonstrated that magnitude dependence of $\varpi_0$ cannot be simultaneously constrained alongside the period--luminosity relation. A similar phenomenon was reported by \cite{ChanBovy2020}. In a similar vein, the variation of the parallax zeropoint with on-sky position is degenerate with the on-sky distance prior variation $p(s|\ell,b)$ \cite[again see][]{ChanBovy2020}. Without additional information (e.g. other tracer populations) the magnitude or on-sky dependence of the zeropoint are not constrained so instead previously determined zeropoint models are used with some additional colour dependence i.e. $\varpi_{0,i}=\varpi^f_{0,i}(G,\ell,b)+\varpi^e_{0,i}(\nu_\mathrm{eff})$ where $i\in\{5,6\}$ denotes whether five- or six-parameter astrometric solutions are considered. For the base model $\varpi^f_{0,i}$, three options are used:
\begin{enumerate}
    \item the zeropoint corrections of \cite{EDR3_ZPT} evaluated at $\nu_\mathrm{eff}=1.25\,\mu\mathrm{m}^{-1}$ (this wavenumber is within the interpolation grid for both five- and six-parameter solutions) accounting for the $15\,\mu\mathrm{as}$ overestimate reported by \cite{Riess2021} and \cite{Zinn2021} for $G<10.8$,
    \item the colour-independent Healpix level 1 corrections from \cite{Groenewegen2021} (also incorporating the inflation of uncertainties he suggested) and
    \item the zeropoint model from \cite{MaizApellaniz2021} evaluated at $\nu_\mathrm{eff}=1.25\,\mu\mathrm{m}^{-1}$.
\end{enumerate}
For the additional modelled colour-dependent zeropoint, $\varpi^e_{0,i}(\nu_\mathrm{eff})$, a quadratic is used with different parameters for the five- and six-parameter solutions such that in summary the model is
\begin{equation}
    \varpi_{0,i}(G,\nu_\mathrm{eff},\ell,b) =
\varpi^f_{0,i}(G,\ell,b)
    +\sum_{j=0}^{j=2} q_{i,j}(\nu_\mathrm{eff}-1.1\,\mu\mathrm{m}^{-1})^j.
\end{equation}
There are then three free parameters $q_{i,j}$ for each of the five- and six-parameter solutions.

\subsubsection{Parallax uncertainty underestimate model}
For the scaling factor of the parallax uncertainties, $f_\varpi$, two quadratics in $G$ and $\nu_\mathrm{eff}$ for the five- and six-parameter solutions are used:
\begin{equation}
    \ln f_{\varpi,i}(G,\nu_\mathrm{eff}) = \sum_{k,l\in\{0,1,2\}} r_{i,k,l}(G-14)^k(\nu_\mathrm{eff}-1.1\,\mu\mathrm{m}^{-1})^l,
\end{equation}
where $i\in\{5,6\}$. This choice is motivated by the Gaia astrometric performance being sensitive to colour and magnitude. As highlighted in Section~\ref{section::data}, the parallax uncertainties may also be underestimated due to AGB photocentre wobble. In Appendix~\ref{appendix::gaia_scanning} an additional parallax-dependent term is included in $f_\varpi$ which does not affect the overall period--luminosity relation fits.

\subsection{Implementation}

The models are implemented in \textsc{Stan} \citep{stan} using the python interface \textsc{PyStan} \footnote{Stan Development Team. 2018. PyStan: the Python interface to Stan, Version 2.17.1.0. \href{http://mc-stan.org}{http://mc-stan.org}.}. The following priors are adopted:\begin{enumerate}
\item
$a_m\sim\mathcal{N}(a_{m,\mathrm{LMC}},0.5)$,
\item
$(b_m,c_m)\sim\mathcal{N}((b_{m,\mathrm{LMC}},c_{m,\mathrm{LMC}}),25\Sigma_{bc,\mathrm{LMC}})$,
\item
$F\sim\mathcal{N}(F_\mathrm{LMC},((3,1,1)\otimes(3,1,1))\Sigma_{F,\mathrm{LMC}})$ where $F=(\sigma_{2.3},m_{\sigma-},m_{\sigma+})$,
\item $q_{i,j}\sim\mathcal{N}(0,1)$,
\item$r_{i,k,l}\sim\mathcal{N}(0,3)$,
\item$\sigma_{\varpi,0}\sim\mathcal{N}(-4.6,1.5)$,
\item$\ln\sigma_{\mu,0,2}\sim\mathcal{N}(0.5,0.5)$,
\item$\ln \vartheta_2\sim\mathcal{N}(-4.6, 1.5)$,
\item
$\ln L_0\sim\mathcal{N}(1.1,0.6)$,
\item$\bs{\tilde{S}}\sim\mathcal{N}(\bs{0},\tau\times\bs{1})$, $\tau\sim\mathcal{C}(0,1)$ (a unit Cauchy prior),
\item and when required $e\sim\mathcal{N}(e_0,0.05e_0)$ where $e_0$ is from \cite{WangChen2019}. 
\end{enumerate}
The LMC fits from Appendix~\ref{appendix::lmc} have been used as weak priors on the slopes $(b_m,c_m)$ and error model parameters $F=(\sigma_{2.3},m_{\sigma-},m_{\sigma+})$. For $(b_m,c_m)$ and $\sigma_{2.3}$ a generous $5$ and $3$ times the LMC fit uncertainty is used respectively as the prior width.
Instead of performing the integration in equation~\eqref{eqn::zeropoint_model}, the logarithm of the true parallax of each star minus the zeropoint offset in magnitude, $-\ln s_i - 0.2\ln(10)(m_\mathrm{abs}(P_i)-m_\mathrm{abs,LMC}(P_i))$, is sampled (accounting for the additional Jacobian factor of $s$ due to sampling in $\ln s$). This combination of parameters minimises the correlations in the likelihood leading to more efficient sampling.

\section{Results}\label{section::results}
\begin{figure*}
    \centering
    \includegraphics[width=\textwidth]{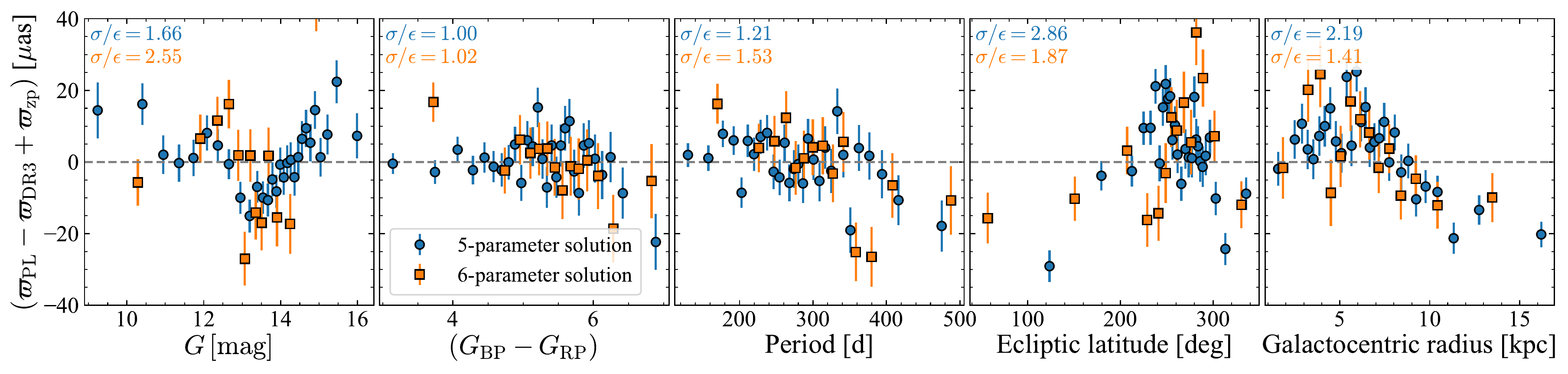}
    \caption{Difference between parallaxes from the fitted O-rich Mira variable $W_{Ks,J-Ks}$ period--luminosity relations (fourth row of Table~\ref{tab:plr_results}) and the zeropoint-corrected Gaia DR3 parallaxes \protect\citep[using the][with an additional colour-dependent term]{EDR3_ZPT}. The median and uncertainty for 30 (15) equally-populated bins for 5(6)-parameter astrometric solutions are shown as blue circles (orange squares). The annotation in each panel shows the standard deviation of the estimates over the typical error (i.e. a measure of any additional bias).}
    \label{fig:bias}
\end{figure*}

\begin{figure*}
    \centering
    \includegraphics[width=\textwidth]{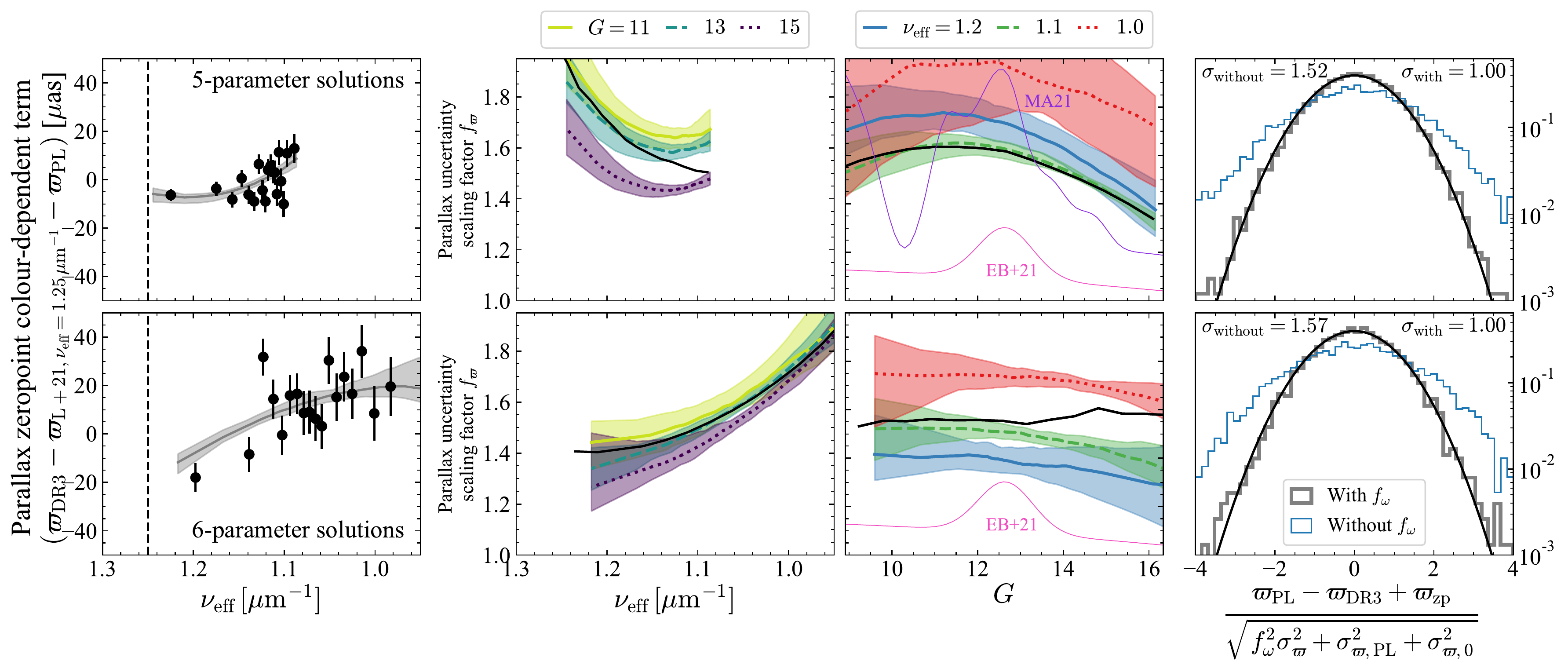}
    \caption{Results of fitting the Gaia EDR3 parallax zeropoint and the parallax uncertainty scaling factor. The results use the Wesenheit $W_{Ks,J-Ks}$ magnitude relation (fourth row of Table~\ref{tab:plr_results}). The top (bottom) row corresponds to Gaia EDR3 5(6)-parameter solutions. The left plots show the fitted colour-dependent parallax zeropoint variation in addition to that reported by \protect\cite{EDR3_ZPT} at $\nu_\mathrm{eff}=1.25\,\mu\mathrm{m}^{-1}$. The black points show the mean difference between the corrected DR3 parallaxes and the parallax computed from the period--luminosity relation for the data. The middle two panels show the parallax uncertainty scaling factor as a function of $\nu_\mathrm{eff}$ and $G$ (models from \protect\cite{ElBadryRix2021} and \protect\cite{MaizApellaniz2021} are shown). The right panel shows the distribution of the parallax residual between the period--luminosity relation and the corrected Gaia DR3 astrometry divided by the combined error with (grey) and without (blue) the parallax scaling factor.}
    \label{fig:systematic}
\end{figure*}

\begin{figure}
    \centering
    \includegraphics[width=\columnwidth]{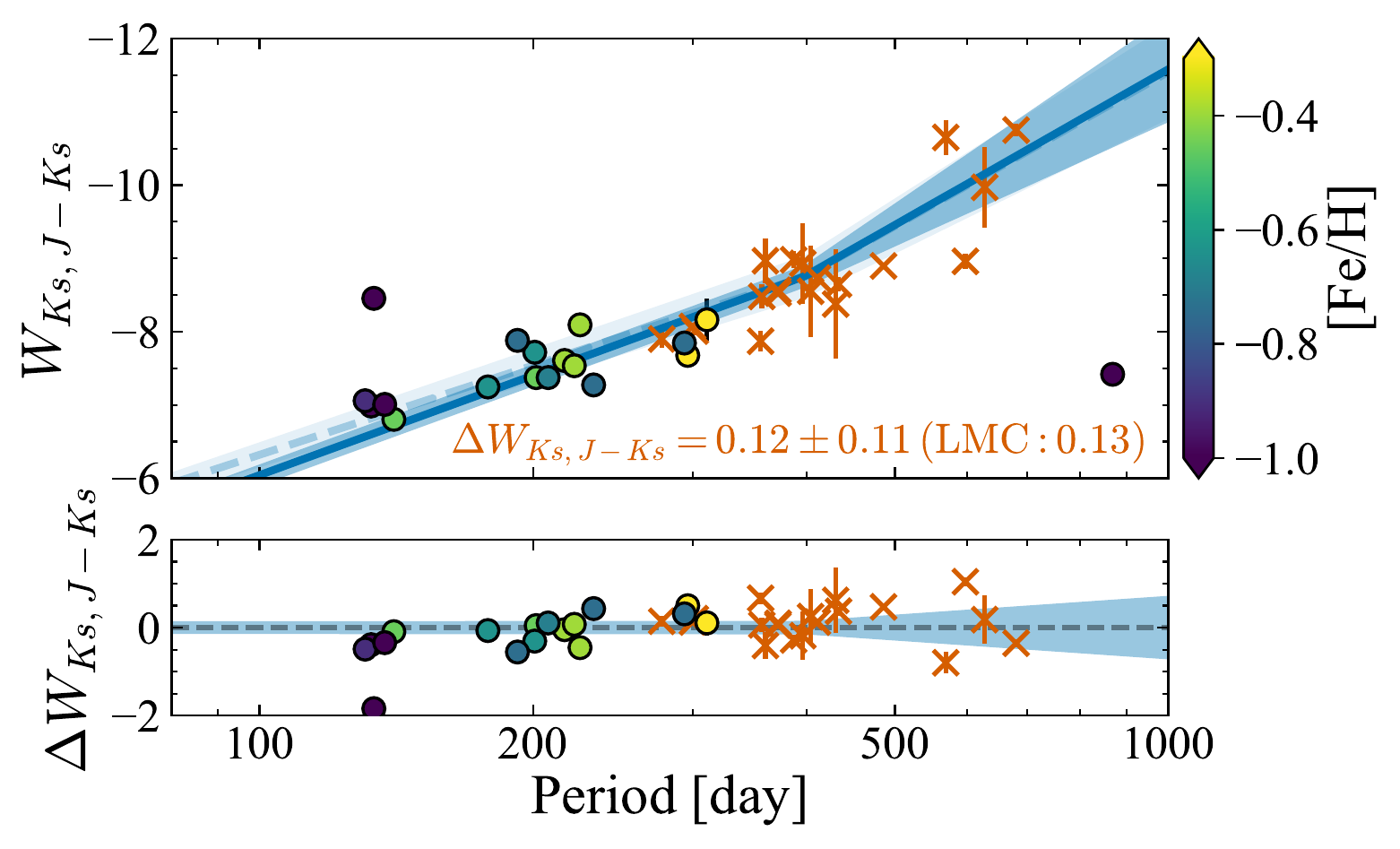}
    \caption{Comparison between the derived period--luminosity relation (using the Wesenheit index $W_{Ks,J-Ks}$) and (i) VLBI parallax measurements for O-rich Mira variables (orange crosses) and (ii) likely O-rich Mira variable globular cluster members (dots coloured by their metallicity). The solid blue line and shaded region gives the Milky Way model and its scatter, whilst the dashed line shows the LMC measurement. The numbers show the median offsets (data -- model, as shown in the lower panel) with respect to the MW (and LMC) relations for VLBI measurements. The uncertainty is the inverse-variance-weighted error from the VLBI parallaxes, the photometric uncertainties and the scatter about the period--luminosity relation (due to using single-epoch observations).}
    \label{fig:vlbi}
\end{figure}

The results of the period--luminosity relation fitting are presented in Table~\ref{tab:plr_results} and the associated parameters for the Gaia EDR3 systematics in Table~\ref{tab:zpt_results}. The default base parallax zeropoint model is option (i) from Section~\ref{sec::zpt} that primarily uses the correction from \cite{EDR3_ZPT}. As previously reported elsewhere \citep[see][]{Iwanek2021}, the gradients, $b$ and $c$, steepen for longer wavelengths. The scatter $\sigma_{2.3}$ also decreases with wavelength. The Wesenheit models typically agree very well with those computed using the single band models (e.g. $W_{Ks,J-Ks}=K_s-e(J-K_s)$ for $a$ gives $a=-7.41$ compared to $a=-7.40$) suggesting circumstellar dust in the O-rich Mira variables is unimportant (if it has a similar reddening law to the interstellar medium). This tallies with the results of \cite{Bladh2015} who showed using a grid of theoretical models that circumstellar dust around O-rich stars is mostly transparent in optical and near-infrared bands. When comparing the Milky Way results to linear fits of the LMC period--luminosity relation (see Table~\ref{tab:period_luminosity_relations_lmc}), consistently fainter zeropoints (higher $a$) of the Milky Way period--luminosity relation are found ($\Delta J, \Delta H, \Delta K_s)_{\log_{10}P=2.3}=(0.19,0.19,0.11)\,\mathrm{mag}$) but these differences are well within the $0.5\,\mathrm{mag}$ prior width.  Typically the gradients ($b$ and $c$) are found to be steeper for the Milky Way relations but it doesn't appear the broad LMC prior is causing any tension (possibly for $c$ for the $J$ and $H$ relations as illustrated in Fig.~\ref{fig:comp} although this may be more linked to the selection of LMC sources). Note that due to the gradient differences, the magnitude difference between the Milky Way and LMC period--luminosity relations decreases with increasing period. As evidenced in Fig.~\ref{fig:comp}, literature quadratic model fits to the LMC Mira variables show smaller offsets with respect to the Milky Way linear fits particularly around the characteristic $200$ day period. It could be that more flexible models produce less tension between the two period--luminosity relations. The differences between the LMC and Milky Way relations in the context of their population differences are discussed further in Section~\ref{sec::pop}.

Fig.~\ref{fig:bias} shows the residuals of the parallaxes predicted from the $W_{Ks,J-Ks}$ relation from Table~\ref{tab:plr_results} compared to the zeropoint-corrected Gaia EDR3 parallaxes. We see in general the satisfactory agreement demonstrating the quality of the period--luminosity relation. However, residuals and trends remain. The left panel of Fig.~\ref{fig:systematic} shows the fitted Gaia EDR3 zeropoint term for this model. For 5-parameter solutions small corrections ($\lesssim5\,\mu \mathrm{as}$) are required on top of the \cite{EDR3_ZPT} corrections. For 6-parameter solutions however, larger corrections are required that typically increase as the sources get redder. This implies the recommended zeropoint corrections evaluated at $\nu_\mathrm{eff}=1.25\,\mu\mathrm{m}^{-1}$ do not apply well to redder sources with 6-parameter solutions and appear to overcorrect the parallaxes. Similar behaviour is found for the other models shown in Table~\ref{tab:zpt_results}. Fig.~\ref{fig:systematic} shows the factor by which the parallax uncertainties must be inflated to account for the observed spread about the period--luminosity relation. In agreement with previous work \citep[e.g.][]{ElBadryRix2021,MaizApellaniz2021,Andriantsaralaza2022} an inflation of the parallax uncertainties is required. The behaviour is relatively flat with colour (although increases quite steeply for very red sources with 6-parameter solutions). For 5-parameter solutions the factor is around $1.3$ for brighter $(G\sim9)$ and fainter ($G\sim16$) sources but for more intermediate ($G\sim12$ as reported in Table~\ref{tab:zpt_results}) the factor increases to around $1.6$. This behaviour mirrors that found by \cite{ElBadryRix2021} using wide binaries although larger factors are found that are more consistent with the results of \cite{MaizApellaniz2021}. A fit using only five-parameter solutions from Gaia produces very similar results for the Gaia systematic parameters and the period--luminosity relations suggesting although the six-parameter solutions appear more biased, they are not affecting the overall fit too strongly.

As shown in Fig.~\ref{fig:bias}, some residuals in the fits remain, particularly as a function of $G$ and on-sky location. In Section~\ref{sec::pop} possible population effects producing such residuals are discussed. However, particularly in the case of the residuals with $G$ where there are features around $G\approx13$, some level of residual at the $10\,\mu\mathrm{as}$ level appears to arise from the Gaia EDR3 zeropoint model. The \cite{Groenewegen2021} and \cite{MaizApellaniz2021} zeropoint corrections have been used as variants of the base model. As seen in Table~\ref{tab:plr_results}, this can produce changes in the period--luminosity zeropoint of $\sim0.1\,\mathrm{mag}$. However, both of these alternatives also produce larger residual features with $G$. The residual scatter is quantified using the inverse-variance-weighted bin-to-bin scatter in the mean divided by the mean uncertainty in the mean residual in each bin ($\sigma/\epsilon$). For the five-parameter solutions binned as a function of $G$, the base $K_s$ model produces $\sigma/\epsilon=1.6$ for the \cite{EDR3_ZPT} model whilst this inflates to  $\sigma/\epsilon=2.3$ and $\sigma/\epsilon=2.4$ for \cite{Groenewegen2021} and \cite{MaizApellaniz2021} models respectively. The largest problems occur around $G\approx12-13$. As noted previously, simultaneously fitting the magnitude (and on-sky dependence) of the parallax zeropoint was found to be degenerate with parameters of the period--luminosity relation. A future approach should adopt a more flexible model for the parallax zeropoint constrained to be small by a careful choice of prior.

Table~\ref{tab:plr_results} also displays results for the \cite{Yuan2013} extinction law. As with the case using the \cite{WangChen2019} extinction law, the Wesenheit magnitude zeropoint is very similar ($\lesssim0.01\,\mathrm{mag}$) to that computed using the single band results suggesting the adopted extinction law doesn't change the conclusions significantly. 
The sensitivity to the RUWE cut (by default $1.4$) has been investigated. Relaxing to RUWE $<2$ produces a slightly steeper fainter $K_s$ relation that is consistent with the RUWE $<1.4$ relation for $P>200$ day but deviates slightly at the shorter period end. Many of the higher RUWE stars are located near the midplane and so potentially are affected by high source density.
Results are also reported for C-rich Mira variables. As done in Appendix~\ref{appendix::lmc} for the C-rich LMC Mira variables, a quadratic period--luminosity relation $m_\mathrm{abs}(P)=a+b(\log_{10}P-2.3)+c(\log_{10}P-2.3)^2$ with a linear scatter $\sigma_\mu(P)=\sigma_{2.3}+m_{\sigma-}(\log_{10}P-2.3)$ is used. C-rich Mira variables are typically not employed as distance indicators due to their larger scatter in the period--luminosity relation compared to the O-rich Mira variables. Here it is found that in the Wesenheit magnitude $W_{Ks,J-Ks}$ the C-rich Mira variables at short periods ($\lesssim300\,\mathrm{days}$) are $\sim0.4\,\mathrm{mag}$ brighter than the O-rich relations (also seen in the LMC, Appendix~\ref{appendix::lmc}) and the scatter is comparable to that of the O-rich Mira variables. At longer periods ($\gtrsim400\,\mathrm{days}$) the period--luminosity relation flattens (or possibly even turns over, see Appendix~\ref{appendix::lmc}).

\begin{table*}
\rowcolors{1}{}{lightgray}
    \caption{Period--luminosity relations for O-rich Mira variables. The period--luminosity relations have the form $a+b(\log_{10}P-2.3)$ for $\log_{10}P\leq2.6$ and $a+0.3b+c(\log_{10}P-2.6)$ for $\log_{10}P>2.6$ with scatter $\sigma=\sigma_{2.3}+m_{\sigma-}(\log_{10}P-2.3)$ for $\log_{10}P\leq2.6$ and $\sigma=\sigma_{2.3}+0.3m_{\sigma-}+m_{\sigma+}(\log_{10}P-2.6)$ for $\log_{10}P>2.6$ (note for the C-rich relation a quadratic relation $a+b(\log_{10}P-2.3)+c(\log_{10}P-2.3)^2$ with a linear scatter $\sigma=\sigma_{2.3}+m_{\sigma-}(\log_{10}P-2.3)$ for all periods is used instead). Here $P$ is in days. $L_0$ is the logarithm of the mean of the exponential of the distance prior scalelength in kpc.
    The first section of rows show results for the 2MASS $JHK_s$ bands and using the Wesenheit indices $W_{x,y-x}=x - e(y-x)$. All of these models use the default setup correcting the Gaia EDR3 parallaxes using the \protect\cite{EDR3_ZPT} zeropoints evaluated at $\nu_\mathrm{eff}=1.25\,\mu\mathrm{m}^{-1}$ as a base model and fitting for an additional colour-dependent term. The second section shows model variations: (i) using the \protect\cite{Yuan2013} extinction coefficients for the $W_{Ks,J-Ks}$ relation, (ii) allowing the extinction coefficient $e$ to vary for the $W_{Ks,J-Ks}$ relation, (iii) using stars with RUWE $<2$ for the $K_s$ relation, (iv) using the \protect\cite{Groenewegen2021} parallax zeropoint correction as a base model for the $K_s$ relation and (v) using the \protect\cite{MaizApellaniz2021} parallax zeropoint correction evaluated at $\nu_\mathrm{eff}=1.25\,\mu\mathrm{m}^{-1}$ as a base model for the $K_s$ relation. The final section gives the $W_{Ks,J-Ks}$ relation for C-rich stars using the \protect\cite{EDR3_ZPT} zeropoints as a base model.}
    \centering
    \begin{tabular}{l|ccccccccc}
    Band/Model & $a$ & $b$ & $c$& $\ln\sigma_{2.3}$& $m_{\sigma-}$&$m_{\sigma+}$&$L_0$&$e$\\
    \hline
$J$&$-5.66\pm0.02$&$-3.56\pm0.06$&$-2.42\pm0.48$&$-1.75\pm0.06$&$-0.02\pm0.04$&$2.35\pm0.30$&$1.51\pm0.05$&$-$\\
$H$&$-6.46\pm0.02$&$-3.84\pm0.06$&$-4.08\pm0.48$&$-1.77\pm0.07$&$0.06\pm0.04$&$1.86\pm0.25$&$1.51\pm0.05$&$-$\\
$K_s$&$-6.85\pm0.02$&$-4.22\pm0.06$&$-5.52\pm0.47$&$-1.84\pm0.07$&$0.04\pm0.04$&$1.58\pm0.28$&$1.51\pm0.05$&$-$\\
$W_{Ks,J-Ks}$&$-7.40\pm0.02$&$-4.52\pm0.06$&$-7.06\pm0.45$&$-1.90\pm0.06$&$0.01\pm0.04$&$1.42\pm0.24$&$1.50\pm0.05$&$0.47$\\
$W_{Ks,H-Ks}$&$-7.40\pm0.02$&$-4.76\pm0.06$&$-7.81\pm0.48$&$-1.91\pm0.07$&$-0.04\pm0.04$&$0.97\pm0.22$&$1.50\pm0.05$&$1.47$\\
$W_{H,J-H}$&$-7.34\pm0.02$&$-4.11\pm0.07$&$-6.13\pm0.49$&$-1.85\pm0.05$&$0.05\pm0.03$&$1.55\pm0.24$&$1.50\pm0.05$&$1.17$\\
\hline
Yuan $e_i$ $W_{Ks,J-Ks}$&$-7.74\pm0.02$&$-4.66\pm0.06$&$-8.04\pm0.46$&$-1.85\pm0.07$&$0.02\pm0.03$&$1.38\pm0.24$&$1.51\pm0.05$&$0.74$\\
Free $e_i$ $W_{Ks,J-Ks}$&$-7.36\pm0.02$&$-4.51\pm0.06$&$-6.95\pm0.46$&$-1.87\pm0.05$&$0.02\pm0.04$&$1.39\pm0.25$&$1.50\pm0.05$&$0.45\pm0.02$\\
RUWE $<2$ $K_s$&$-6.80\pm0.02$&$-4.37\pm0.06$&$-5.11\pm0.38$&$-1.88\pm0.06$&$-0.01\pm0.03$&$1.72\pm0.27$&$1.61\pm0.04$&$-$\\
G21 $K_s$&$-6.73\pm0.02$&$-4.12\pm0.06$&$-5.53\pm0.47$&$-1.79\pm0.06$&$0.06\pm0.04$&$1.56\pm0.29$&$1.42\pm0.05$&$-$\\
MA22 $K_s$&$-6.76\pm0.02$&$-4.18\pm0.06$&$-5.54\pm0.46$&$-1.83\pm0.06$&$0.04\pm0.03$&$1.61\pm0.28$&$1.43\pm0.05$&$-$\\
C-rich $W_{Ks,J-Ks}$&$-7.73\pm0.09$&$-4.00\pm0.51$&$0.59\pm1.07$&$-1.72\pm0.15$&$0.48\pm0.04$&$-$&$2.64\pm0.12$&$0.47$\\
    \end{tabular}
    \label{tab:plr_results}
\end{table*}

\begin{table}
    \centering
\setlength{\tabcolsep}{4.5pt}
\rowcolors{1}{}{lightgray}
\caption{
Parallax zeropoint and uncertainty model results for the models shown in Table~\ref{tab:plr_results}. $\varpi_{i,\mathrm{zp}}$ gives the sky-averaged Gaia EDR3 parallax zeropoint in addition to the assumed model for $i$-parameter astrometric solutions (at $\nu_\mathrm{eff}=1.15\,\mu m^{-1}$ for 5-parameter solutions and $\nu_\mathrm{eff}=1.05\,\mu m^{-1}$ for 6) in units of $\mu\mathrm{as}$. $f_{i,\varpi}$ gives the scaling of the parallax errors for the $i$-parameter solutions (again at the representative colours and magnitude $G=12$). The assumed base parallax zeropoint model is by default the \protect\cite{EDR3_ZPT} correction at $\nu_\mathrm{eff}=1.25\,\mu m^{-1}$ except for G21 that uses the \protect\cite{Groenewegen2021} parallax zeropoint correction and MA22 that uses \protect\cite{MaizApellaniz2021} parallax zeropoint correction evaluated at $\nu_\mathrm{eff}=1.25\,\mu\mathrm{m}^{-1}$.}

\begin{tabular}{l|cccc}
Band/Model&$\varpi_{5,\mathrm{zp}}$&$\varpi_{6,\mathrm{zp}}$&$f_{5,\varpi}$&$f_{6,\varpi}$\\
\hline
$J$&$-2\pm1$&$21\pm2$&$1.52\pm0.04$&$1.62\pm0.02$\\
$H$&$-2\pm1$&$22\pm3$&$1.56\pm0.04$&$1.61\pm0.02$\\
$K_s$&$-3\pm1$&$17\pm3$&$1.58\pm0.04$&$1.61\pm0.02$\\
$W_{Ks,J-Ks}$&$-4\pm2$&$15\pm2$&$1.61\pm0.04$&$1.60\pm0.02$\\
$W_{Ks,H-Ks}$&$-6\pm2$&$10\pm3$&$1.60\pm0.03$&$1.59\pm0.02$\\
$W_{H,J-H}$&$-5\pm1$&$18\pm2$&$1.58\pm0.05$&$1.62\pm0.02$\\
\hline
Yuan $e_i$ $W_{Ks,J-Ks}$&$-6\pm1$&$10\pm3$&$1.62\pm0.04$&$1.59\pm0.02$\\
Free $e_i$ $W_{Ks,J-Ks}$&$-4\pm1$&$15\pm3$&$1.60\pm0.04$&$1.60\pm0.03$\\
RUWE $<2$ $K_s$&$-4\pm1$&$15\pm2$&$1.61\pm0.03$&$1.59\pm0.02$\\
G21 $K_s$&$-25\pm1$&$0\pm3$&$1.57\pm0.05$&$1.59\pm0.02$\\
MA22 $K_s$&$-1\pm1$&$9\pm3$&$1.60\pm0.04$&$1.61\pm0.02$\\
C-rich $W_{Ks,J-Ks}$&$26\pm9$&$16\pm15$&$1.49\pm0.30$&$0.35\pm0.18$\\
    \end{tabular}
    \label{tab:zpt_results}
\end{table}

\subsection{Comparison with VLBI parallaxes}
An alternative to the astrometric distances of Mira variables from Gaia are interferometric measurements from very long-baseline interferometry (VLBI). As VLBI is able to resolve AGB stars, any systematics from photocentre wobble are minimal (see Section~\ref{section:photocentre}). In combination with Hipparcos parallaxes, \cite{Whitelock2008} used the available VLBI measurements to calibrate the $K$-band period--luminosity relation. Since then, several more AGB stars have had VLBI measurements. \cite{Andriantsaralaza2022} has inspected the Gaia DR3 astrometry of AGB stars with VLBI measurements. Fig.~\ref{fig:vlbi} displays the absolute $W_{Ks,J-Ks}$ measurements against period for the recent VLBI compilations of AGB stars from \cite{Xu2019} and \cite{Hirota2020}, preferentially using the results from \cite{Hirota2020} in the case of duplicates. The periods are from VSX \citep{Watson2006} and magnitudes from 2MASS. Only O-rich Mira variables as defined by the selection in Section~\ref{section::data} are displayed. FV Boo is removed as it appears to be a clear outlier as noted by \cite{Kamezaki2016} and there are concerns it displays additional variability due to potentially being in a binary system \citep{Kamezaki2016}. The inverse-variance-weighted offset of the absolute Wesenheit magnitudes computed using VLBI parallaxes with respect to the period--luminosity relation is $(0.12\pm0.11)\,\mathrm{mag}$. Here the error is the inverse-variance-weighted error from the photometric uncertainties, the VLBI parallax uncertainties and the scatter model due to using single epoch observations. Although the measurements are consistent, the VLBI measurements are slightly fainter than the Gaia-derived Milky Way trend, possibly as they are a dustier or a more metal-rich population compared to the Gaia-selected O-rich Mira variables \citep[also seen in][]{Whitelock2008}. A concern is that many of the 2MASS measurements are saturated for these bright stars. \cite{Whitelock2000} and \cite{Whitelock2008} provide $JHK$ measurements in the SAAO system. Transformation to the 2MASS system is not simple for these very red sources but using the relations in \cite{Koen2007} the offset with respect to the derived period--luminosity relation is $(0.15\pm0.05)\,\mathrm{mag}$. However, it should be noted that \cite{Koen2007} find brighter stars appear to have larger differences between SAAO $K$ and 2MASS $K_s$ ($K_s$ smaller than $K$) which could explain some of this difference.

\subsection{Population variations}\label{sec::pop}
It has been found that the Milky Way O-rich Mira variable relations derived here are typically slightly fainter than those derived for the LMC (see Appendix~\ref{appendix::lmc}) particularly at the short period end due to a steeper gradient. One interpretation of this result is that there is variation of the O-rich Mira period--luminosity relation with stellar population, in particular with the age and metallicity of the population.
Typically, it has been found that population effects are quite minimal for the Mira variables, particularly in the near- and mid-infrared \citep[\protect{$K_s$, $[3.6]$ and $[4.5]$,}][]{Whitelock2008,Goldman2019,Menzies2019} or using bolometric magnitudes \citep[e.g.][]{Andriantsaralaza2022}. However, there are suggestions from theoretical results that there can be more significant variations in the period--luminosity relations \citep{Wood1990,Qin2018} particularly for the bluer bands, $J$ and $H$, that are also investigated here. 

\subsubsection{Comparison with theoretical models}
Fundamentally, it is expected that a given mass and radius combination will give rise to the same fundamental period. \cite{Wood1990} demonstrated using a linear calculation how the period of a Mira variable is related to the luminosity $L$, metallicity $Z$ and mass as $M$ as $P\propto L^{1.59}Z^{0.46}M^{-1.55}$. If it is assumed that an AGB star will only pulsate with Mira-like oscillations when it reaches a certain radius (or narrow radial range) for its given mass, this gives us a relationship between bolometric magnitude $M_\mathrm{bol}$ and metallicity at fixed radius $\Delta M_\mathrm{bol}=0.72\Delta\log_{10}Z$ \citep[see also figure 12 of][for a similar calculation with a very similar result]{Trabucchi2019}. As noted by \cite{Wood1990}, the corresponding change in near-infrared magnitudes with metallicity is smaller than the change in bolometric magnitude. Assuming Mira variables of fixed radius but different metallicities are black-bodies with varying effective temperatures $\Delta \log_{10} T_\mathrm{eff}\approx0.072\Delta\log_{10}Z$ the magnitude differences are $(\Delta M_J,\Delta M_H,\Delta M_{Ks})=(0.68,0.52,0.42)\Delta \log_{10}Z$. Taking the typical $Z_\mathrm{LMC}=0.5Z_\mathrm{MW}$, the magnitude differences are $(\Delta M_J,\Delta M_H,\Delta M_{Ks})=(0.20,0.16,0.13)$ in rough agreement with the zeropoint differences found. 

It is anticipated that linear calculations will differ most strongly from non-linear calculations in the computation of period at a given mass and radius \citep{Trabucchi2021} making these arguments valid irrespective or whether linear or non-linear calculations are considered. However, \cite{Trabucchi2019} has shown that, particularly for the fundamental mode, the composition (metallicity, C/O ratio) can affect the period at fixed mass and radius. For instance, making a star more metal-rich (increasing from typical LMC to typical Milky Way metallicity) or making a star carbon-rich (increasing C/O from $0.55$ to $\sim3$) decreases the period by $\sim10\percent$ (for a linear calculation). Therefore, period is not solely a function of mass and radius. In a similar vein, \cite{Feast1996} has questioned the validity of the assumption that a star of given mass reaches Mira-like oscillations at fixed radius independent of its metallicity as it is related to the mass loss. For a given initial mass and metallicity, an AGB star could reach the Mira pulsation stage with a different mass-radius combination that produces a similar period. However, there is evidence to suggest metallicity-dependence on mass loss is not a significant effect \citep[see][for a summary]{Hofner2018}.

Using $P\propto L^{1.59}Z^{0.46}M^{-1.55}$ and the period-mass-radius relation, the dependence of the effective temperature can be derived as $T_\mathrm{eff}\propto P^{-0.1}Z^{-0.073}M^{0.014}$ demonstrating that at fixed period the effective temperature is a weak function of the mass and more dependent upon metallicity. This then suggests even when the mass evolution at a given metallicity is poorly known, the metallicity of a Mira variable of fixed period will be related to its effective temperature and hence infrared colours \citep[this is corroborated by the fuller calculation of][that is considered later and that shows $J$, $H$ and $K_s$ at fixed period all have similar age dependence such that the gradient of $J-K_s$ with age is $\lesssim0.002\,\mathrm{mag}/\mathrm{Gyr}$]{Qin2018}. Using the blackbody model from before, the colour difference is found to be $\Delta(J-K_s)\approx0.26\Delta\log_{10}Z=-3.56\log_{10}T_\mathrm{eff}$. This is in agreement with PARSEC isochrones \citep{Bressan2012,Marigo2017} which suggest $\mathrm{d}(J-K_s)/\mathrm{d}[\mathrm{Fe}/\mathrm{H}]\approx0.2$. For the LMC sample, the mean colour $(J-K_s)\approx1.11$ at $\log_{10}P=2.3$ whilst for the Milky Way sample it is $\sim1.2$, which using the simplistic approach would translate into a $\sim0.4\,\mathrm{dex}$ metallicity shift. 

It seems from simple considerations that the derived differences between the LMC and Milky Way relations are consistent with linear pulsation calculations. However, the \cite{Wood1990} formulae have been criticized by \cite{Feast1992} as they fail to simultaneously explain the period-colour relation in the Milky Way/LMC and the period--metallicity relation observed in globular cluster Mira variables \citep{FeastWhitelock2000P}. Fig.~\ref{fig:vlbi} displays possible globular cluster members taken from the main Milky Way sample defined as within $3$ half-light radii of a known globular cluster \citep{Harris2010} with proper motions in each component consistent at the $4\sigma$ level with those determined by \cite{BaumgardtVasiliev2021}. It is clear this generous cross-match introduces a couple of non-members. A globular cluster period--metallicity gradient is visible where there is a collection of metal-poor stars at around $140$ day periods and a collection of more metal-rich stars at $300$ day periods. This is slightly puzzling but it should be noted that some globular clusters show Mira variables with a range of periods \citep{Matsunaga2007} suggesting we are seeing the effects of age-metallicity correlations and/or the impact of multiple populations in globular clusters.

\begin{figure}
    \centering
    \includegraphics[width=\columnwidth]{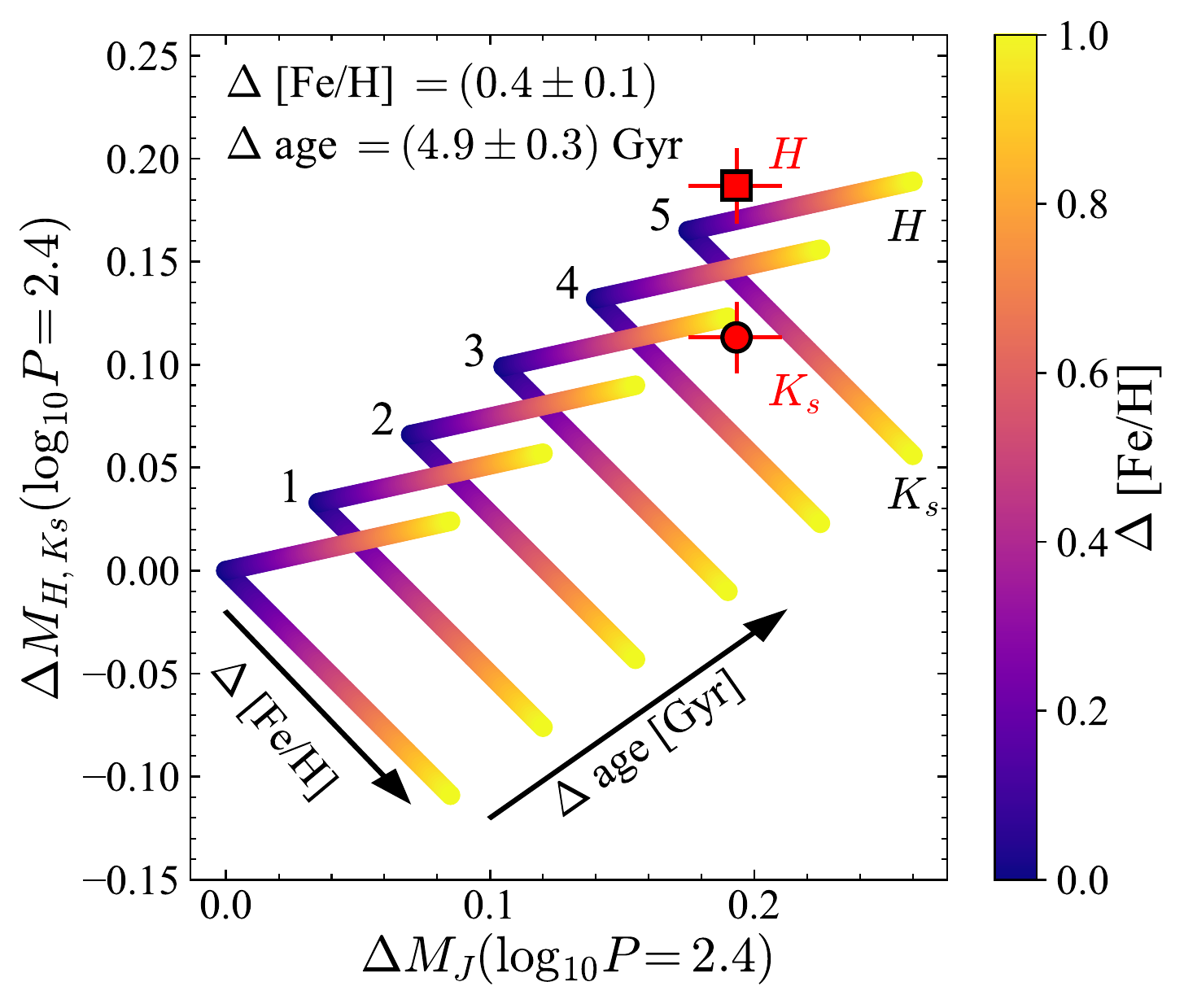}
    \caption{Differences between MW and LMC O-rich Mira variable period--luminosity relations at $\log_{10}P/\,\mathrm{day}=2.4$ in the $JHK_s$ bands compared to the linear pulsation models reported by \protect\cite{Qin2018}. A set of models with different age differences (as numbered in Gyr) and $\feh$ (coloured) are shown. The upwards diagonal sequences depict $\Delta M_{H}$ as a function of $\Delta M_J$ whilst the downwards diagonal sequences depict $\Delta M_{Ks}$. Given the measured differences the MW O-rich Mira variable population is found to be $(0.4\pm0.1)\,\mathrm{dex}$ more metal-rich and $(4.9\pm0.3)\,\mathrm{Gyr}$ older than the LMC population considered.}
    \label{fig:popvar}
\end{figure}

The previous arguments explained in simple terms why both magnitude and colour differences with varying metallicity at fixed period are to be expected for Mira variables. This can be elucidated further with a more sophisticated model. \cite{Qin2018} have used the linear pulsation models from \cite{Wood2014} combined with a relation for mass as a function of age, metallicity and helium abundance from \cite{Nataf2012} and the bolometric corrections from \cite{Casagrande2014} to derive gradients of $JHK_s$ magnitude with these quantities at fixed period ($\log_{10}P=2.4$, although they report similar gradients for other periods in the near infrared bands). These authors caution that the models are approximate and do not seem to explain the differences between Mira variables in the Galactic bulge and the LMC. Indeed, at fixed age and helium abundance, the models predict brighter $K_s$ with metallicity in contrast to the previous discussion. Nonetheless, in the absence of other models, they are used here. Again, although the period for a given mass and radius combination is affected by the linear approximation \citep[e.g.][]{Trabucchi2021}, the gradient of magnitude with age and metallicity at fixed period is more related to the gross stellar evolutionary properties. The models from \cite{Qin2018} are used to infer the age and metallicity difference between the Milky Way population and LMC population (see Appendix~\ref{appendix::lmc}) as shown in Fig.~\ref{fig:popvar}. Here it is assumed the helium abundance is similar in both systems. The combination of $J$ and $H$ differences provides little leverage for breaking age/metallicity differences but when combined with the comparatively smaller $K_s$ difference the LMC O-rich Mira variable population is found to be younger by $(4.9\pm0.3)\,\mathrm{Gyr}$ and more metal-poor by $(0.4\pm0.1)\,\mathrm{dex}$, somewhat consistent with expectation. There is evidence for a gap in the star formation history of the LMC and an increase in the star formation rate in the last $\sim1\,\mathrm{Gyr}$ based on the properties of its star clusters \citep{Jensen1988}, its chemical evolution \citep[e.g.][]{Hasselquist2021} and its photometrically-derived star formation history \citep{Javiel2005}.

Further evidence for variation in the zeropoint with metallicity (or more generally stellar population) comes from the globular clusters. Fig.~\ref{fig:vlbi} demonstrates that there is a weak tendency for the globular cluster members to get brighter as a function of metallicity relative to the LMC and Milky Way relations (or putting it another way, the globular clusters alone suggest a flatter period--luminosity slope). The lack of metal-rich shorter-period and metal-poor longer-period globular cluster Mira variables makes this conclusion somewhat uncertain. Using a globular cluster-calibrated period--luminosity relation, \cite{Feast2002} find a distance modulus for the LMC $\sim0.1\,\mathrm{mag}$ further than modern estimates suggest and \cite{Whitelock2008} find the $K_s$ period--luminosity relation for globular cluster members is $\sim0.1$ brighter than the LMC relation \citep[using the][LMC distance modulus]{Pietrzynski2019}, but in both cases the uncertainties were large. Finally, in Appendix~\ref{appendix::lmc} the period--luminosity relations for the Sagittarius dwarf spheroidal galaxy (Sgr dSph) and the Small Magellanic Cloud (SMC) are estimated. It is found that typically the (relatively few) O-rich Mira variables in these systems are slightly brighter than their presumably more metal-rich counterparts in the LMC in all bands particularly for periods greater than $250$ days \citep[corroborating the results of][]{Ita2004}. The steep period--luminosity relations typically found for the SMC mean for stars with periods less than $200$ days the SMC Mira variables are fainter than those in the LMC but these stars are comparatively rare.

\subsubsection{Population gradients within the samples}
We have seen how differences in period--luminosity relations between systems can be explained by population differences. However, the populations in the LMC and Milky Way are not homogeneous so similar gradients should be observed within these systems.

Fig.~\ref{fig:systematic} shows the variation of the zeropoint-corrected Gaia EDR3 parallax residual with respect to the estimates from the $W_{Ks,J-Ks}$ model of Table~\ref{tab:plr_results}. We see there is a tendency for the outer parts of the Galaxy to have larger Gaia parallaxes (smaller distances) than the period--luminosity relations suggest. This implies that for the outer disc the absolute $W_{Ks,J-Ks}$ needs to be fainter. Using the \cite{Qin2018} relations, inside-out formation \citep[a negative age gradient with radius,][]{Frankel2019,Grady2019} would imply $W_{Ks,J-Ks}$ gets brighter with Galactocentric radius, but a negative radial metallicity gradient produces the opposite effect although with a too weak $0.03\,\mathrm{mag}/\mathrm{dex}$ gradient. Neither age nor metallicity effects appear to explain the observations, although the exact slope reported by \cite{Qin2018} depends on the uncertain bolometric corrections for cool stars \citep{Casagrande2014} and \cite{Qin2018} themselves find inconsistencies between the theoretical models and the expectations for Mira variables in the Galactic bulge. The effect we are seeing could be driven by C-rich contamination that is more prevalent in the outer-disc. There are some very red stars ($H-K_s>0.7$) even after extinction correction. Typically removal of these redder sources makes the long period end of the period--luminosity relation brighter (note the bias in Fig.~\ref{fig:bias} at long periods which is somewhat alleviated by removing very dusty sources) but the trends with Galactocentric radius remain. A further cause could be incorrect extinction correction but there is no trend in the parallax residuals against extinction. It is clear from Fig.~\ref{fig:systematic} that systematic trends in $G$ and on-sky position are present (the inner and outer Galaxy samples have different mean $G$ magnitudes) and so potentially the cause of the Galactocentric radius trend is remaining systematics in the Gaia parallaxes and not due to any population differences. 

As previously highlighted, the metallicity of giant stars correlates well with their colour (\citealt{Qin2018} suggest that colours at fixed period are insensitive to age variations,
$\lesssim0.002\,\mathrm{mag}/\mathrm{Gyr}$, and nearly completely depend upon helium abundance and metallicity). 
Here, the impact of a colour term in the period--luminosity relations is investigated. Table~\ref{tab:plr_results} gives the result of fitting the Wesenheit magnitude $W_{Ks,J-Ks} = K_s-e(J-K_s)$ with $e$ a free parameter finding $e=(0.45\pm0.02)$ fully consistent with the estimate from interstellar extinction considerations \citep[0.47,][]{WangChen2019}. This gives no evidence that there is additional colour dependence and in turn metallicity dependence to the O-rich period--luminosity relation. However, this simple approach uses the extincted $J$ and $K_s$ magnitudes in the modelling. Instead including an additional extinction-corrected colour term $b_{JK}(J-K_s)$ in the $K_s$ period--luminosity relation, the best-fitting gradient is found as $b_{JK}=(0.34\pm0.05)$ giving evidence that the period--luminosity relation is fainter for redder (more metal-rich) stars. However, the remaining colour-magnitude-spatial correlations in the Gaia zeropoints make this conclusion uncertain.

As discussed in Appendix~\ref{appendix::lmc}, there is also evidence in the LMC sample for a metallicity gradient to the period--luminosity relation with more metal-rich stars being fainter although this interpretation is somewhat complicated by age-metallicity correlations. However, again assuming colours are age-insensitive, the $(J-K_s)$ colour gradient to the $K_s$ period--luminosity relation is $b_{JK}=(0.45\pm0.07)$ or using $\mathrm{d}(J-K_s)/\mathrm{d}[\mathrm{Fe}/\mathrm{H}]\approx0.2$ the gradient with metallicity is $(0.09\pm0.02)\,\mathrm{dex}^{-1}$. This is in rough agreement with the differences found between the MW and LMC systems as a whole and consistent with the population gradient in the Milky Way sample. 

A further check of metallicity dependence of the period--luminosity relation is through analysis of the Galactic bulge Mira variables \citep{Groenewegen2005,Qin2018}. The period--luminosity relation can be calibrated under the assumption that the spatial distribution peaks around the now well-determined distance of Sgr A* \citep{GravityCollaboration2021}. However, these bulge stars are more sensitive to extinction assumptions and modelling the distance distribution requires good knowledge of the selection function. Finally, in the Galactic disc, the period--luminosity relation could be inspected as a function of kinematics which acts as a proxy for age/metallicity. \cite{Alvarez1997} reported differences in the period--luminosity relation for different kinematically-defined populations using Hipparcos data. Both of these avenues require further investigation that is deferred to future work. In conclusion, there is evidence from both the mean difference between the LMC and Milky Way and from differences within the LMC and Milky Way samples of a metallicity gradient to the period--luminosity relations for O-rich Mira variables with the more metal-rich stars intrinsically fainter than the metal-poor as expected from theoretical studies. 

\section{Consequences for the Hubble constant}\label{section::h0}
Our period--luminosity relations for Milky Way O-rich Mira variables provide useful anchors for the Type Ia supernova Hubble diagram and in turn a measurement of the Hubble constant. Currently the only SNIa host with observed Mira variables is NGC 1559 \citep{Huang2020} so Mira-based Hubble constant measurements are limited primarily by the uncertainty on the properties of the single supernova. However, over the coming years more observations of Mira variables in other SN Ia host galaxies are expected, so reducing the sources of uncertainty in the period--luminosity calibrations will become increasingly important. Here measurements of the Hubble constant are provided largely following the analysis of \cite{Huang2020} but replacing their period--luminosity relations with those derived here. In addition to the Milky Way relations, the LMC period--luminosity relations and Mira variables in the water maser host galaxy NGC 4258 are used as further anchors.

NGC 1559 hosted the Type Ia supernova SN 2005df with peak magnitude $m_B=(12.14\pm0.11)\,\mathrm{mag}$ \citep{Scolnic2018}. Given a distance modulus to NGC 1559, $\mu_{1559}$, the Hubble constant is estimated as
\begin{equation}
\log_{10}H_0 = \tfrac{1}{5}(m_B+5a_B+25) - \tfrac{1}{5}\mu_{1559},
\end{equation}
where $a_B=(0.71273\pm0.00176)$ is the SNIa magnitude-redshift intercept as measured by \cite{Riess2016}. It is beyond the scope of this work to combine the Type Ia supernovae modelling with the anchors in a probabilistic model as done by \cite{Riess2022} but the adopted $a_B$ encompasses the range of fits from \cite{Riess2022} and alters $H_0$ by $\sim0.2\,\mathrm{km\,s}^{-1}\mathrm{kpc}^{-1}$. The model for the Mira variables in NGC 1559 as presented by \cite{Huang2020} is first described and then used to derive the estimate of $H_0$.

\subsection{Basic model and data}

The NGC 1559 Mira variables are taken from \cite{Huang2020} and the NGC 4258 Mira variables are from \cite{Huang2018}. For both samples, mean magnitudes (and for NGC 1559 uncertainties) in the Hubble WFC3 $F160W$ band are provided along with period estimates. Both samples are defined to have peak-to-trough $F160W$ amplitude between $0.4$ and $0.8$ (to reduce C-rich contamination as discussed later). NGC 4258 has an additional colour cut ($F125W-F160W<1.3$ equivalent to $J-H<2.2$ using the colour transformations from the X-Shooter spectra as described below) which is relatively mild as for the LMC Mira variable sample it only removes $2$ of $907$ Mira variables with $P<300$ days (independent of whether extinction corrections are applied). For the NGC 4258 sample, there are further cuts on  $F814W$ detection and variability to define a `Silver' and `Gold' sample respectively. For the NGC 1559 sample these colour and variability cuts are not possible due to the lack of multiband data. However, as a quality cut, sources in NGC 1559 with crowding corrections $>0.25\,\mathrm{mag}$ are removed. 

The $F160W$ magnitudes are corrected for foreground extinction of $E(B-V)=0.0298$ for NGC 1559 and $E(B-V)=0.0163$ for NGC 4258 in \cite{SFD} units using the extinction coefficients from \cite{WangChen2019} and the $F160W$ uncertainties are broadened by a $16\percent$ uncertainty in $E(B-V)$ and a $2.5\percent$ uncertainty in the $F160W$ coefficient (the systematic uncertainty on the derived NGC 1559 and NGC 4258 distance moduli arising from the uncertainty in the extinction is $\sim0.002\,\mathrm{mag}$ so negligible compared to other sources of uncertainty). This ignores any extinction within the systems. The uncertainties on the periods of the Mira variables are ignored as they are not provided and for near-linear models period uncertainties are equivalent to an additional intrinsic magnitude spread (for approximately constant period uncertainties).

For each galaxy's Mira variable sample, a two-component Gaussian mixture model is fitted to the residuals of the $F160W$ magnitudes with respect to the period--luminosity relation (shifted by the distance modulus $\mu$) as
\begin{equation}
\begin{split}
p(F160W|P) =\sum_{j=1}^{j=2}\vartheta_j\mathcal{N}(F160W|&a+b(\log_{10}P-2.3)+\mu_j,\\&f^2\sigma_{F160W}^2+\sigma_{0,j}^2).
\end{split}
\end{equation}
All considered Mira variables have $P<400\,\mathrm{day}$ so only a linear model is considered. 
The mixture model allows for a contribution from outliers that do not follow a tight period--luminosity relation.
An initial consideration is that the Milky Way (and LMC) period--luminosity relations are derived in the 2MASS $JHK_s$ bands whilst the extragalactic Mira variable observations have been made in the Hubble WFC3 $F160W$ band \citep[effective wavelength of
$1.528\,\mu\mathrm{m}$
compared to $J$ of $1.235\,\mu\mathrm{m}$
and $H$ of $1.662\,\mu\mathrm{m}$,][]{Huang2018,Huang2020}. Following \cite{Huang2020} a colour term is used to convert 2MASS $H$ magnitudes into $F160W$ magnitudes. $43$ stars in the O-rich Mira sample with periods $<400$ days are taken from the second release of the X-Shooter Spectral Library \citep{Gonneau2021}. Using the filters provided by the SVO filter service \citep{svo1,svo2}, the expected magnitudes of these stars in the 2MASS filters and $F160W$ are found. The expected $J$ and $H$ 2MASS magnitudes are on average $0.07$ and $0.13$ mag brighter than measured in agreement with the comparison from \cite{Gonneau2021}. The broad-band colours are extinction corrected using the procedure described in Section~\ref{section::ext_corr} using
the interpolated $F160W$ coefficient $A_{F160W}/A_V=0.1556$ from \cite{WangChen2019}. The relationship between the $F160W$ and 2MASS bands is found to be $F160W = H + c_{JH}(J-H)$ with $c_{JH}=(0.379\pm0.012)$ which agrees well with the colour coefficient from \cite{Huang2020}. Using Table~\ref{tab:plr_results}, this implies a period--luminosity relation for the $F160W$ band of
\begin{equation}
    F160W_\mathrm{abs}(P)=(-6.16\pm0.02)+(-3.73\pm0.05)(\log_{10}P-2.3),
\end{equation}
for O-rich Mira variables with $P<400\,\mathrm{days}$.
The unknown $F160W$ period--luminosity relation is modelled probabilistically by allowing the parameters $h=(a,b)$ and $c_{JH}$ to vary and including a `prior' term of the form
\begin{equation}
p(\bar h|h, c_{JH})=-\tfrac{1}{2}(h-\bar h)^\mathrm{T}\Sigma_h^{-1}(h-\bar h).
\label{eqn:fprior}
\end{equation}
Here $\bar h=\langle h_H+c_{JH}(h_J-h_H)\rangle$, $\Sigma_{f,a,a}=\mathrm{Var}(a_H)$ as the uncertainty in the zeropoint is assumed to be wholly driven by distance uncertainties, $\Sigma_{h,a,b}=\mathrm{Cov}(a_H+c_{JH}(a_J-a_H),b_H+c_{JH}(b_J-b_H))$ and $\Sigma_{h,b,b}=\mathrm{Var}(b_H+c_{JH}(b_J-b_H)$. A prior $c_{JH}\sim\mathcal{N}(0.379,0.012^2)$ is adopted along with flat priors on $h$. $(a_i,b_i)$ are from the Milky Way fits, the LMC fits or a combination of both. The models of the scatter about the period--luminosity relation from the fits of the Milky Way and LMC data are not used as the NGC 1559 and NGC 4258 data are multi-epoch mean magnitudes whilst for the Milky Way only single-epoch data are available. Therefore, a simple constant scatter about the period--luminosity relation is adopted.
In total, nine parameters are fitted for: two distance moduli $\mu_j$ and widths $\sigma_{0,j}$ of the Gaussian components (no uncertainties are available for the NGC 4258 sample so the width models both intrinsic and observational spread), their relative weight ($\vartheta_1=1-\vartheta_2$), a scaling ($f$) of the reported uncertainties ($\sigma_{F160W}$), the colour term ($c_{JH}$) and the two parameters $h=(a,b)$ of the (linear) period--luminosity relation. Logarithmic priors are used for all intrinsically positive parameters. The condition  $\sigma_{0,1}<\sigma_{0,2}$ is to identify the outlier as the 2nd component. Further priors are adopted on the mixing parameter $\ln\vartheta_1\sim\mathcal{N}(0,1)$ (with $0<\vartheta_1<1$ and $\vartheta_1+\vartheta_2=1$) and the error scaling $\ln f\sim\mathcal{N}(0,1)$. The model is sampled from using \textsc{emcee} \citep{emcee}.

The period--luminosity relation anchors (as defined in equation~\eqref{eqn:fprior}) are selected as the O-rich Mira variable period--luminosity relations reported in Table~\ref{tab:plr_results} for the Milky Way and Table~\ref{tab:period_luminosity_relations_lmc} for the LMC. Using the Milky Way relation for the NGC 4258 sample gives $\mu_{4258}=(29.34\pm0.05)$, $\mu_{4258}=(29.36\pm0.04)$ and  $\mu_{4258}=(29.36\pm0.05)$ for the `Bronze', `Silver' and `Gold' samples respectively in good agreement with (though slightly lower than) the water maser distance of $\mu_{4258,\mathrm{maser}}=(29.398\pm0.032)$. Due to the similarity of the results, from now on the `Bronze' sample is used. The agreement with the water maser distance suggests the level of C-rich contamination is low in NGC 4258 and the metallicities of the Mira variables in the Milky Way are similar to those in NGC 4258. If instead the O-rich period--luminosity relation for the central LMC sample from \cite{Yuan2018} is used (as given in Table~\ref{tab:period_luminosity_relations_lmc}), it is found that $\mu_{4258}=(29.53\pm 0.06)$ which is $\sim2\sigma$ higher than the water maser distance modulus suggesting the metallicities of the LMC Mira variables are lower than those in NGC 4258 \citep{Bresolin2011}.

Initially the NGC 1559 sample is assumed to be purely composed of O-rich Mira variables and this assumption is relaxed below. These examples are illustrative as it is expected that the selection of Mira variables will introduce C-rich contamination, so the results should not be taken as realistic estimates of $H_0$. In the top section of Table~\ref{tab:h0} the results using the Milky Way O-rich relation and the LMC O-rich relation are reported. For the Milky Way O-rich relation a $\sim5\,\mathrm{km\,s}^{-1}\mathrm{Mpc}^{-1}$ higher $H_0$ is found than when using the LMC O-rich relation (here the LMC relation for the \citealt{Yuan2018} subsample is used which is $\sim0.05\,\mathrm{mag}$ fainter than the relations for the full LMC sample) due to the different period--luminosity zeropoints (driven by population effects). When combining both in the modelling, different $\mu_j$, $\sigma_{0,j}$, $f$ and $\vartheta_j$ are used for each galaxy, and for NGC 4258 the prior $\mu_1\sim\mathcal{N}(29.398,0.032^2)$ is used \citep{Reid4258}. For the pure O-rich period--luminosity relation case, the combined NGC 4258, MW and LMC fits give an average value of $H_0$ between the estimates from the MW and LMC alone.

\subsection{C-rich contamination}

\begin{figure}
    \centering
    \includegraphics[width=\columnwidth]{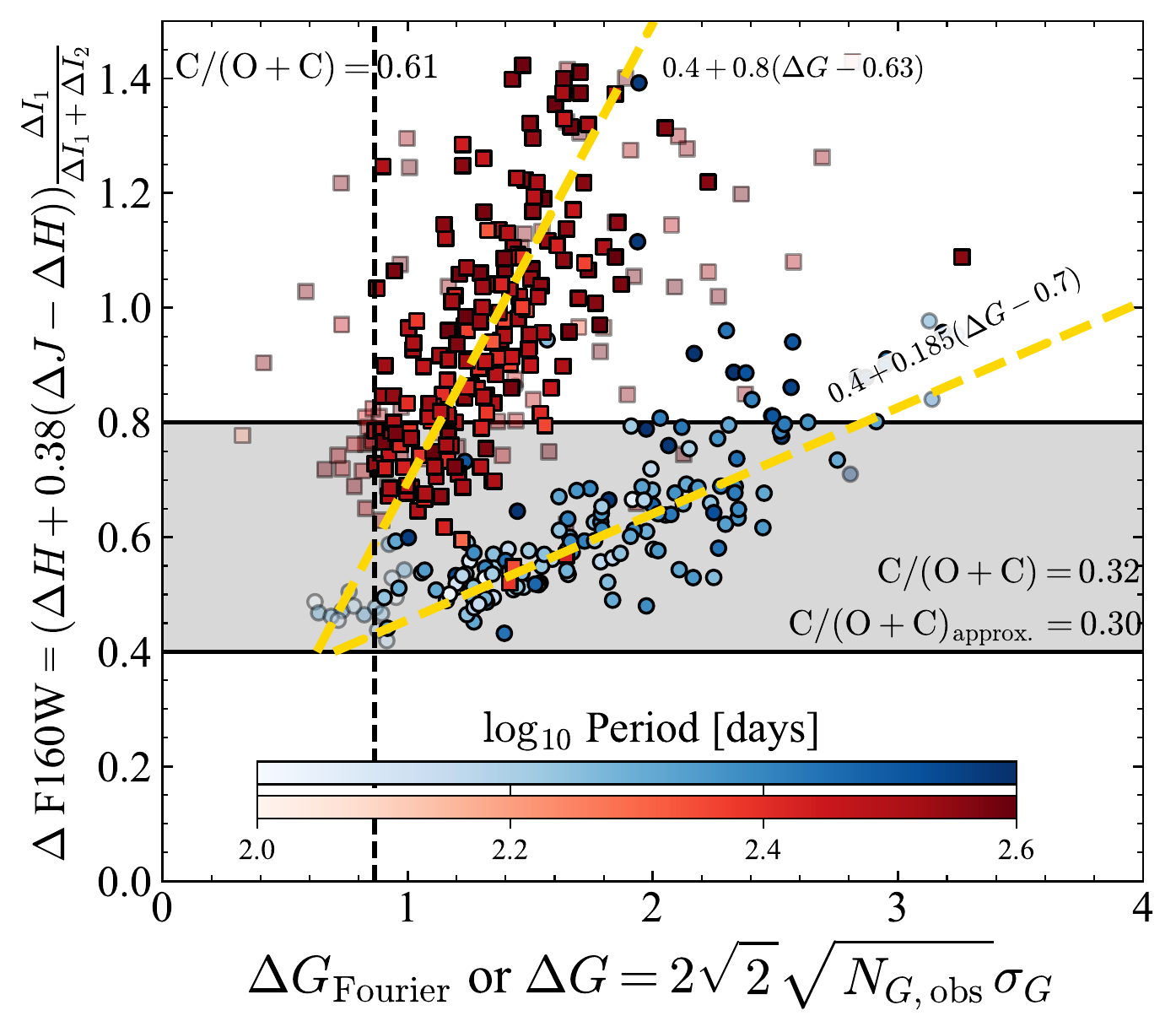}
    \caption{LMC Mira variable amplitudes: $G$-band Fourier amplitude, $\Delta G_\mathrm{Fourier}$, or amplitude measured from the Gaia DR3 uncertainties, $\Delta G$ (fainter points), against the approximate $F160W$ amplitude $\Delta F160W$ (computed from the modelled $J$ and $H$ amplitude scaled by the ratio of the primary period amplitude to the total amplitude from the OGLE $I$-band light curves). The blue circles are classified as O-rich whilst the red squares are C-rich using the classification from \protect\cite{Lebzelter2018}. The colour shades correspond to the period. All stars have periods less than $400\,\mathrm{days}$. The approximate trends of the two types are shown by the yellow dashed lines. For the full sample, C-rich Mira variables make up $61\percent$ of the sample, whilst for the selection $0.4<\Delta F160W<0.8$ they make up $32\percent$ or $30\percent$ using the approximate $\Delta F160W$ computed from $\Delta G$ (yellow dashed lines). The vertical dashed line is $\Delta G=\Delta G_\mathrm{thresh}=0.865\,\mathrm{mag}$ which defines the lower boundary for Mira variables from Gaia photometry.}
    \label{fig::amp_relation}
\end{figure}
The first set of models ignored the selection of the Mira variables simply assuming that the samples were fair representations of the O-rich Mira population. However, there is significant but uncertain contamination from C-rich Mira variables in these samples.
Limiting to periods less than $400$ days mitigates the impact of C-rich contamination considerably but some contamination remains that typically makes the mean magnitude at fixed period fainter, but also flattens the period--luminosity relation and increases the scatter at longer periods. As C-rich Mira variables have higher near-infrared amplitudes than O-rich Mira variables in this period range, \cite{Huang2020} imposed a cut of $0.4<\Delta F160W<0.8$ on their sample to isolate the O-rich Mira variables. Here $\Delta F160W$ is the peak-to-trough amplitude over a single cycle and neglects any longer term periodic trends common for Mira variables. This cut reduces C-rich contamination but some contaminants remain. First, the impact of this cut on the MW and LMC samples is estimated and then models of the period--luminosity relation with appropriate C-rich contamination levels are generated.

The LMC Mira variable sample from \cite{Yuan2018} have well-sampled I-band light curves from OGLE for which \cite{Soszynski2009} have provided amplitudes $I_1$ and $I_2$ for two identified periods.
\cite{Yuan2018} used the $I$-band light curves to model the more sparsely sampled $JHK_s$ light curves reporting the mean, maximum and minimum $JHK_s$ magnitudes. The single-cycle amplitude in $JHK_s$ is approximated as e.g. $\Delta J = (J_\mathrm{min}-J_\mathrm{max})I_1/(I_1+I_2)$ where the ratio between the amplitudes of the two periodic trends is assumed to be similar in all (near-infrared) bands. Using these approximations, the single cycle $\Delta F160W$ is found as $\Delta H+0.38(\Delta J-\Delta H)$ (see previous subsection). For the main Milky Way sample of Mira variables from Gaia, only amplitude indicators in the Gaia passbands are available. Gaia's observing window ($22$ months) is relatively short compared to long period trends in Mira variables so the Gaia amplitudes are assumed to correspond most closely with single-cycle amplitudes. Fig.~\ref{fig::amp_relation} displays $\Delta G$ as defined in equation~\eqref{eqn::delta_G} against $\Delta F160W$ for the LMC sample from \cite{Yuan2018}. We see $\Delta G$ correlates with $\Delta F160W$ but follows different relations for O-rich ($\Delta F160W\approx0.4+0.185(\Delta G-0.7)$) and C-rich ($\Delta F160W\approx0.4+0.617(\Delta G-0.63)$) Mira variables. As shown by \cite{Iwanek2021}, O-rich Mira variables have a steeper fall-off in amplitude with wavelength than C-rich Mira variables. For the full LMC sample, $61\percent$ of the Mira variables with periods $<400\,\mathrm{days}$ are C-rich, whilst restricting to $0.4<\Delta F160W<0.8$ (using the infrared amplitudes) reduces this to $32\,\percent$. These numbers are in good agreement with those reported by \cite{Huang2020}. Using the approximate $\Delta F160W$ computed from $\Delta G$, $30\,\percent$ of selected stars are C-rich, thus validating the approximate relations. Repeating this analysis for the Milky Way Mira sample, $2.0\percent$ of the sample without spatial cuts is C-rich which reduces to $0.6\percent$ using $0.4<\Delta F160W<0.8$, whilst removing the bulge and $|b|<3\,\mathrm{deg}$ results in a similar reduction from $3.4\percent$ to $0.8\percent$. Clearly even with the $\Delta F160W$ cut, for more metal-poor systems the C-rich contamination can be significant.

Period--luminosity relations like those in Section~\ref{section::model} and Appendix~\ref{appendix::lmc} have been fitted to the contaminated LMC and Milky Way Mira variables with $P<400$ days and $0.4<\Delta F160W < 0.8$ (using the previously derived approximate relations).  Instead of a linear relation, a quadratic period--luminosity relation of the form
\begin{equation}
m_\mathrm{abs,contam}(P)=a+b(\log_{10}P-2.3)+b_2(\log_{10}P-2.3)^2,
\end{equation}
is used
due to increasing C-rich contamination with increasing period causing a down-turning for periods $>300$ days (also exhibited by a contaminated SMC sample). The spread about the period--luminosity relation is modelled as
\begin{equation}
\sigma_{\mu,\mathrm{contam}}(P)=\sigma_{2.3}+
\begin{cases}
0,&\mathrm{if} \log_{10}P<2.3,\\
p_\sigma(\log_{10}P-2.3)\\+q_\sigma(\log_{10}P-2.3)^2,&\mathrm{otherwise},
\end{cases}
\end{equation}
to capture the sharp increase in the scatter for periods $>300$ days due to the C-rich contamination. The results of these fits to the Milky Way stars with $|b|>3\,\mathrm{deg}$ and outside the bulge region, and to the central LMC sample of Mira variables with measurements from \cite{Yuan2018} are given in Table~\ref{tab:crich_contaminated_relations}.

The modelling of the previous section has been repeated with these C-rich contaminated models (using $f=(a,b,b_2)$) and the results are reported in Table~\ref{tab:h0}.
This modelling assumes the contamination level is the same in NGC 1559 as in the MW and/or LMC. For the MW, the C-rich contamination is so low that adopting the contaminated model makes a very small difference to $H_0$. However, for the LMC C-rich contamination changes $H_0$ by $\sim10\,\mathrm{km\,s}^{-1}\mathrm{\,Mpc}^{-1}$. This shows that the effects of C-rich contamination are comparable, if not larger, than the effects of age/metallicity on the period--luminosity relations. Inclusion of the NGC 4258 measurements (already argued to have a low C-rich contamination level based on comparison with the water maser distance) brings $H_0$ down to values more consistent with the MW O-rich model or the MW (weakly) C-rich contaminated model.

\begin{table}
    \centering
\setlength{\tabcolsep}{3.5pt}
\rowcolors{1}{}{lightgray}
\caption{C-rich contaminated period--luminosity relations. Quadratic relations of the form $m_\mathrm{abs,contam}(P)=a+b(\log_{10}P-2.3)+b_2(\log_{10}P-2.3)^2$ have been fitted to Mira variable samples defined by $0.4<\Delta F160W<0.8$. C/(O+C) is the fraction of C-rich Mira variables in each system (also called $\eta$ in the modelling). The $F160W$ relations are derived using $F160W=H+(0.38\pm0.01)(J-H)$.}

\begin{tabular}{l|cccc}
System/Band&C/(O+C)&$a$&$b$&$b_2$\\
\hline
MW $J$&$0.008$&$-5.65\pm0.02$&$-3.34\pm0.08$&$-1.57\pm0.40$\\
MW $H$&$0.008$&$-6.45\pm0.02$&$-3.61\pm0.08$&$-1.52\pm0.39$\\
MW $F160W$&$0.008$&$-6.15\pm0.02$&$-3.51\pm0.06$&$-1.55\pm0.28$\\
LMC $J$&$0.324$&$-5.89\pm0.03$&$-2.19\pm0.15$&$+5.71\pm0.97$\\
LMC $H$&$0.324$&$-6.62\pm0.03$&$-2.91\pm0.12$&$+2.31\pm0.78$\\
LMC $F160W$&$0.324$&$-6.34\pm0.02$&$-2.64\pm0.09$&$+3.59\pm0.62$\\
\hline
\end{tabular}
\label{tab:crich_contaminated_relations}
\end{table}

\subsection{Variable C-rich contamination}

\begin{table*}{}
\centering
\rowcolors{1}{}{lightgray}
\caption{Hubble constant measurements using Mira variables in the SNIa host galaxy NGC 1559 and a range of different anchors. $H_0$ in units of $\mathrm{km\,s}^{-1}\mathrm{Mpc}^{-1}$. $\mu_{1559}$ is the distance modulus of NGC 1559, $\mu_{4258}$ is the distance modulus of NGC 4258, $a$ is the zeropoint of the $F160W$ period--luminosity relation evaluated at $\log_{10}P=2.3$, $b$ is the slope of the $F160W$ period--luminosity relation with $\log_{10}P$ and when given $b_2$ the quadratic term (i.e the relation is $a+b(\log_{10}P-2.3)+b_2(\log_{10}P-2.3)^2$). For reference, the distance modulus to the water maser in NGC 4258 is $\mu_{4258,\mathrm{maser}}=(29.398\pm0.032)\,\mathrm{mag}$ \protect\citep{Reid4258}, the $H_0$ measurement from \protect\cite{Planck} is $(67.4\pm0.5)\mathrm{km\,s}^{-1}\mathrm{Mpc}^{-1}$, the recent Cepheid-based $H_0$ estimate from \protect\cite{Riess2022} is $(73.04\pm1.04)\mathrm{km\,s}^{-1}\mathrm{Mpc}^{-1}$ or $(73.01\pm0.99)\mathrm{km\,s}^{-1}\mathrm{Mpc}^{-1}$ 
 for those Cepheids in clusters \protect\citep{Riess2022_CLUSTER}, the recent tip of the giant branch $H_0$ estimate from \protect\cite{Freedman2021} is $69.8\pm0.6 \mathrm{(stat)}\pm 1.6 \mathrm{(sys)}\mathrm{km\,s}^{-1}\mathrm{Mpc}^{-1}$ and the combination of Cepheid-based and tip of the giant branch from \protect\cite{Riess2022} is $(72.53\pm0.99)\mathrm{km\,s}^{-1}\mathrm{Mpc}^{-1}$. The top section uses pure O-rich Mira variable period--luminosity relations whilst the middle section uses Mira variable period--luminosity relations for contaminated samples. These are illustrative limits and the results should not be considered as recommended measurements. The final section uses a variable C-rich fraction for each system with the recommended final measurement in bold.}
        \begin{tabular}{l|cccccc}
Anchor&$H_0$&$\mu_{1559}$&$\mu_{4258}$&$a$&$b$&$b_2$\\
\hline
           MW O-rich& $71.6\pm4.0$& $31.43\pm0.05$& $  -$ & $-6.16\pm0.04$& $-3.73\pm0.10$& $  -$\\
          LMC O-rich& $66.7\pm3.8$& $31.58\pm0.06$& $  -$ & $-6.35\pm0.06$& $-3.51\pm0.15$& $  -$\\
NGC 4258 + MW +LMC O-rich& $69.9\pm3.9$& $31.48\pm0.05$& $29.40\pm0.03$ & $-6.22\pm0.04$& $-3.64\pm0.08$& $  -$\\
\hline
   MW C-rich contam.& $71.2\pm4.1$& $31.44\pm0.06$& $  -$ & $-6.15\pm0.04$& $-3.51\pm0.11$& $-1.51\pm0.56$\\
  LMC C-rich contam.& $76.5\pm4.5$& $31.28\pm0.07$& $  -$ & $-6.34\pm0.06$& $-2.67\pm0.18$& $ 3.28\pm1.09$\\
NGC 4258 + MW + LMC C-rich contam.& $70.4\pm4.0$& $31.47\pm0.05$& $29.39\pm0.03$ & $-6.22\pm0.04$& $-3.36\pm0.09$& $-1.19\pm0.45$\\
\hline
MW + LMC variable contam.& $74.6\pm4.5$& $31.34\pm0.07$& $  -$ & $-6.14\pm0.05$& $-3.54\pm0.11$& $-1.72\pm0.57$\\
\textbf{NGC 4258 + MW + LMC variable contam.}& $73.7\pm4.4$& $31.37\pm0.07$& $29.38\pm0.03$ & $-6.15\pm0.04$& $-3.54\pm0.12$& $-1.69\pm0.58$\\
    \end{tabular}
    \label{tab:h0}
\end{table*}

\begin{figure*}
    \centering
    \includegraphics[width=0.8\textwidth]{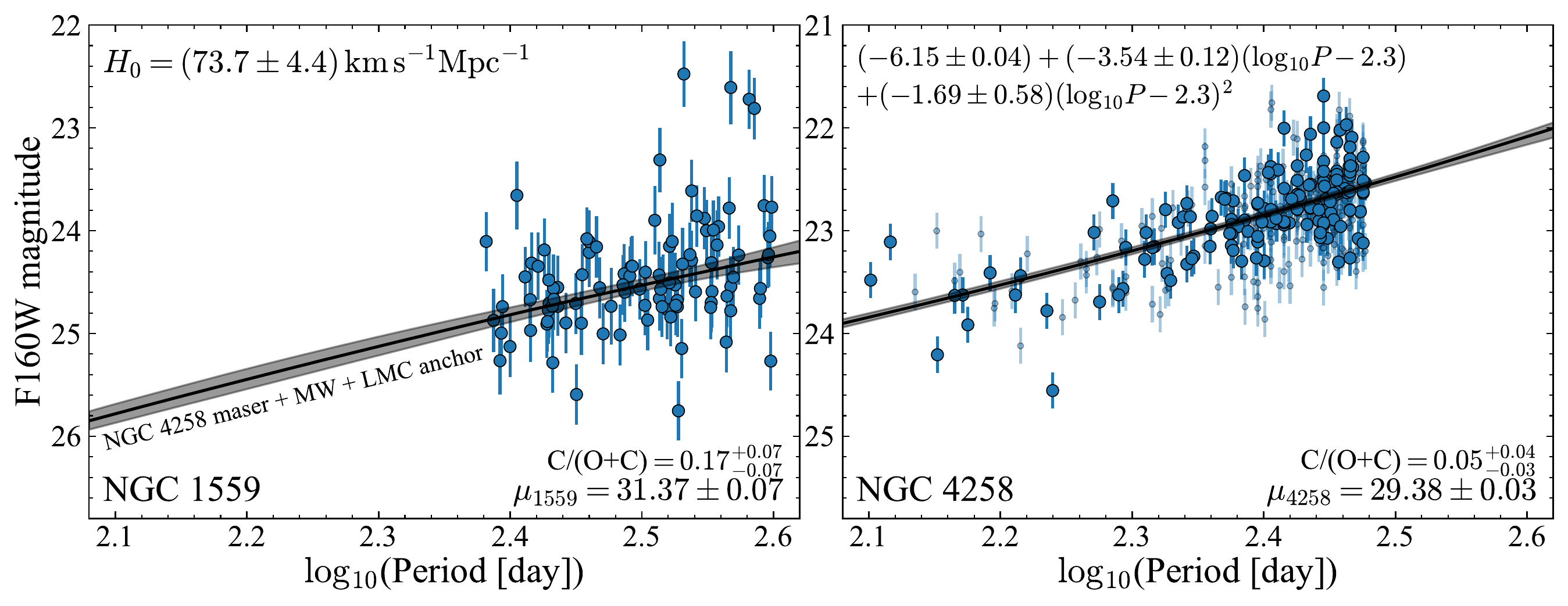}
    \caption{Period--magnitude diagrams for Mira variables in NGC 1559 (left) and NGC 4258 (right -- smaller faint points are non-`Gold' Mira variables). The black line shows the best-fitting period--luminosity relation (as given in the left panel) shifted by the distance estimate for each galaxy and evaluated at the estimated C-rich contamination fraction (as given in each panel) with the grey interval showing the $1\sigma$ uncertainty. The period--luminosity relation has been anchored using the Gaia EDR3 Milky Way results from this paper together with the LMC Mira variables anchored with the detached eclipsing binary distance from \protect\cite{Pietrzynski2019} and NGC 4258 Mira variables anchored using the distance to the NGC 4258 maser from \protect\cite{Reid4258}. The resulting Hubble constant using SN 2005df is $(73.7\pm4.4)\,\mathrm{km\,s}^{-1}\mathrm{Mpc}^{-1}$.}
    \label{fig:hubble}
\end{figure*}

In reality, the C-rich contamination of the NGC 4258 and NGC 1559 sample is unknown and the best choice of period--luminosity relation lies somewhere between the pure O-rich and C-rich contaminated cases. To incorporate this into the modelling, the NGC 4258 and NGC 1559 samples are considered to have individual unknown C-rich contamination factors, $\eta=N_C/(N_C+N_O)$, where $N_i$ is the number of $i$-rich stars. The parameters of the quadratic period--luminosity relations, $a, b$ and $b_2$, are assumed to vary linearly with $\eta$. The Milky Way and LMC C-rich contaminated samples then give two (probabilistic) points on this relation with $\eta_\mathrm{MW}=0.008$ and $\eta_\mathrm{LMC}=0.324$ respectively (the binomial uncertainty in these numbers is not considered). In this way, the range of different environments has been reduced down to a single parameter, $\eta$. As discussed in Section~\ref{sec::pop}, at the most basic level both age and metallicity play a role in determining the period--luminosity relation so a single parameter is an oversimplification. Both age and metallicity also govern the degree of C-rich contamination. Age largely influences the periods of Mira variables and so over a limited period range $\eta$ can be considered as a proxy for metallicity \citep{Brewer,Hamren2015} and age is then considered as affecting the period distribution rather than the shape of the period--luminosity relation. As $\eta$ is increased (metallicity decreased), the period--luminosity relation of the O-rich Mira variables becomes brighter, the C-rich contamination increases and the contaminated period--luminosity relation gets fainter and flatter (as C-rich Mira variables are more common for younger, longer period populations).
In the absence of clear O-rich/C-rich discrimination on a star-by-star basis, C-rich contamination can be measured from the shape of the period--luminosity relation, and in turn the metallicity of the environment measured and the period--luminosity zeropoint more precisely known. 

This approach is incorporated into the modelling by taking $h=(a, b, b_2)$ as the period--luminosity relation parameters for a completely uncontaminated sample, $\eta=0$, and introducing parameters of the gradients of $h$ with $\eta$, $g=\mathrm{d}h/\mathrm{d}\eta$. The term in equation~\eqref{eqn:fprior} is then adjusted to instead be
\begin{equation}
p(\bar h_i|\eta_i,h,g,c_{JH})=-\tfrac{1}{2}(\bar h_i-\eta_i g - h)^\mathrm{T}\Sigma_{hi}^{-1}(\bar h_i-\eta_i g - h).
\end{equation}
for system $i$ (Milky Way and LMC) with measured $\bar h_i=\langle h_H+c_{JH}(h_J-h_H)\rangle$ and covariance $\Sigma_{hi}$ as before. For the NGC 1559 Mira variables, the period--luminosity relation parameters are $f+\eta_{1559}g$ (and similar for NGC 4258). The prior $\eta\sim\mathcal{N}(0,0.3^2)$ is used as it is anticipated the C-rich contamination in both NGC 1559 and NGC 4258 is lower than that in the LMC. In theory, the scatter parameters $\sigma_{0,j}$ could also be made to vary with $\eta$ in a similar way. This may give more handle on $\eta$ as the scatter about the period--luminosity relation increases substantially with increased C-rich contamination. However, mean $F160W$ magnitudes for NGC 4258 and NGC 1559 are considered, whilst only single-epoch data are available for the Milky Way sample making it awkward to estimate the expected mean scatter at fixed contamination.

This procedure is quite similar to that of \cite{Huang2020}, who used a fixed gradient linear model fitted to the mean magnitude in a set of period bins to find the variation in the zeropoint. The zeropoint variation was then matched on to the corresponding variation in the LMC scaling by an unknown contamination factor, $\alpha$, that gave the fraction of the zeropoint shift between the contaminated and uncontaminated LMC sample that must be applied to the sample. 

The variable C-rich contamination models are fitted to NGC 1559 alone, and NGC 1559 and NGC 4258 together and the results are reported in Table~\ref{tab:h0}. There is good agreement between the two models. In agreement with the previous O-rich models of NGC 4258, it is found that $\eta_{4258}<0.13$ at $95\percent$ confidence. For NGC 1559, it is found that $\eta_{1559}=(0.17\pm0.07)$ or $\eta_{1559}/\eta_\mathrm{LMC}=(0.52\pm0.22)$, in very good agreement with \cite{Huang2020} who find the adjustment of the zeropoint must be a fraction $(0.58\pm0.18)$ of the LMC zeropoint adjustment. The final estimate of $H_0=(73.7\pm4.4)\,\mathrm{km\,s}^{-1}\mathrm{Mpc}^{-1}$ is in agreement with the analysis of \cite{Huang2020} who found $H_0=(73.3\pm4.0)\,\mathrm{km\,s}^{-1}\mathrm{Mpc}^{-1}$. Despite also using the Milky Way Mira variables in the analysis, the uncertainty here is $0.4\,\mathrm{km\,s}^{-1}\mathrm{Mpc}^{-1}$ larger. This is probably because of the difference in the handling of the C-rich contamination and the marginalization over the period--luminosity relation gradient. If instead the parallax zeropoint corrections from \cite{Groenewegen2021} or \cite{MaizApellaniz2021} as described in Section~\ref{sec::zpt} are used, larger $H_0$ of $(76.3\pm4.3)$ and $(75.8\pm4.2)\,\mathrm{km\,s}^{-1}\mathrm{kpc}^{-1}$ respectively are found. However, these zeropoint models are disfavoured as they lead to larger parallax residuals as a function of $G$. Nonetheless, these estimates are within the reported uncertainties, and point to the importance of an improved understanding of the Gaia parallax systematics for refining these estimates. In accord with many recent estimates of the Hubble constant based on near-Universe tracers, the estimate is higher than the \cite{Planck} estimate but only at the $\sim1.4\sigma$ level. Although some uncertainty arises from the modelling of the Mira variable period--luminosity relation and contamination, the dominant uncertainty is from the peak luminosity of SN2005df so further measurements of Mira variables in Type Ia host galaxies are required.

\section{Conclusions}\label{section::conclusions}
Preliminary period--luminosity calibrations have been presented for Milky Way O-rich Mira variables in the 2MASS $JHK_s$ bands using astrometric data from Gaia Data Release 3. The relations have been derived using a probabilistic model incorporating a flexible distance prior and models for the Gaia EDR3 parallax zeropoint and uncertainty underestimates. Period--luminosity relations have been estimated for $JHK_s$ magnitudes and extinction-free Wesenheit indices, and also estimated for the C-rich Mira population. 
The corresponding relations for the Large and Small Magellanic Clouds, and the Sagittarius dwarf spheroidal galaxy have also been estimated.

The Mira variables provide an interesting regime for testing the Gaia astrometry. Although the large angular size is a potential concern for a handful of the nearest brightest Mira variables, it appears the Gaia EDR3 astrometry is accurate for these very red, bright stars. 
A full investigation of the theoretical expected astrometric performance of Gaia EDR3 for AGB stars has been performed indicating that, despite the intrinsic photocentre wobble of these stars, on average the measured parallaxes should be unbiased but the uncertainties are likely underestimated for $\varpi\gtrsim0.5\,\mathrm{mas}$ and $G\lesssim11$.
It is found that the parallax zeropoint corrections from \cite{EDR3_ZPT} evaluated at $\nu_\mathrm{eff}=1.25\,\mu\mathrm{m}^{-1}$ approximately capture the behaviour of the zeropoint for redder five-parameter sources (to within $\sim5\,\mu\mathrm{as}$) but overcorrect the parallaxes for redder six-parameter sources. The Gaia EDR3 parallax uncertainties are typically underestimated with the largest correction factor ($\sim1.6$) required at $G\approx12.5$. The modelling approach adopted here does not fully capture the magnitude and spatial dependence of the parallax offset with respect to the period--luminosity relation although disentangling whether these are systematic or population effects is not possible with the approach. A future study should utilise a more sophisticated parallax offset model to account for these effects. Although the simple Mira variable selection criteria used here have been demonstrated to successfully isolate the required population, it is anticipated that more sophisticated selections for Mira variables using Gaia data will be developed. Further improvements and refinements to the period--luminosity relation of Mira variables are expected with future Gaia data releases. Some of the limitations of the current Gaia astrometric solution (e.g. not using epoch photometry) have been assessed as having a minimal impact. However, future Gaia data releases will improve the calibrations and extend the baseline providing improved astrometry for the inspected sources.

Mira variables have significant promise as a competitive distance ladder calibrator due to their lack of bias to younger stellar populations and their brightness in the infrared. However, this will require a solid understanding of any population effects, typically believed to be small in the infrared. The local O-rich Mira variables have been shown to be fainter than their Large Magellanic Cloud counterparts at fixed $250$ day period by $(0.19,0.19,0.11)$ in the $J$, $H$ and $K_s$ bands respectively arising primarily from a steeper derived slope for the Milky Way period--luminosity relation. This difference is larger than previous observational work has reported \citep[][report a $K_s$ offset of $(0.02\pm0.07)$ mag using the same gradient for the Milky Way and LMC period--luminosity relations]{Whitelock2008} but in some accord with the expected behaviour of metallicity dependence in theoretical models. 
Evidence for similar metallicity-dependent period--luminosity variation within these systems has also been presented. In particular, both the Milky Way and LMC samples favour a colour term in the period--luminosity relation of $\mathrm{d}M_{K_s}/\mathrm{d}(J-K_s)\approx0.4$ that suggests the redder, possibly more metal-rich Mira variables are fainter than their bluer, possibly more metal-poor counterparts. Future work should incorporate more flexible models for the period--luminosity relation to determine the extent to which a rigid assumed functional form is leading to the results of this work.

Using period--luminosity relations derived from the Gaia DR3 data, the Mira variable sample in the SNIa host galaxy NGC 1559 \citep{Huang2020} has been used to measure the distance modulus to this galaxy and in turn estimate the Hubble constant from the SNIa analysis of \cite{Riess2016}. The level of C-rich contamination of the NGC 1559 Mira variable sample is significant and leads to fainter zeropoints (higher $H_0$) but its strength can be constrained from the modelling due to the effect increasing C-rich contamination has on flattening the period--luminosity relation. By joint modelling the NGC 1559 and NGC 4258 Mira variables, and using the NGC 4258 water maser and the contaminated Milky Way and LMC Mira variable samples as anchors, the Hubble constant has been estimated as $H_0=(73.7\pm4.4)\,\mathrm{km\,s}^{-1}\mathrm{Mpc}^{-1}$. Although the Mira-based Hubble constant uncertainty is currently dominated by there being only a single SNIa host galaxy with Mira observations, the results suggest the population effects on the Mira period--luminosity relation are significant and must be better understood to make Mira variables a precision distance estimator for Hubble constant measurements.

\section*{Data Availability}
All data used in this work are in the public domain. A catalogue of the inspected sample along with derived properties is available at \url{https://www.homepages.ucl.ac.uk/~ucapjls/data/gaia_dr3_mira_plr.fits}.

\section*{Acknowledgements}
I thank the referee for their detailed comments and the support of the Royal Society (URF\textbackslash R1\textbackslash191555). This paper made use of the Whole Sky Database (wsdb) created by Sergey Koposov and maintained at the Institute of Astronomy, Cambridge by Sergey Koposov, Vasily Belokurov and Wyn Evans with financial support from the Science \& Technology Facilities Council (STFC) and the European Research Council (ERC). This software made use of the Q3C software \citep{q3c}. This research has made use of the SVO Filter Profile Service (http://svo2.cab.inta-csic.es/theory/fps/) supported from the Spanish MINECO through grant AYA2017-84089. This work has made use of data from the European Space Agency (ESA) mission
{\it Gaia} (\url{https://www.cosmos.esa.int/gaia}), processed by the {\it Gaia}
Data Processing and Analysis Consortium (DPAC,
\url{https://www.cosmos.esa.int/web/gaia/dpac/consortium}). Funding for the DPAC
has been provided by national institutions, in particular the institutions
participating in the {\it Gaia} Multilateral Agreement. This research has made use of the International Variable Star Index (VSX) database, operated at AAVSO, Cambridge, Massachusetts, USA. This paper made use of
\textsc{numpy} \citep{numpy},
\textsc{scipy} \citep{scipy}, 
\textsc{matplotlib} \citep{matplotlib}, 
\textsc{seaborn} \citep{seaborn}, \textsc{pandas} \citep{pandas1}
\textsc{astropy} \citep{astropy:2013,astropy:2018}, and
\textsc{Stan} \citep{stan}.

%%%%%%%%%%%%%%%%%%%%%%%%%%%%%%%%%%%%%%%%%%%%%%%%%%

%%%%%%%%%%%%%%%%%%%% REFERENCES %%%%%%%%%%%%%%%%%%

% The best way to enter references is to use BibTeX:

\bibliographystyle{mnras}
\bibliography{bibliography}

%%%%%%%%%%%%%%%%%%%%%%%%%%%%%%%%%%%%%%%%%%%%%%%%%%

%%%%%%%%%%%%%%%%% APPENDICES %%%%%%%%%%%%%%%%%%%%%

\appendix

\section{Gaia DR3 LPV candidate catalogue completeness}\label{appendix::completeness}
\begin{figure*}
    \centering
    \includegraphics[width=0.95\textwidth]{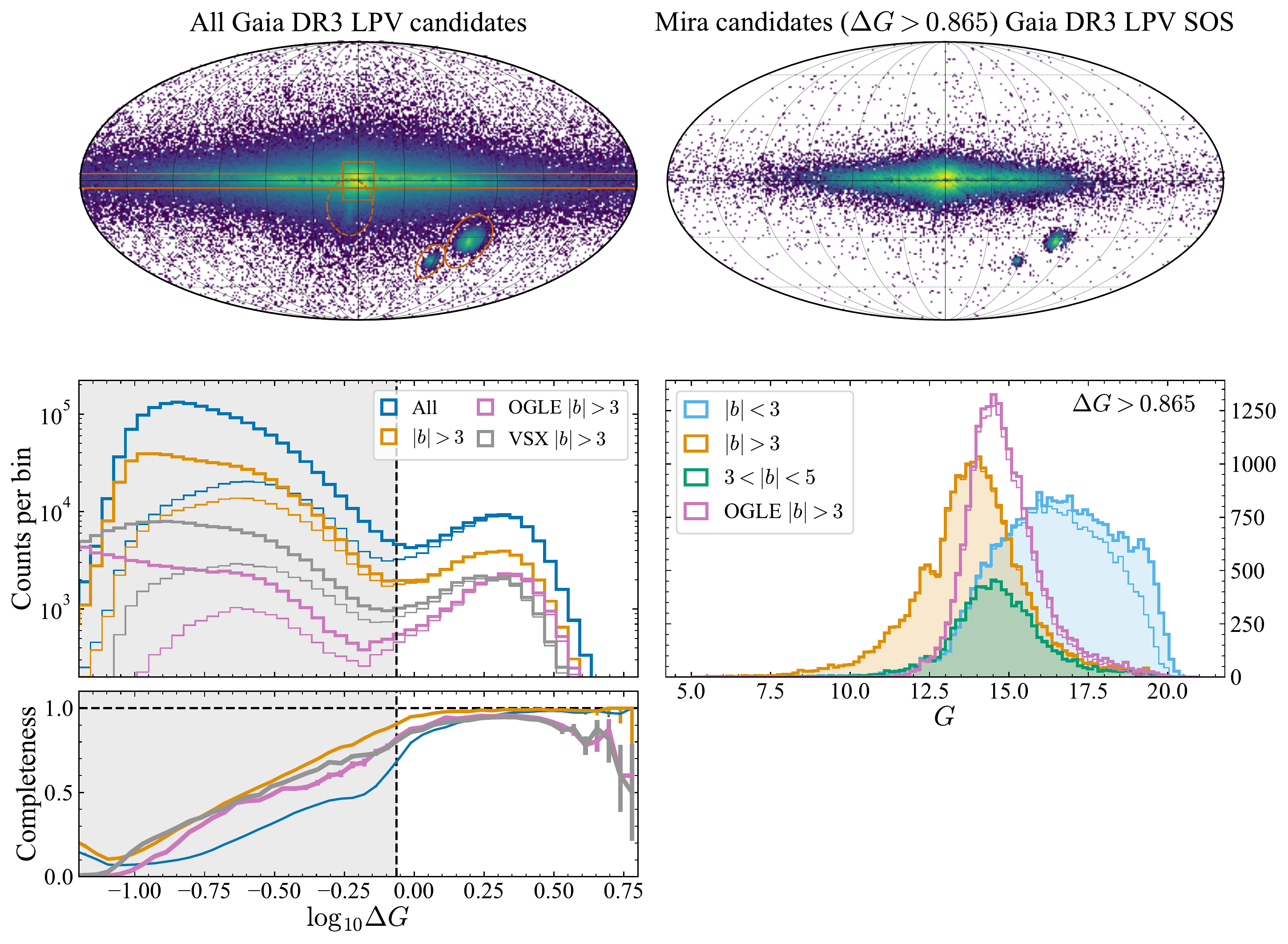}
    \caption{Approximate completeness of the Gaia DR3 Mira variable candidates catalogue. The top left panel shows the on-sky distributions of all Gaia DR3 LPV candidates from \protect\cite{Lebzelter2022} with the on-sky selections employed in Section~\ref{section::data} marked in orange (the dashed Sgr region is combined with a distance cut). The top right panel shows only those LPVs with reported periods and with $\Delta G>\Delta G_\mathrm{thresh}=0.865$ such that they are likely Mira variables. The middle left panel shows the number counts of the full LPV candidates excluding those in the LMC and SMC in bins of $G$ amplitude computed from the mean photometry, $\Delta G$, (thick lines) compared to the subset with periods (thin lines; blue: full sample and orange: $|b|>3\,\mathrm{deg}$). The pink and grey lines show the counts of OGLE and VSX LPVs in the Gaia DR3 source catalogue with $|b|>3\,\mathrm{deg}$ with the thinner lines showing those with reported periods in the Gaia DR3 LPV candidates catalogue. The bottom left panel shows the ratio of these number counts giving the approximate completeness of the subset with periods. The lower right panel shows the $G$ distributions of the different subsets of the full Mira variable set ($\Delta G>G_\mathrm{thresh}=0.865$ as marked by the shading in lower left panels) excluding the LMC and SMC and those with reported periods as thin lines.}
    \label{fig:completeness}
\end{figure*}
The set of stars utilised in this work is the subset of the full Gaia DR3 LPV candidates catalogue from \cite{Lebzelter2022} with reported periods and high amplitudes. In Section~\ref{section::data}, Mira candidates are defined as having $\Delta G>\Delta G_\mathrm{thresh}=0.865$, where $\Delta G$ is the scatter from the mean photometric uncertainty (the Fourier amplitude $\Delta G_\mathrm{Fourier}$ is also used) and periods are required in the period--luminosity modelling. The parent catalogue of all Gaia DR3 LPV candidates is also obviously a subset of all Milky Way LPV candidates. In this appendix, the completeness of the utilised sample is briefly assessed.

In Fig.~\ref{fig:completeness} the full Gaia DR3 LPV candidate catalogue is shown along with the subsample with reported periods and $\Delta G>\Delta G_\mathrm{thresh}=0.865$. The $\Delta G$ distribution of the full sample and those with periods is shown which clearly shows the peak at high amplitude due to Mira variables. Both samples exclude the LMC and SMC regions. The completeness of the Mira variable subset is assessed as the ratio of the number counts of the two samples. This is only a valid estimate of the completeness if all LPVs observed by Gaia are in the LPV candidates catalogue and the reason for no reported period isn't because the LPV candidate is spurious. If the former isn't true, the completeness will be overestimated whilst if there are many spurious LPVs in the full catalogue, the completeness will be underestimated. Note that the completeness of the entire Gaia catalogue must then be considered for a full assessment of completeness but this is only important for $G\approx20.5$ \citep{CantatGaudin2022}. For $\Delta G>\Delta G_\mathrm{thresh}$, the completeness of the Mira variable sample relative to the full catalogue is above $90\percent$ for $|b|>3\,\mathrm{deg}$ and drops slightly at the $\Delta G$ boundary when including $|b|<3\,\mathrm{deg}$. The number counts with $G$ are shown for the high-amplitude $\Delta G>\Delta G_\mathrm{thresh}$ set indicating that the period requirement only affects stars with $|b|<3\,\mathrm{deg}$ and $G>17$.

To assess the completeness of the full Gaia DR3 LPV candidates catalogue, samples of LPVs from OGLE \citep{Soszynski2009,Soszynski2013,Iwanek2022} and VSX \citep{Watson2006} are matched to the Gaia DR3 source catalogue with a $1\,\mathrm{arcsec}$ cross-match radius. The distributions of these samples are shown in Fig.~\ref{fig:completeness} along with that for the subset with periods from the Gaia DR3 LPV SOS catalogue. This suggests the completeness is similar to the previous estimate but slightly lower at the $\gtrsim80\percent$ level for $\Delta G>\Delta G_\mathrm{thresh}$. The previous completeness estimate must be too high because the parent LPV candidate catalogue does not contain all the known LPVs that Gaia sees due to the quality cuts described in Section~\ref{section::data}. At high $\Delta G$ the completeness with respect to VSX and OGLE is lower possibly as some highly variable sources are deemed spurious in the Gaia pipeline although there are few sources here. This analysis suggests that for high-amplitude sources ($\Delta G>0.865$) the LPV candidates catalogue is about as complete as expected given the overall completeness of Gaia, and the subset with periods is $\gtrsim90\percent$ complete for $|b|>3\,\mathrm{deg}$ and $\Delta G>\Delta G_\mathrm{thresh}$.

\section{Gaia astrometry for AGB stars}\label{appendix::gaia_scanning}

Section~\ref{section::data} discusses why there is good reason to believe the Gaia EDR3 astrometry forthe majority of AGB stars is unbiased although poorly estimated uncertainties due to current limitations of the instrument model. Although AGB stars have turbulent convective envelopes leading to significant perturbations of the photocentre, the current single CCD Gaia astrometric uncertainties for red sources are typically still larger than the size of the photocentre wobble. Even assuming systematics are not significant and that we can average over the $\sim18$ CCD observations per transit, the combined uncertainty is in the very best case scenario of the same order as the photocentre displacement. It therefore is unlikely that the Gaia EDR3 astrometric solutions for the bulk of the Mira variable sample are significantly biased. However, in this appendix the astrometric solutions for AGB stars are investigated in significantly more detail by modelling the expected photocentre wobble and the resulting Gaia astrometric parameter recovery.

\subsection{Gaia astrometric solution}
First, the tools for astrometric modelling and approximately reproducing the Gaia astrometric pipeline are presented \citep[see][and the python package \texttt{astrometpy}, \citealt{Penoyre2020}, for more details]{Lindegren2012,Lindegren2021,Everall2021}. Here, simple approximations to the full astrometric equations \citep[e.g.][]{kovalevsky_seidelmann_2004} are used that assume Gaia is on a circular orbit at $L_2=(1+(M_{\earth}/(3M_{\sun}))^{1/3})\,\mathrm{AU}$ from the Sun. As we are creating mock solutions and recovering the parameters with the same equations, having fast, easy-to-calculate equations are more important than high accuracy. In equatorial coordinates, the relative position of a star on the sky at time $t$ (in years) is given by
\begin{equation}
\begin{pmatrix}
\Delta\alpha\cos\delta\\
\Delta\delta
\end{pmatrix}=
\mathsf{M}(t)
\cdot
\mathbf{A},
\end{equation}
where 
the astrometric parameters are
\begin{equation}
    \mathbf{A}=\Big((\Delta\alpha\cos\delta)_0,
(\Delta\delta)_0,
\varpi,
\mu_\alpha*,
\mu_\delta\Big),
\end{equation}
with $\varpi$ the parallax (in mas) and $(\mu_\alpha*,\mu_\delta)$ the proper motions in $(\Delta\alpha\cos\delta,\Delta\delta)$ (in mas/yr). 
The design matrix is given by
\begin{equation}
\mathsf{M}(t)=
\begin{pmatrix}
1&0&\Pi_\alpha(t)&(t-t_\mathrm{ref})&0\\
0&1&\Pi_\delta(t)&0&(t-t_\mathrm{ref})\\
\end{pmatrix}.
\end{equation}
Here the parallax column of the design matrix is given by
\begin{equation}
\begin{pmatrix}
\Pi_\alpha(t)\\
\Pi_\delta(t)
\end{pmatrix}
=
L_2 \mathbf{R}_\gamma
\begin{pmatrix}
\sin(2\pi(t-t_\mathrm{E})-\lambda)\\
-\cos(2\pi(t-t_\mathrm{E})-\lambda)\sin\beta
\end{pmatrix},
\end{equation}
where $(\lambda, \beta)$ are ecliptic coordinates, $t_\mathrm{E}\approx0.2160\,\mathrm{yr}$ is the approximate vernal equinox and $t_\mathrm{ref}=2016\,\mathrm{yr}$ is the reference epoch for Gaia EDR3.
$\mathbf{R}_\gamma$ is a rotation matrix between the local ecliptic coordinates and local equatorial coordinates with angle
\begin{equation}
\tan\gamma = \frac{\cos\alpha\sin e}{\cos\delta\cos e + \sin\alpha\sin\delta\sin e},
\end{equation}
where $e=23.436\,\mathrm{deg}$ is the angle of obliquity.

Gaia rotates on its axis scanning the sky along great circles and slowly precesses to cover the entire celestial sphere. Astrometric measurements along the scan direction are significantly more precise than measurements across the scan direction by a factor of $\sim5.65$ \citep{Lindegren2012} and only sources with $G<13$ have across-scan measurements used in their astrometric solutions. Each scan of a source is recorded by the $9$ CCDs through each of the two fields-of-view. Therefore, there are effectively $18$ astrometric measurements per transit. To simulate the Gaia observations both the scanning times $\{t_i\}$ and the scanning directions $\{\phi_i\}$ (measured eastwards of equatorial North) for each on-sky location must be known. The \texttt{scanninglaw} package \citep{Green2018,Boubert2020,Everall2021} provides an interface to the nominal Gaia EDR3 scanning law (\url{http://cdn.gea.esac.esa.int/Gaia/gedr3/auxiliary/commanded_scan_law/}) incorporating known gaps in the data taking \citep{Lindegren2021}. For each source, the astrometric position
\begin{equation}
x_i = \begin{pmatrix}\sin\phi_i&\cos\phi_i\end{pmatrix}\cdot\begin{pmatrix}
\Delta\alpha\cos\delta\\
\Delta\delta
\end{pmatrix},
\end{equation}
is recorded.
Across-scan observations are included by setting $\phi_i\leftarrow\phi_i+\pi/2$. The observation is replicated $18$ times for the $9$ CCDs in the two fields of view.

\begin{figure}
    \centering
    \includegraphics[width=\columnwidth]{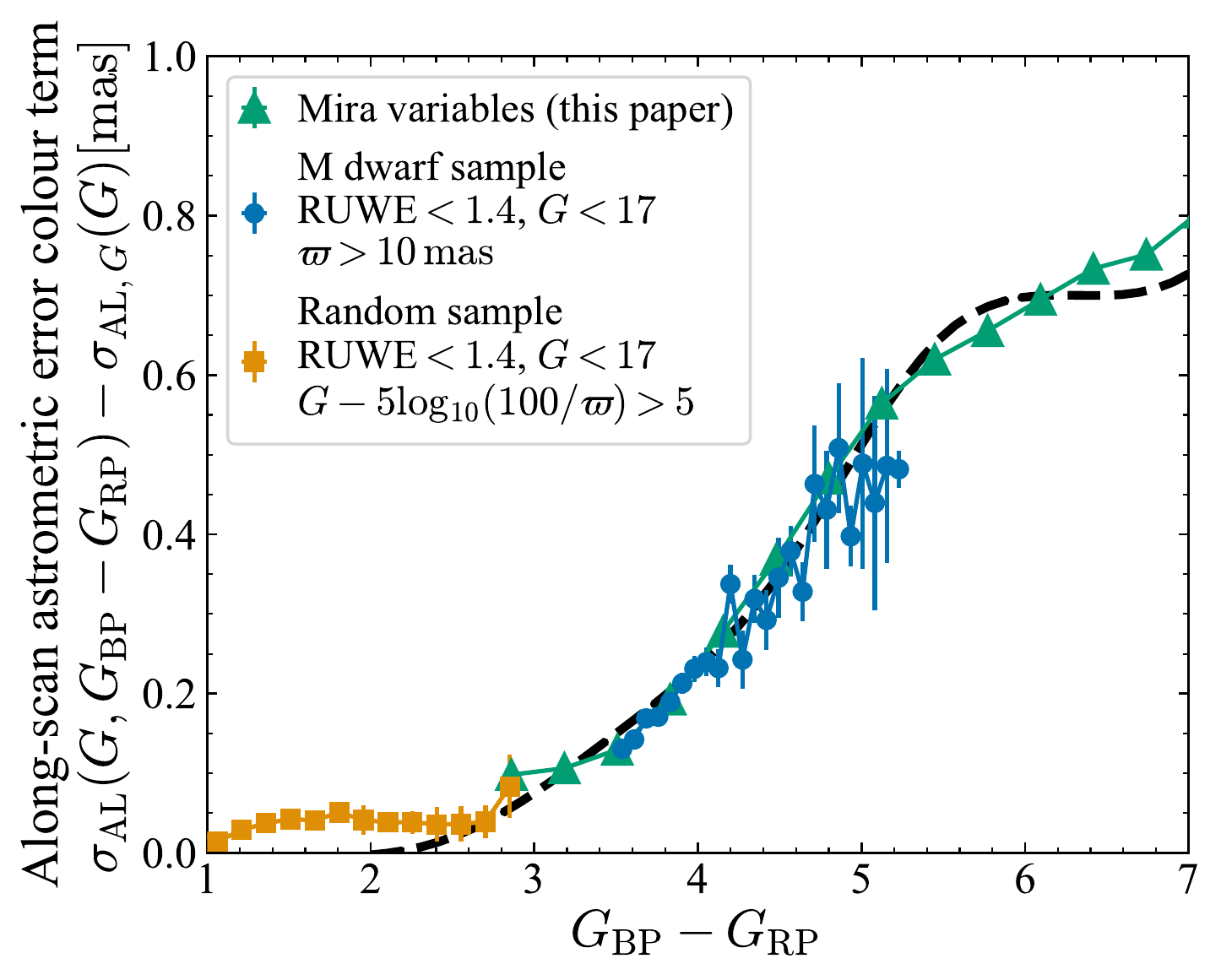}
    \caption{
    Astrometric uncertainty colour term: the along-scan astrometric uncertainty (estimated from the parallax uncertainty, $\sigma_\varpi$, and number of good along-scan observations, $N$, as $0.53\sqrt{N}\sigma_\varpi$) with the median component as a function of $G$ subtracted (from figure A.1 of \protect\cite{Lindegren2021}) as a function of colour. The median (and uncertainty in the median) of three samples are shown: the Mira variable sample used in this work as green triangles, an M dwarf sample as blue circles and a random main sequence sample as orange squares. The colour term is insignificant for $(G_\mathrm{BP}-G_\mathrm{RP})<3$ but for $(G_\mathrm{BP}-G_\mathrm{RP})>3$ the astrometric uncertainty increases as a function of colour as seen in both the M dwarf and Mira variable sample (so is not related to the photocentre wobble of the AGB stars). The black dashed line is a spline approximation used in the modelling.
    }
    \label{fig:astrometric_colour}
\end{figure}

The along-scan measurement uncertainties, $\sigma_\mathrm{AL}$ ($=\sigma_x$ for along-scan measurements and $5.65\sigma_x$ for across-scan), are assumed to be functions of $G$ and $G_\mathrm{BP}-G_\mathrm{RP}$. The $G$ dependence, $\sigma_{\mathrm{AL},G}(G)$, is extracted from figure A.1. of \cite{Lindegren2021}. As described by \cite{BelokurovBinary}, the approximation $\sigma_\mathrm{AL}=0.53\sqrt{N}\sigma_\varpi$ (where $N$ is the number of good along-scan astrometric measurements) accurately reproduces this trend. To find the colour dependence of the astrometric uncertainty, a sample of M dwarf stars is extracted from Gaia DR3 using $(G_\mathrm{BP}-G_\mathrm{RP})>3.5$, RUWE $<1.4$ and $\varpi>10\,\mathrm{mas}$. It is anticipated that the point-source astrometric solution will be appropriate for their astrometry such that any uncertainty is inherent to Gaia and not a result of other factors (e.g. photocentre wobble as in the case of AGB stars). The along-scan uncertainty for this sample is computed as $0.53\sqrt{N}\sigma_\varpi$ and then the contribution from $\sigma_{\mathrm{AL},G}(G)$ subtracted off to find the colour dependent term, $\sigma_{\mathrm{AL},G_\mathrm{BP}-G_\mathrm{RP}}$. In Fig.~\ref{fig:astrometric_colour}, this residual astrometric uncertainty colour term is shown along with equivalent for a random control sample of main sequence stars (with RUWE$<1.4$, $G<17$ and $G-5\log_{10}(100/\varpi)>5$ and the Mira variable sample used in this work. For the control sample, essentially no additional colour term is required whilst for the redder sources the astrometric uncertainty increases with increasing $G_\mathrm{BP}-G_\mathrm{RP}$. Note that both the Mira variable sample and the M dwarf sample exhibit the same trends, suggesting the astrometric uncertainty is driven mostly by intrinsic Gaia limitations and not intrinsic photocentre wobble. The black dashed line is a simple spline fit for $\sigma_{\mathrm{AL},G_\mathrm{BP}-G_\mathrm{RP}}$. For the mock observations, the uncertainties are included by scattering by $\sigma_{\mathrm{AL},G}(G)+\sigma_{\mathrm{AL},G_\mathrm{BP}-G_\mathrm{RP}}$ (inflated by a factor $5.65$ for the across-scan observations).

Given the observations $\bs{x}$ and covariance matrix $\Sigma_x=\mathrm{diag}(\bs{\sigma}_{x}^2)$, the astrometric equations are solved for the astrometric parameters using the usual weighted least-squares scheme. The $i$th row of the along-(across-)scan design matrix $\mathsf{M}_\mathrm{s}$ is
\begin{equation}
(\mathsf{M}_\mathrm{s})_i =\begin{pmatrix}\sin\phi_i&\cos\phi_i\end{pmatrix}\cdot
\mathsf{M}(t_i)
\end{equation}
and
\begin{equation}
\mathbf{A} = (\mathsf{M}_\mathrm{s}^\mathrm{T}\Sigma_x^{-1}\mathsf{M}_\mathrm{s})^{-1} \mathsf{M}_\mathrm{s}^\mathrm{T}\Sigma_x^{-1}\mathbf{x},
\end{equation}
and
\begin{equation}
\Sigma_A = (\mathsf{M}_\mathrm{s}^\mathrm{T}\Sigma_x^{-1}\mathsf{M}_\mathrm{s})^{-1}.
\end{equation}

As described in \cite{Lindegren2012}, a weighting scheme and adjustment of the noise are incorporated in the full Gaia astrometric solution. A set of weights $\bs{w}$ are first determined by finding the residuals with respect to an unweighted fit normalized by the uncertainties and then utilising equation (66) of \cite{Lindegren2012} which penalises large residuals. After this, the uncertainties $\bs{\sigma}$ are summed in quadrature with an additional term $\epsilon\bs{I}$, the astrometric excess noise, to ensure the sum of the squared residuals normalized by the square of the noise is approximately the number of degrees of freedom (no. of astrometric parameters minus the number of observations with weights $>0.2$). $\epsilon$ is found through an iterative procedure as described by \cite{Lindegren2012}. The astrometric fit is then re-performed using $\bs{\sigma}\leftarrow\bs{w}^{-1/2}\sqrt{\bs{\sigma}^2+\epsilon^2\bs{I}}$ and the $\bs{w}$ and $\epsilon$ redetermined. This iteration is repeated four times.

\subsection{AGB models}
Simple approximate models for the AGB photocentre are adopted and calibrated to the hydrodynamic simulations of \cite{Chiavassa2018}. For each on-sky dimension, $(x,y)=(\Delta\alpha\cos\delta, \Delta\delta)$, the photocentre is assumed to follow a Gaussian process e.g. $x(t)\sim\mathcal{GP}(0,K(t,t'))$. The kernel is chosen to be a sum of two kernels that represent short timescale wobbles (timescale of order weeks to months due to small convective cells in the upper atmosphere) and longer timescale wobbles (timescale of years). The Gaussian process package \texttt{celerite2} \citep{celerite1,celerite2} is used which implements a fast inversion for kernels $K$ that are sums of (complex) exponentials. A particular case is the damped simple harmonic oscillator kernel (SHO), $K_\mathrm{SHO}(\sigma,\rho,\tau)$ with standard deviation $\sigma$, period $\rho$ and damping timescale $\tau$. For simplicity, the damping timescale is set equal to the period, $\tau=\rho$. The full kernel is then given by
\begin{equation}
K = K_\mathrm{SHO}(a_\mathrm{l}R_\star/\sqrt{2}, P, P)+ K_\mathrm{SHO}(a_\mathrm{s}R_\star/\sqrt{2}, 0.1 P, 0.1 P),
\end{equation}
where the short timescale (of order weeks) is assumed to be a tenth of the longer timescale, the amplitude of the long and short timescale terms are $a_\mathrm{l}$ and $a_\mathrm{s}$ respectively and $R_\star$ is the stellar radius.

\begin{figure*}
    \centering
    \includegraphics[width=\textwidth]{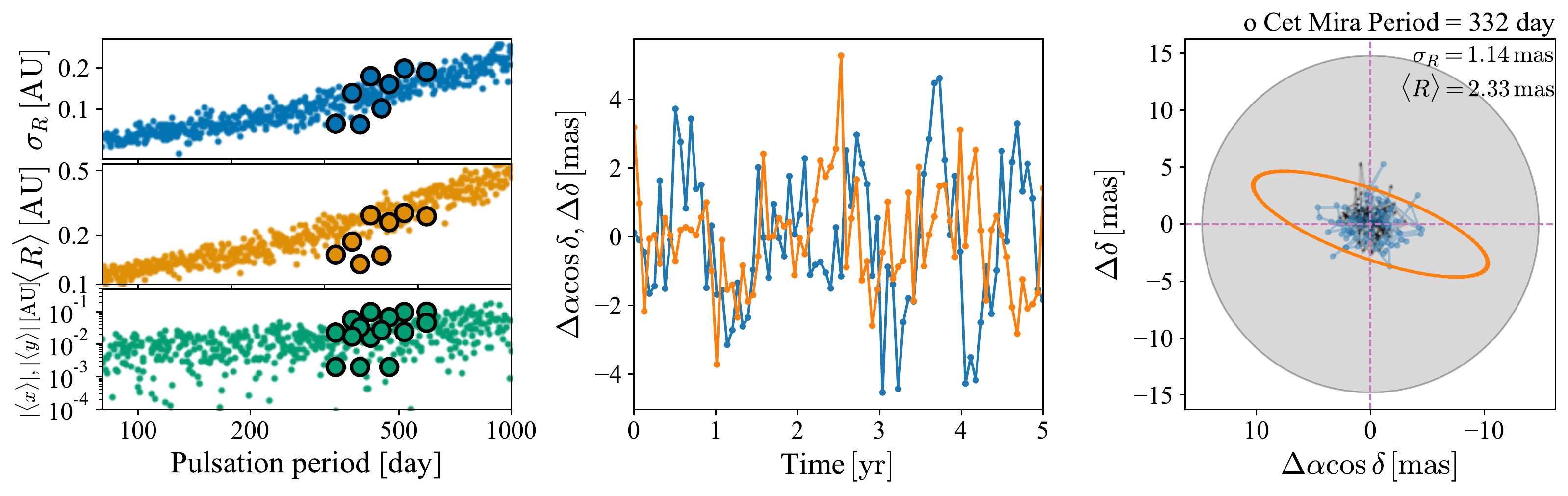}
    \caption{Approximate AGB models. The left set of panels show statistics of the centroid offset, $R=\sqrt{x^2+y^2}=\sqrt{(\Delta\alpha\cos\delta)^2+(\Delta\delta)^2}$, for the AGB models presented by \protect\cite{Chiavassa2018} as large outlined points and the approximate models used here as small points. The central panel shows one draw of the two components of the AGB centroid offset for a model of the prototypical Mira variable. The right panel shows the approximate model size of Mira along with its parallactic motion in orange (no proper motion is shown here), the centroid motion over $5$ years for the closest AGB model from \protect\cite{Chiavassa2018} in black and the centroid from the approximate models used here in blue. The statistics for these models are printed in the top right corner. For comparison, \protect\cite{Chiavassa2011} reported the measured $\sigma_R$ and $\langle R \rangle$ of Mira as $0.5\,\mathrm{mas}$ and $1.2\,\mathrm{mas}$ respectively.}
    \label{fig:mira_model_astrometry}
\end{figure*}

Draws from the Gaussian process prior are made for a set of times $\{t_i\}$ for each on-sky dimension independently (as shown in the central panel of Fig.~\ref{fig:mira_model_astrometry}). The parameters of the Gaussian process kernel are calibrated using the results presented by \cite{Chiavassa2018}. To reproduce their calculations, $80$ evenly-spaced time samples over $5$ years are used from which $\langle x\rangle$, $\langle y\rangle$, $\langle R\rangle=\langle \sqrt{x^2+y^2}\rangle$ and the standard deviation of $\sqrt{x^2+y^2}$, $\sigma_R$, are measured. Setting $a_\mathrm{l}=0.14$, $a_\mathrm{s}=0.1$ and $R_\star = 1.4\,\mathrm{AU}((P-350\,\mathrm{d})/400\,\mathrm{d}+1.4)$ gives a good match to the simulations presented by these authors as shown in the left panels of Fig.~\ref{fig:mira_model_astrometry}. In this way, the models are parametrized solely by the period, $P$. In the centre and right panels of Fig.~\ref{fig:mira_model_astrometry}, an example model is shown with $P=332\,\mathrm{day}$ appropriate for the prototypical Mira variable, Mira, with a comparison to its parallax ellipse and a close model from \cite{Chiavassa2018}. $\langle R\rangle$ and $\sigma_R$ are a factor of two larger than the measured photocentre wobble for Mira \citep{Chiavassa2011} as seen in Fig.~\ref{fig:centroid_wobble}.

\begin{figure*}
    \centering
    \includegraphics[width=\textwidth]{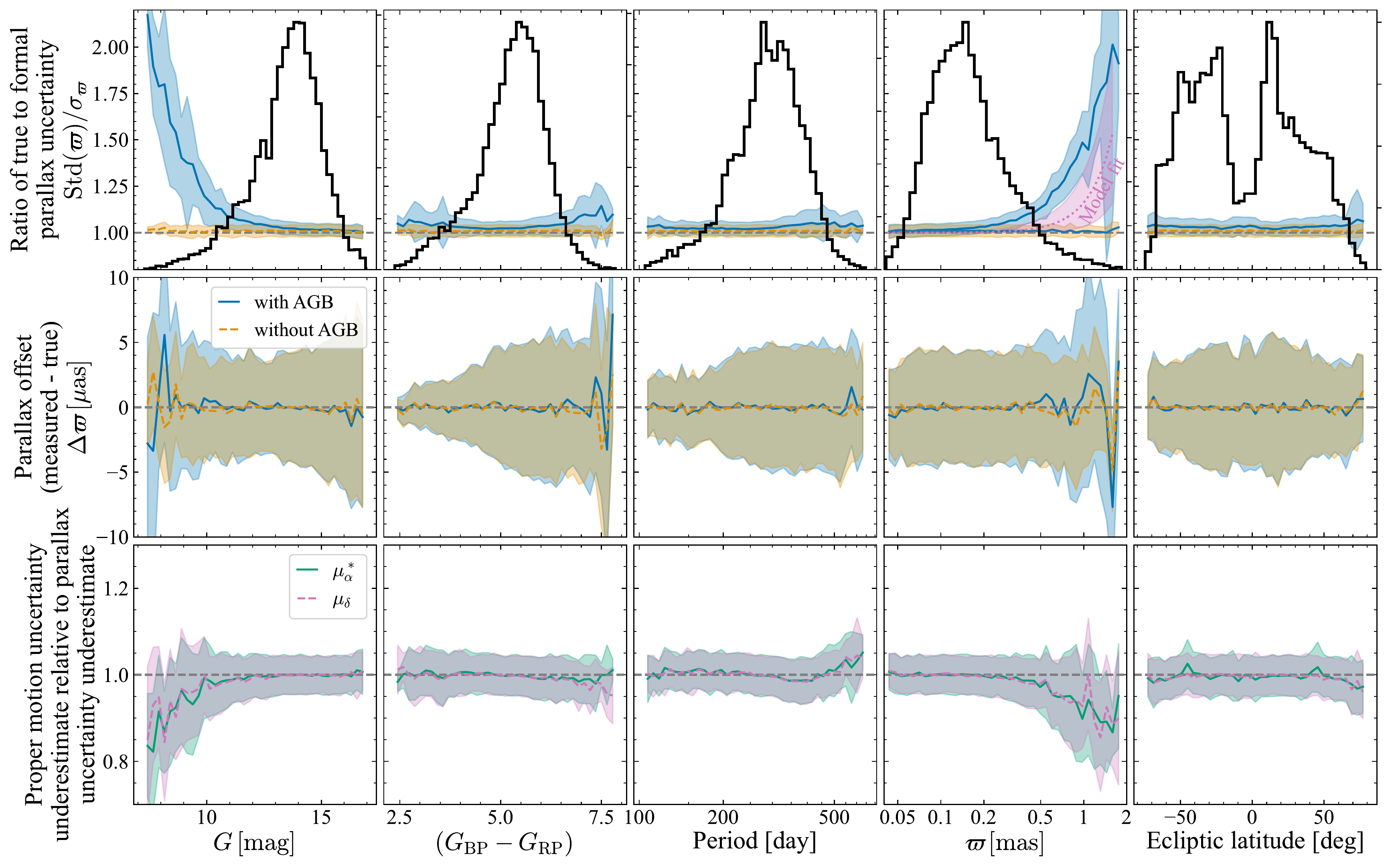}
    \caption{Quality of Gaia EDR3 astrometry for simulated Mira variable sources. The top row shows the median and $\pm1\sigma$ spread 
    of the ratio of the true parallax uncertainties (estimated from the standard deviation of the recovered parallaxes for $500$ realisations per star) to the formal parallax uncertainties. The middle row shows similar for the offset between the mean parallax averaged over realisations relative to the true parallax. The blue solid lines show results including the AGB models whilst the orange dashed lines show results without the AGB model included where the point source model is appropriate. In the top parallax panel, the result of fitting a parallax uncertainty inflation factor in the period--luminosity modelling of the form $f_\varpi(\varpi) = 1+\mathrm{exp}((5\log_{10}(100/\varpi)-b_\varpi)/a_\varpi)$ is shown in dotted pink. This gives an approximate match to the expectation from the AGB models. The black lines are histograms showing the relative fraction of objects at each coordinate. For nearby, bright stars the parallax uncertainties are underestimated but the resulting parallaxes are on average unbiased. The bottom row shows the ratio of the proper motion true to formal uncertainty relative to similar for the parallax (the top row). For bright nearby stars, the formal parallax errors are more underestimated than the formal proper motion errors.
    }
    \label{fig:orich_astrometry}
\end{figure*}

For the set of Mira variables considered in this work (both C-rich and O-rich), mock astrometry is simulated by combining the AGB models (evaluated using the period and distance from the LMC period--luminosity relation of each Mira variable) with the point-source astrometric solution (evaluated for the Mira variable on-sky position, proper motion and distance) and astrometric uncertainties (evaluated using the Mira variable $G$ and $G_\mathrm{BP}-G_\mathrm{RP}$). $500$ sets of mock astrometry are generated per star. The results of recovering the astrometric parameters using the Gaia astrometric pipeline are shown in Fig.~\ref{fig:orich_astrometry} for O-rich stars.
Both the median parallax difference with respect to the truth and the scatter of the parallaxes averaged over samples divided by the expected parallax error are shown. As a comparison, the recovery is also shown with the AGB photocentre wobble set to zero, which produces unbiased parallaxes with well-estimated uncertainties. When including the AGB models, the parallaxes remain on average unbiased (as expected if there is on average no correlation between the direction of the parallax ellipse and the photocentre wobble direction) but the parallax uncertainties are underestimated for $G\lesssim11$ and $\varpi>0.5\,\mathrm{mas}$. The proper motions are similarly unbiased and have similar underestimated uncertainties for nearby, bright stars although the underestimate is smaller than for the parallaxes.

Rerunning the period--luminosity modelling described in Section~\ref{section::model} with an additional multiplicative factor in the parallax inflation term, $f_\varpi(\varpi)$ such that $f_\varpi(G,\nu_\mathrm{eff},\varpi)\leftarrow f_\varpi(G,\nu_\mathrm{eff})f_\varpi(\varpi)$ with
\begin{equation}
f_\varpi(\varpi) = 1+\mathrm{exp}(-(5\log_{10}(100/\varpi\,[\mathrm{mas}])-b_\varpi)/a_\varpi),
\end{equation}
where $\varpi$ is the modelled parallax produces the pink curve in Fig.~\ref{fig:orich_astrometry} with $a_\varpi=(0.8\pm0.3)$ and $b_\varpi=(8.5\pm0.6)$. This is in some agreement with the expectation from the simulated AGB modelling and agrees with the analysis of bright AGB stars with VLBI from \cite{Andriantsaralaza2022} suggesting the photocentre wobble has a measurable effect for the closest stars in the sample by producing a lower parallax uncertainty than expected from comparison to the period--luminosity relation. Assuming the model expectations are an accurate representation of the sample, some of the inflation factor could be being absorbed in the colour dependence. Inclusion of this additional term does not affect the parameters of the period--luminosity relation quoted in Table~\ref{tab:plr_results} within the uncertainties.

\subsection{Epoch photometry for astrometry}
As noted by \cite{Mowlavi2018}, \cite{Lebzelter2022} and in Section~\ref{section::data}), Gaia DR2 and Gaia EDR3 use mean colours and not epoch photometry for the image parameter determination in the astrometric solutions. For six-parameter solutions, the mean colour is fitted alongside the astrometry. This leads to two effects for each CCD measurement: (i) the centroid shift with varying effective wavenumber is not corrected, and (ii) the astrometric uncertainty is misestimated.

The typical gradient of the centroid shift with effective wavenumber is estimated as $\mathrm{d}\delta/\mathrm{d}\nu_\mathrm{eff}=2.1\,\mathrm{mas}\,\mu\mathrm{m}$ given by the mean of $\texttt{pseudocolour\_error}\sqrt{\texttt{phot\_g\_n\_obs}}/\sigma_{\mathrm{AL}}$ for the six-parameter solutions where estimating $\sigma_\mathrm{AL}$ from the parallax uncertainties or the mean reported curves in \cite{Lindegren2021} makes little difference (see Section~\ref{section::data}). This estimate agrees with figure A.7 from \cite{Lindegren2021} and the earlier estimate from \cite{deBruijne}. The centroid shifts for each CCD are correlated. We ignore this complication and every time we require a CCD centroid shift we make a random draw from a Gaussian with width $\mathrm{d}\delta/\mathrm{d}\nu_\mathrm{eff}$.

The analysis of the previous section is repeated ignoring the effects of the AGB wobble. The semi-amplitude in $G$ is computed as $\sqrt{2}\texttt{std\_dev\_mag\_g\_fov}$ and the semi-amplitude in $(G_\mathrm{BP}-G_\mathrm{RP})$ as $\Delta(G_\mathrm{BP}-G_\mathrm{RP})=\sqrt{2}(\texttt{std\_dev\_mag\_bp}^2+\texttt{std\_dev\_mag\_rp}^2-2\times0.9\times \texttt{std\_dev\_mag\_bp}\times\texttt{std\_dev\_mag\_rp})^{1/2}$ to account for the fact that the BP and RP observations are highly correlated (but not perfectly hence $0.9$ correlation coefficient). The light curves then follow sinusoids of the period of each datum with randomly assigned periods and the given semi-amplitudes. The centroid shifts are added for each CCD observation and the astrometric error is assigned from the mean $G$ and $(G_\mathrm{BP}-G_\mathrm{RP})$.

Both effects lead to underestimates in the parallax uncertainty. The centroid offsets with colour produce a small underestimate of $\sim2\percent$ increasing slightly to $3\percent$ for the highest $\Delta(G_\mathrm{BP}-G_\mathrm{RP})$. This seems a very minor effect although it should be stressed the correlations between different CCD observations haven't been considered here. Combining the centroid offsets with neglecting the individual epoch uncertainties leads to an underestimate of the uncertainty of $5\percent$ increasing to $10\percent$ for the highest $\Delta(G_\mathrm{BP}-G_\mathrm{RP})$. The inclusion of epoch uncertainties has the potential to reduce the parallax uncertainties by at most $10\percent$. These effects are small and will be incorporated in the error inflation model adopted in the main body of the paper.

\section{O-rich Mira variable period--luminosity relations in LMC, SMC and Sgr dSph}\label{appendix::lmc}
\begin{figure}
    \centering
    \includegraphics[width=.919\columnwidth]{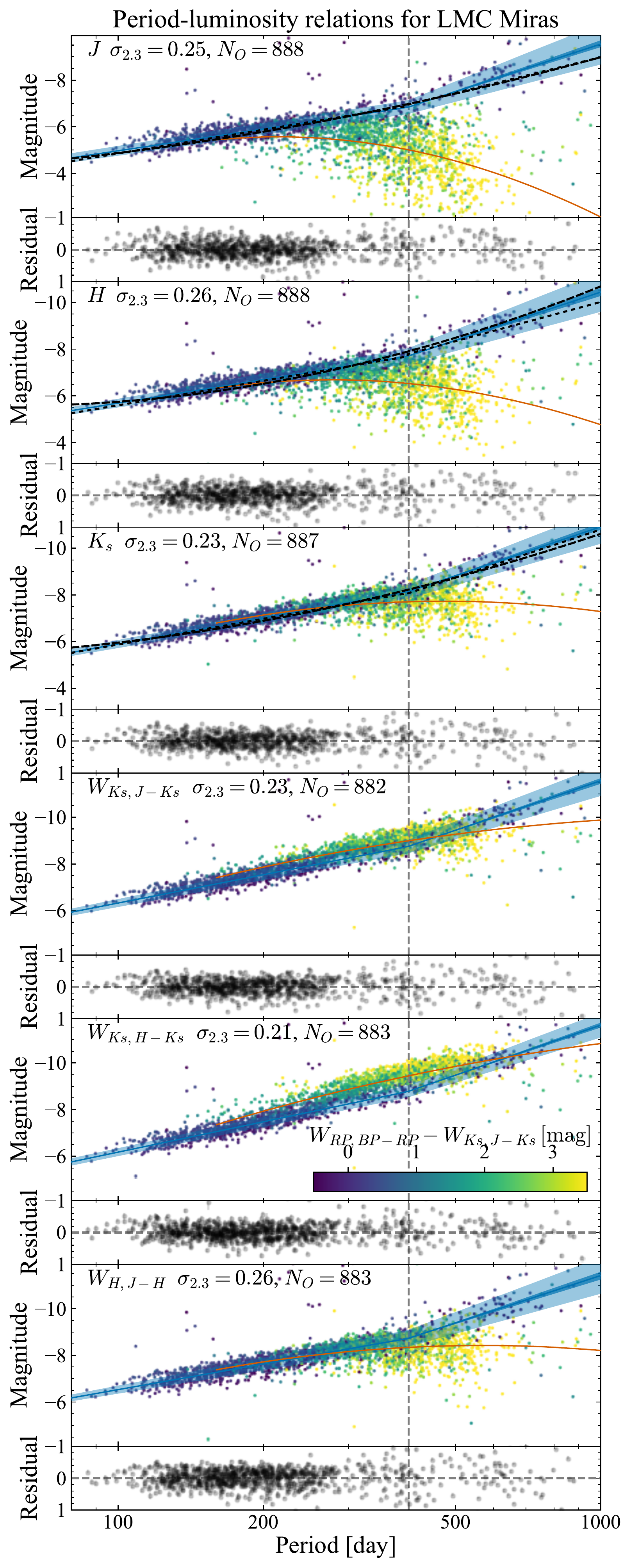}
    \caption{Period--luminosity relations for LMC Mira variables. Each set of panels corresponds to a different photometric band. Points are coloured by the difference of two Wesenheit indices, $W_\mathrm{RP,BP-RP}=G_\mathrm{RP}-1.3(G_\mathrm{BP}-G_\mathrm{RP})$ and $W_{Ks,J-Ks}=K_s-0.686(J-K_s)$, which separates O-rich and C-rich Mira variables \protect\citep{Lebzelter2018}.
    The blue (orange) line shows the best fit O-rich (C-rich) relations with $1\sigma$ bracket. The fainter blue band is the fitted scatter about the relation, which is quoted at $\log_{10}P=2.3$. Residuals of the O-rich Mira magnitudes with respect to the models are shown below each main panel. The long-dashed and short-dashed black lines are relations from \protect\cite{Yuan2018} and \protect\cite{Ita2011}. The vertical grey line marks $400$ days and the number of O-rich Mira variables used in each band is given in each panel.}
    \label{fig:pl_relations}
\end{figure}

In this appendix, period--luminosity relations for Mira variables in the Large and Small Magellanic Clouds and the Sagittarius dwarf spheroidal galaxy are given. These provide useful comparisons for the local Milky Way Mira variables and the LMC results also serve as priors in the local Milky Way model fits. The probabilistic model is similar to that used in the main body of the paper. However, a slightly different sampling procedure is employed.

For the LMC and SMC Mira variables, the dataset is composed of all long period variables from \cite{Soszynski2009} classified as Mira variables (based on a $\Delta I$ cut) and the Gaia DR3 long period variable candidates from \cite{Lebzelter2022} within $15 (5)\,\mathrm{deg}$ of the centre of the LMC (SMC) and with Fourier peak-to-peak amplitudes, $\Delta G_\mathrm{Fourier}$, (twice the \texttt{amplitude} column) greater than $0.865\,\mathrm{mag}$ \citep{Grady2019}. For both datasets, we further restrict to those stars with $G$-band amplitudes, $\Delta G>0.865\,\mathrm{mag}$. For duplicates between the two catalogues, the entries from the OGLE catalogue are preferentially retained. The combined LMC catalogue is further complemented with Gaia, 2MASS and SAGE \citep[][$3.6$, $4.5$, $5.8$ and $8.0\,\mu\mathrm{m}$]{SAGE} photometry (using a cross-matching radius of $0.4\,\mathrm{arcsec}$ for Gaia and 2MASS and $1\,\mathrm{arcsec}$ for SAGE). For those stars with Gaia DR3 BP/RP spectra, the unsupervised classification scheme from \cite{Sanders2023} based upon the UMAP algorithm is used and described in Section~\ref{section::data}. For stars fainter than $G>17.65$, BP/RP spectra are unavailable so for these stars the supervised classification algorithm from \cite{Sanders2023} is used which uses Gaia colours, 2MASS colours, periods and amplitudes.
The LMC sample of O-rich Mira variables consists of $888$ stars with 2MASS magnitudes, around $701$ of which have Spitzer magnitudes.
For the SMC sample, there are $58$ O-rich Mira variables.

The magnitudes are corrected for foreground extinction using the maps from \cite{Skowron2021} if the resolution is $<7\,\mathrm{arcmin}$ otherwise using the \cite{SFD} maps. \cite{Skowron2021} provides $E(V-I)$ computed from the red clump colours. This is converted to $E(B-V)$ in \cite{SFD} units using $E(V-I)=1.082E(B-V)_\mathrm{SFD}$ using the \cite{SFD} recalibration from \cite{SchlaflyFinkbeiner2011} of $E(B-V)=0.86E(B-V)_\mathrm{SFD}$ and the extinction law of \cite{WangChen2019} with $A_I/E(B-V)$ interpolated from their reported $\lambda^{-2.07}$ law at the effective wavelength of the OGLE I band \citep[using the SVO filter service,][]{svo1,svo2}. For other bands, the coefficients provided by \cite{WangChen2019} are used \citep[which utilise the Spitzer coefficients computed by][]{Chen2018}. \cite{Skowron2021} provide uncertainties on $E(V-I)$ using an asymmetric Gaussian distribution. The uncertainty on $E(V-I)$ is taken as the mean of the plus/minus uncertainties \citep[a $16\percent$ uncertainty in $E(B-V)_\mathrm{SFD}$ is assumed when using the results of][]{SFD}. Furthermore, the uncertainty in the coefficients $A_i/E(B-V)$ reported by \cite{WangChen2019} is propagated.

The Sgr dSph sample is composed of those stars in the Gaia sample described in Section~\ref{section::data} within $10\,\mathrm{deg}$ of Sgr dSph and between $20$ and $35\,\mathrm{kpc}$ as assessed by the LMC O-rich period--luminosity relations (found below). This leaves $103$ O-rich Mira variables. There are $2$ stars with periods greater than $400$ days.

The three samples are fitted with simple models for the extinction-corrected magnitude $m_i$ of the $i$th band vs. period $P$ (in days) relation using equation~\eqref{eqn::period_luminosity_split} described by three parameters with a scatter given by equation~\eqref{eqn::scatter_period_luminosity_split} described by a further three parameters.
In addition to this, a simple Gaussian outlier model is adopted and described by the simplex $\vartheta_i$ with $\vartheta_1+\vartheta_2=1$ and additional (large) scatter about the period--luminosity fit ($\sigma_{0,2}$). The Gaia long period variable periods can have significant uncertainties. The uncertainties are marginalized over using Monte Carlo samples in the logarithm of the frequency (to avoid negative values) and assume an uncertainty of $1$ day for the OGLE data (for which no uncertainties are provided). We restrict to stars with periods between $80$ and $1000$ days. 1 star with frequency uncertainty greater than $100\percent$ has been removed from the whole LMC sample. The total likelihood is given by the product of the likelihoods of the period $P$ and extinction-corrected magnitude $m$ for each star given corresponding uncertainties $\sigma_P$ and $\sigma_m$ (accounting for photometric uncertainty and extinction correction uncertainty):
\begin{equation}
\begin{split}
p(P,m|&\sigma_P,\sigma_{m}) \propto \sum_j \mathcal{N}(\log_{10}P_j|q,\sigma_{q}^2)\times\\
    &\sum_{k=1}^{k=2}\vartheta_k\mathcal{N}(m_i|m_{\mathrm{abs}}(P_j),\sigma_{\mu}(P_j)^2+\sigma_{m}^2+\sigma_{k,0}^2),
\end{split}
\end{equation}
where $j$ indexes the Monte Carlo sum and $k$ indexes the component (`true' Mira variable or outlier). $\mathcal{N}(x|\mu,\sigma^2)$ is a Gaussian distribution in $x$ with mean $\mu$ and variance $\sigma^2$. $\sigma_{0,1}$ is set to zero. As the uncertainties in the periods are explicitly considered, the underlying true period distribution is also fitted for, and is assumed to be Gaussian with mean $q$ and variance $\sigma_q^2$. For each band, there are ten fitting parameters: $a,b,c,\sigma_{2.3},m_{\sigma-},m_{\sigma+},\vartheta_2,\sigma_{0,2},q,\sigma_q$. Logarithmic priors are used for $\sigma_{2.3}$, $\vartheta_2$, $\sigma_{0,2}$ and $\sigma_q$. Further priors on $\sigma_{2.3}$ and $m_{\sigma-}$ are adopted to ensure $\sigma(P)>0$ for all of the Monte Carlo samples. The posterior is sampled from using the \textsc{emcee} algorithm \citep{emcee}. Note this procedure differs slightly from the model fitting procedure employed in the main body of the paper as here there is a parameter for each datum's `true' period whilst in the main body the marginalization is performed analytically in an approximate way. After fitting, the period--luminosity zeropoints $a$ are shifted by the distance modulus of galaxies considered and propagate the uncertainty. The distance modulus of the LMC is taken as $(18.477\pm 0.026)\,\mathrm{mag}$ from \cite{Pietrzynski2019}, the distance modulus of the SMC as $(18.977\pm0.032)\,\mathrm{mag}$ from \cite{Graczyk2020} and the distance modulus of the Sgr dSph as $(17.11\pm0.08)\,\mathrm{mag}$ from \cite{FergusonStrigari2020}.

\begin{figure*}
    \centering
    \includegraphics[width=\textwidth]{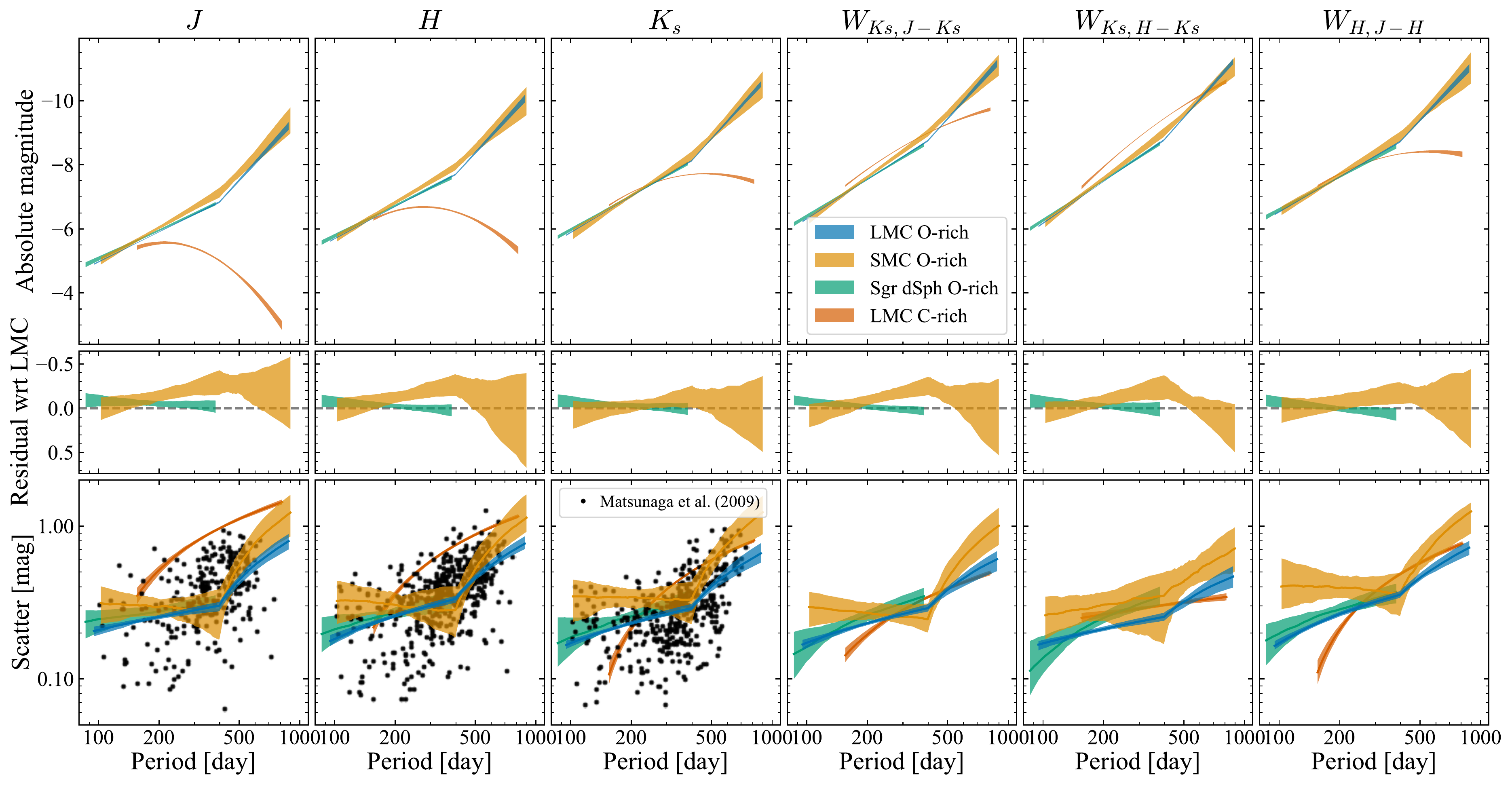}
    \caption{Period--luminosity relations and their associated scatter. The top row of panels show the period--luminosity $\pm1\sigma$ brackets for the bands as labelled above the plot and different systems as displayed in the legend. The central row shows the residuals of the SMC and Sgr dSph O-rich relations with respect to the LMC O-rich relation. The bottom panels show the scatter in the period--luminosity relations with period and the points are from the nuclear stellar disc Mira variable sample of \protect\cite{Matsunaga2009}.}
    \label{fig:plr_lmc_smc_sgr}
\end{figure*}

The results for the O-rich Mira variables are given in Table~\ref{tab:period_luminosity_relations_lmc}. Fig.~\ref{fig:pl_relations} displays the results for the LMC whilst Fig.~\ref{fig:plr_lmc_smc_sgr} shows a comparison of the LMC, SMC and Sgr results. For Sgr there are very few O-rich Mira variables with periods greater than $400$ days so only  a linear relation has been fitted to these data ($c=b$). Table~\ref{tab:period_luminosity_relations_crich} also presents results for C-rich variables. For the C-rich variables a quadratic period--luminosity relation $m_\mathrm{abs}(P)=a+b(\log_{10}P-2.3)+c(\log_{10}P-2.3)^2$ and a linear scatter model $\sigma_\mu(P)=\sigma_{2.3}+m_\sigma(\log_{10}P-2.3)$ are used. Typically for the O-rich Mira variables, the fitted Wesenheit relations are within $\sim0.05\,\mathrm{mag}$ of the relations computed from the separate bands across the range of periods, whilst for long-period C-rich Mira variables the difference can be $0.5$ to $1\,\mathrm{mag}$ due to the presence of circumstellar dust \citep{Ita2011} with a different extinction law to the interstellar dust.

We see from Fig.~\ref{fig:plr_lmc_smc_sgr} that Sgr and the LMC have very similar O-rich period--luminosity relations in all bands with Sgr possibly $\sim0.05\,\mathrm{mag}$ brighter. The SMC relations are brighter than both the LMC and Sgr relations around $P\sim300\,\mathrm{day}$ although they are steeper so at short periods ($P<200\,\mathrm{day}$) there is the suggestion they are fainter. As discussed in Section~\ref{sec::pop}, this may be related to metallicity effects as the SMC is $\sim0.5\,\mathrm{dex}$ more metal-poor than Sgr and LMC which have similar metallicities \citep[see][for a recent compilation of the APOGEE data for these systems]{Hasselquist2021}. From the lower panels of Fig.~\ref{fig:plr_lmc_smc_sgr}, we see that the scatter about the O-rich period--luminosity relation is very similar for all three systems, and similar to that reported by \cite{Matsunaga2009} for a sample of Mira variables towards the Galactic Centre (likely nearly all O-rich). In all bands, an increasing scatter with period is found for the O-rich Mira variables and for $P>400\,\mathrm{day}$ a steeper increase is required. Note that the modelling has not accounted for the scatter produced by the single-epoch photometry, which could explain this effect. As other authors have discussed, the intrinsic scatter is smallest for $K_s$ for the 2MASS bands (approximately $10\percent$ distance errors) and $[3.6]$ for the Spitzer bands ($8\percent$). As with the O-rich Mira variables, the scatter for the C-rich Mira variables increases with period, and we see that in general the scatter for the C-rich Mira variables is larger than the O-rich and increases significantly for the bluer bands ($J$). For the Wesenheit magnitudes, the scatter for O-rich and C-rich Mira variables is very similar. Using the O-rich/C-rich classification from \cite{Lebzelter2018} based upon the location of stars in the diagram of $W_\mathrm{RP,BP-RP}-W_{Ks,J-Ks}$ vs. $K_s$ results in changes in the O-rich zeropoints of $\lesssim0.005\,\mathrm{mag}$.

\begin{table*}

\rowcolors{1}{}{lightgray}
    \caption{Period--luminosity relations for LMC O-rich Mira of the form $a+b(\log_{10}P-2.3)$ for $\log_{10}P\leq2.6$ and $a+0.3b+c(\log_{10}P-2.6)$ for $\log_{10}P>2.6$ with scatter $\sigma=\sigma_{2.3}+m_{\sigma-}(\log_{10}P-2.3)$ for $\log_{10}P\leq2.6$ and $\sigma=\sigma_{2.3}+0.3m_{\sigma-}+m_{\sigma+}(\log_{10}P-2.6)$ for $\log_{10}P>2.6$. The Wesenheit indices $W_{x,y-x}=x-A_x/E(y-x)(y-x)$ use the extinction law from \protect\cite{WangChen2019} such that $A_{Ks}/E(J-K_s)=0.473$, $A_{Ks}/E(H-K_s)=1.472$ and $A_{H}/E(J-H)=1.170$. The uncertainties in $a$ include the uncertainty in the adopted distance modulus to each galaxy. The `LMC Centre' section uses the mean magnitudes for the \protect\cite{Yuan2018} LMCNISS sample. Note as these are mean magnitudes rather than single-epoch magnitudes, the scatter model is significantly narrower.}
    \centering
    \begin{tabular}{ll|cccccc}
    \hline
System&Band&$a$&$b$&$c$&$\ln\sigma_{2.3}$&$m_{\sigma-}$&$m_{\sigma+}$\\
\hline
\textbf{LMC}
&$J$&$-5.90\pm0.03$&$-3.11\pm0.07$&$-6.87\pm0.41$&$-1.37\pm0.03$&$0.15\pm0.05$&$1.48\pm0.31$\\
&$H$&$-6.69\pm0.03$&$-3.34\pm0.07$&$-6.86\pm0.43$&$-1.35\pm0.03$&$0.25\pm0.05$&$1.28\pm0.28$\\
&$K_s$&$-7.01\pm0.03$&$-3.73\pm0.06$&$-6.99\pm0.35$&$-1.46\pm0.03$&$0.20\pm0.04$&$1.10\pm0.30$\\
&$[3.6]$&$-7.41\pm0.03$&$-3.97\pm0.06$&$-7.37\pm0.30$&$-1.72\pm0.03$&$0.11\pm0.03$&$0.81\pm0.22$\\
&$[4.5]$&$-7.51\pm0.03$&$-3.83\pm0.06$&$-7.64\pm0.26$&$-1.60\pm0.04$&$0.12\pm0.05$&$0.50\pm0.17$\\
&$[5.8]$&$-7.68\pm0.03$&$-3.83\pm0.07$&$-7.81\pm0.26$&$-1.57\pm0.03$&$0.12\pm0.04$&$0.46\pm0.17$\\
&$[8.0]$&$-7.86\pm0.03$&$-3.90\pm0.07$&$-8.50\pm0.28$&$-1.49\pm0.03$&$0.11\pm0.05$&$0.47\pm0.18$\\
&$W_{Ks,J-Ks}$&$-7.53\pm0.03$&$-4.05\pm0.06$&$-6.99\pm0.34$&$-1.47\pm0.03$&$0.20\pm0.04$&$0.89\pm0.27$\\
&$W_{Ks,H-Ks}$&$-7.48\pm0.03$&$-4.32\pm0.06$&$-7.10\pm0.28$&$-1.55\pm0.03$&$0.15\pm0.04$&$0.64\pm0.22$\\
&$W_{H,J-H}$&$-7.63\pm0.03$&$-3.65\pm0.07$&$-6.77\pm0.40$&$-1.33\pm0.03$&$0.31\pm0.05$&$1.08\pm0.23$\\

\hline
\textbf{LMC Centre}&
$J$&$-5.90\pm0.03$&$-3.48\pm0.09$&$-5.90\pm0.54$&$-1.93\pm0.08$&$0.01\pm0.07$&$0.93\pm0.44$\\
&$H$&$-6.63\pm0.03$&$-3.54\pm0.11$&$-8.95\pm1.04$&$-1.88\pm0.07$&$0.19\pm0.06$&$1.54\pm0.74$\\
&$K_s$&$-6.96\pm0.03$&$-3.73\pm0.10$&$-6.79\pm0.46$&$-2.05\pm0.12$&$0.13\pm0.06$&$0.37\pm0.39$\\

\hline
\textbf{SMC}
&$J$&$-6.04\pm0.06$&$-3.58\pm0.39$&$-6.47\pm1.33$&$-1.24\pm0.15$&$-0.04\pm0.28$&$2.57\pm1.07$\\
&$H$&$-6.82\pm0.07$&$-3.84\pm0.41$&$-5.56\pm1.24$&$-1.15\pm0.17$&$-0.03\pm0.37$&$2.58\pm1.12$\\
&$K_s$&$-7.04\pm0.07$&$-4.00\pm0.42$&$-6.40\pm1.30$&$-1.08\pm0.14$&$-0.02\pm0.33$&$2.48\pm0.98$\\
&$W_{Ks,J-Ks}$&$-7.62\pm0.06$&$-4.63\pm0.33$&$-5.89\pm1.07$&$-1.28\pm0.14$&$-0.06\pm0.20$&$1.98\pm0.75$\\
&$W_{Ks,H-Ks}$&$-7.58\pm0.07$&$-4.83\pm0.40$&$-5.79\pm1.25$&$-1.16\pm0.24$&$0.11\pm0.35$&$1.08\pm0.93$\\
&$W_{H,J-H}$&$-7.68\pm0.08$&$-4.00\pm0.45$&$-6.34\pm1.41$&$-0.91\pm0.16$&$-0.06\pm0.43$&$2.36\pm1.04$\\

\hline
\textbf{Sgr}
&$J$&$-5.95\pm0.09$&$-2.97\pm0.21$&$-$&$-1.31\pm0.08$&$0.10\pm0.13$&$-$\\
&$H$&$-6.72\pm0.09$&$-3.17\pm0.19$&$-$&$-1.29\pm0.08$&$0.19\pm0.14$&$-$\\
&$K_s$&$-7.04\pm0.09$&$-3.59\pm0.20$&$-$&$-1.37\pm0.07$&$0.20\pm0.14$&$-$\\
&$W_{Ks,J-Ks}$&$-7.55\pm0.09$&$-3.86\pm0.18$&$-$&$-1.35\pm0.08$&$0.30\pm0.14$&$-$\\
&$W_{Ks,H-Ks}$&$-7.51\pm0.09$&$-4.15\pm0.19$&$-$&$-1.39\pm0.08$&$0.35\pm0.15$&$-$\\
&$W_{H,J-H}$&$-7.62\pm0.09$&$-3.43\pm0.22$&$-$&$-1.24\pm0.08$&$0.28\pm0.16$&$-$\\

\hline
    \end{tabular}
    \label{tab:period_luminosity_relations_lmc}
\end{table*}

\begin{figure*}
    \centering
    \includegraphics[width=\textwidth]{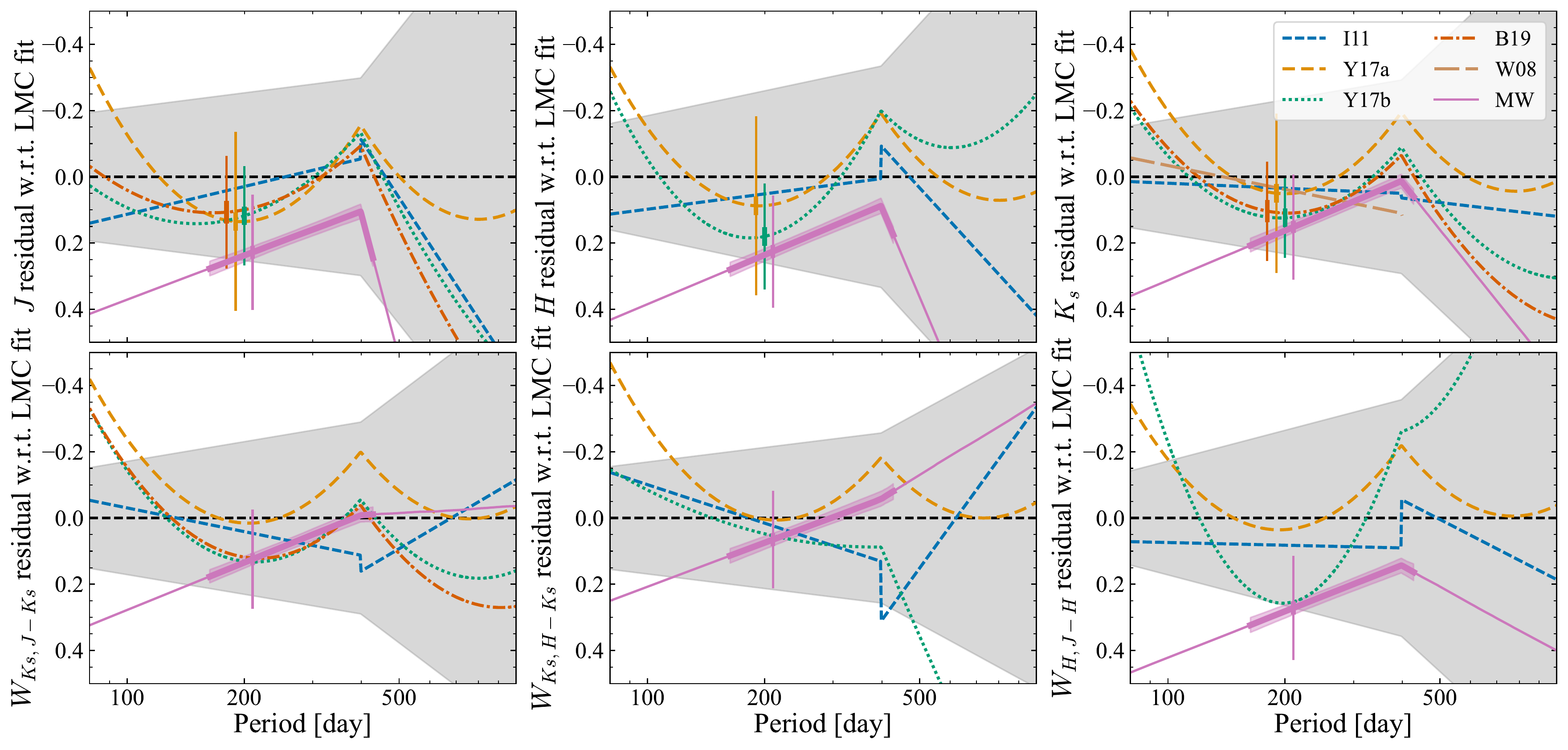}
    \caption{Comparison between the LMC period--luminosity relations fitted in this work (see Table~\ref{tab:period_luminosity_relations_lmc}) and those in the literature. The grey band is the fitted dispersion of the models about the period--luminosity relation. For comparison, results from \protect\citet[][I11]{Ita2011},
    \protect\citet[][Y17a]{Yuan2017},
    \protect\citet[][Y17b]{Yuan2018} and
    \protect\citet[][B19]{Bhardwaj2019}, and the Milky Way period--luminosity relations derived here (MW, the thicker part of the line shows the region covered by the 5th to 95th percentile of the data and the shaded region is the $\pm1\sigma$ uncertainty) and by \protect\citet[][W08, derived in the SAAO $K$ band]{Whitelock2008} are shown. The literature Wesenheit relations are simply computed from the single-band relations. The narrow errorbars show the scatter in each model at $P=200$ days (offset for clarity; note Y17b and B19 use multi-epoch mean magnitudes and hence the spread is smaller), and the thicker errorbars show the scatter in the mean period--luminosity relation (not including any scatter from the reference LMC fit).
    }
    \label{fig:comp}
\end{figure*}

\begin{table*}

\rowcolors{1}{}{lightgray}
    \caption{As Table~\ref{tab:period_luminosity_relations_lmc} but for LMC C-rich Mira variables. A quadratic period--luminosity relation $a+b(\log_{10}P-2.3)+c(\log_{10}P-2.3)^2$ and a linear model for the scatter $\sigma=\sigma_{2.3}+m_\sigma(\log_{10}P-2.3)$ are adopted.}
    \centering
    \begin{tabular}{l|ccccc}
    \hline
Band&$a$&$b$&$c$&$\ln\sigma_{2.3}$&$m_\sigma$\\
\hline
$J$&$-5.56\pm0.05$&$-0.46\pm0.34$&$7.64\pm0.88$&$-0.67\pm0.05$&$1.52\pm0.12$\\
$H$&$-6.56\pm0.04$&$-1.74\pm0.24$&$6.20\pm0.67$&$-1.03\pm0.06$&$1.31\pm0.08$\\
$K_s$&$-7.14\pm0.03$&$-3.15\pm0.13$&$4.24\pm0.38$&$-1.53\pm0.05$&$0.97\pm0.04$\\
$[3.6]$&$-7.60\pm0.05$&$-5.21\pm0.28$&$0.67\pm0.55$&$-1.29\pm0.06$&$0.15\pm0.05$\\
$[4.5]$&$-7.70\pm0.06$&$-5.47\pm0.37$&$-1.77\pm0.70$&$-0.76\pm0.05$&$-0.33\pm0.07$\\
$[5.8]$&$-7.88\pm0.06$&$-5.08\pm0.43$&$-4.94\pm0.80$&$-0.56\pm0.03$&$-0.44\pm0.04$\\
$[8.0]$&$-8.06\pm0.07$&$-5.40\pm0.54$&$-6.40\pm0.97$&$-0.47\pm0.03$&$-0.53\pm0.05$\\
$W_{Ks,J-Ks}$&$-7.86\pm0.03$&$-4.45\pm0.15$&$2.26\pm0.38$&$-1.62\pm0.06$&$0.48\pm0.04$\\
$W_{Ks,H-Ks}$&$-7.93\pm0.04$&$-5.67\pm0.21$&$2.15\pm0.44$&$-1.33\pm0.05$&$0.12\pm0.05$\\
$W_{H,J-H}$&$-7.72\pm0.03$&$-3.11\pm0.14$&$3.46\pm0.42$&$-1.56\pm0.07$&$0.91\pm0.06$\\
    \hline
    \end{tabular}
    \label{tab:period_luminosity_relations_crich}
\end{table*}

\subsection{Choice of functional form}
When fitting the period--magnitude relations there is significant freedom over the choice of functional form to use. Here, a broken continuous linear relation has been chosen as it provides better fits than a quadratic relation over the full period range given the same number of parameters (for the $K_s$ band there is $\Delta$ log-likelihood of $\sim3$). Adopting a quadratic relation beyond the break-point does not give a significant  improvement in $\Delta$AIC and $\Delta$BIC for the $JHK_s$ fits ($\lesssim1$). The small improvement appears to be due to a slight downturn at long periods possibly due to the presence of very dusty objects. A further advantage of the adopted functional form is that the linear relation for $\log_{10}P<2.6$ can be more directly compared to the work of other authors using only the shorter period Mira variables which are more reliable distance indicators \citep{Huang2020}. Finally, the continuity of the relation at $\log_{10}P=2.6$ is physically appropriate unless the onset of hot-bottom burning is very abrupt and the long and short period Mira variables can be treated as entirely different types of star. Similarly, with the period--scatter relation a continuous broken linear relation has been adopted. Slightly better fits are obtained using non-continuous transition in the scatter at $\log_{10}P=2.6$. Adopting a period--scatter relation linear below $\log_{10}P=2.6$ and constant above, there are $\Delta$ log-likelihood improvements of around $10$. The break-point in the period-magnitude and period--scatter relations can also be fitted as a free parameter. For all bands, the best-fitting break-point is less than $\log_{10}P=2.6$ and typically the best fit is $\log_{10}P=2.42$--$2.58$ in agreement with the results from \cite{Bhardwaj2019}. The $\log_{10}P=2.6$ or $P\approx400\,\mathrm{day}$ break-point is retained to align better with previous work. Despite these variations to the model that could offer some small improvements to the fits for the LMC, the default model is chosen due to its easy comparison with previous measurements and the continuous properties of the functional forms.

\subsection{Comparison with previous estimates}
Fig.~\ref{fig:comp} shows a comparison between the LMC O-rich Mira variable period--luminosity relations and previous relations reported in the literature and the Milky Way relations derived in the main body of the paper. On the whole, the relations are very similar to those previously reported except the relations tend to be steeper at the long-period end than previous results \citep[the same is true in the Spitzer bands when comparing the results of Table~\ref{tab:period_luminosity_relations_lmc} with][]{Ita2011}. At the short period end ($P<400$ days) there is very good agreement with the \cite{Ita2011} relations, possibly because they also use a linear relation in this regime and utilise a large fraction of 2MASS data. The quadratic relations from \cite{Yuan2017}, \cite{Yuan2018} and \cite{Bhardwaj2019} tend to be fainter around the characteristic period of $\log_{10}P=2.3$ than the relations, particularly in the $H$ band. This $H$-band discrepancy of $\sim0.18\,\mathrm{mag}$ with respect to the \cite{Yuan2018} approach. Some of the discrepancy could arise from using different parametrizations of the period--luminosity relation: linear vs. quadratic. In Fig.~\ref{fig:comp} we see the quadratic relations almost `envelope' the linear relations over the $100$ to $400$ day period range suggesting no difference on average. However, using the \cite{Yuan2018} linear period--luminosity relation still results in $\sim0.13\,\mathrm{mag}$ difference in $H$. This could have important implications for any Hubble $F160W$ calibrations. The cause for such a discrepancy is now investigated.

The sample utilised by \cite{Yuan2018} is taken from the LMCNISS survey \citep{Macri2015} which surveyed the central 18 square degrees of the LMC, whilst the sample extends across the entire LMC ($\sim 360\,\mathrm{deg}^2$). One cause for concern is that the large angular extent of the sample is introducing biases, particularly as Mira variables possibly trace the LMC disc. The prescription of \cite{vanderMarelCioni2001} is followed assuming the stars lie in a disc with position angle $159.59\,\mathrm{deg}$ and inclination $33.14\,\mathrm{deg}$ \citep{Mackey2016}, and the centre of the LMC is at $(\alpha,\delta)=(82.25, -69.5)\mathrm{deg}$ \citep{vanderMarelCioni2001}. This causes the period--luminosity relations to shift $\sim 0.015\,\mathrm{mag}$ brighter independent of the band, within the reported error on the LMC zeropoint \citep{Pietrzynski2019}.

A further concern is that there are population differences between the central LMC and the extended LMC Mira variable samples, possibly linked to age or metallicity effects (see Section~\ref{sec::pop}). The model has instead been applied to the 2MASS magnitudes of the \cite{Yuan2018} sample (removing those flagged as having possibly unreliable phases) finding that the period--luminosity relations are $(0.05, 0.05, 0.02)\,\mathrm{mag}$ fainter for $(J,H,K_s)$ than using the extended sample. Using the \cite{Yuan2018} mean magnitudes (derived from LMCNISS measurements on the 2MASS system) there are similar shifts $(0.06, 0.05)\,\mathrm{mag}$ fainter in $(H,K_s)$ but essentially no shift in $J$ (the uncertainties from the LMCNISS zeropoint calibration are ($0.011,0.018,0.014$) in $(J,H,K_s)$). These results are reported in Table~\ref{tab:period_luminosity_relations_lmc}. This explains part of the discrepancy shown in Fig.~\ref{fig:comp} but not the whole effect (still a $\sim0.07\,\mathrm{mag}$ discrepancy in $H$). Using the relations from \cite{Qin2018}, the magnitude difference between the inner LMC population and the more extended population is consistent with the central regions being older and/or more metal-rich than the outer parts. The LMC is believed to have a negative metallicity gradient \citep{Grady2021,Choudhury2021}. For the full sample, each Mira variable is assigned a metallicity based on its on-sky location using the maps of \cite{Grady2021}. Their estimates are used for stars with $(J-H)<1$, $(J-K_s)<1.25$ and $K_s>12.5$ and the maps are binned in $\sim0.5\times0.5\,\mathrm{deg}^2$ pixels and smoothed using a Gaussian with a width of one pixel. It should be noted that \cite{Grady2021} find a stronger radial metallicity gradient than \cite{Choudhury2021}. When an additional metallicity term, $b_Z([\mathrm{Fe}/\mathrm{H}]+0.6)$, in the period--luminosity relations is used, weak evidence for a metallicity dependence $b_Z=((0.08\pm0.04),
(0.06\pm0.04),
(0.05\pm0.04)
)\,\mathrm{mag}/\mathrm{dex}$ in $(J,H,K_s)$ is found.  This suggests that metal-rich Mira variables are fainter than the metal-poor counterparts. This is approximately consistent with the theoretical results from \cite{Qin2018} although a negative metallicity gradient with magnitude for $K_s$ is not found. This may be because such an approach fails to account for any age dependence and there may well be age-metallicity correlations within the LMC. If an additional colour term $b_{JK}(J-K_s)$ in the $K_s$ period--luminosity relation is fitted for, the best fit is $b_{JK}=(0.45\pm0.07)$. Again, this suggests redder (more metal-rich) Mira variables are fainter than their metal-poor counterparts. The reason for a more definite signal compared with the metallicity fits is perhaps because the star-by-star variation is considered rather than the average variation due to the on-sky position.

Using the O-rich/C-rich definitions from the Gaia DR3 data \citep{Sanders2023}compared to \cite{Soszynski2009} results in no significant difference in the period--luminosity relation (shifts of $\lesssim0.01\,\mathrm{mag}$). Following \cite{Ita2011} and utilising the IRSF magnitudes transformed to the 2MASS bands \citep{Kato2007} when available also produces very little difference ($\sim0.015\,\mathrm{mag}$ in all bands).

One cause for the remaining discrepancy of $\sim 0.1$ between the measurements and those of \cite{Yuan2018} could be the different handling of extinction. The fits on the \cite{Yuan2018} sample with 2MASS magnitudes have been rerun using the \cite{Haschke2011} extinction maps and a \cite{Cardelli1989} extinction law using the coefficients reported by \cite{Bhardwaj2019} but find differences $\lesssim0.01\,\mathrm{mag}$. Completely neglecting extinction produces a shift fainter of $(0.11, 0.02, 0.02)$ in $(J,H,K_s)$ when compared to the fits using the \cite{Yuan2018} mean magnitudes. \cite{Yuan2018b} report biases of $0.03\,\mathrm{mag}$ can arise from the computation of the mean magnitude when doing piecewise template fitting compared to flux-means from sinusoidal fits although here the same magnitudes as \cite{Yuan2018} have been used.

%%%%%%%%%%%%%%%%%%%%%%%%%%%%%%%%%%%%%%%%%%%%%%%%%%

% Don't change these lines
\bsp	% typesetting comment
\label{lastpage}
\end{document}